\documentclass[prd,nofootinbib,showpacs,preprint,floatfix]{revtex4}
\usepackage{graphicx}
\usepackage{epsfig}
\usepackage{amsmath}
\usepackage{amsfonts}
\usepackage{amssymb}
\usepackage{url}
\usepackage{hyperref}
\usepackage{subfigure}
\usepackage{color}
\usepackage{cancel}
\newcommand{\beqa}{\begin{eqnarray}}
\newcommand{\eeqa}{\end{eqnarray}}
\newcommand{\beq}{\begin{equation}}
\newcommand{\eeq}{\end{equation}}

\newcommand{\nn}{\nonumber}
\newcommand{\bmt}{\begin{pmatrix}}
\newcommand{\emt}{\end{pmatrix}}
\usepackage[toc,page]{appendix}
\usepackage{comment}
\newcommand{\be}{\begin{equation}}
\newcommand{\ee}{\end{equation}}
\newcommand{\bea}{\begin{eqnarray}}
\newcommand{\eea}{\end{eqnarray}}

\begin{document}
\title{Model independent investigation of  rare semileptonic  $b \to u l \bar{\nu}_l$ decay processes }
\author{Suchismita Sahoo, Atasi Ray and Rukmani Mohanta }
\affiliation{\,School of Physics, University of Hyderabad,
              Hyderabad - 500046, India  }
\begin{abstract}
Motivated by the recent observation of lepton  universality violation in the flavour changing charged current transitions $b \to c l \bar{\nu}_l$, we intend to scrutinize the lepton non-universality  effects in  rare semileptonic $B$ meson  decays involving the quark level transitions $b \to u l \bar{\nu}_l$. In this regard, we envisage the model-independent approach and consider the  generalized effective Lagrangian  in the presence of new physics and constrain the new  parameters by using the experimental branching fractions of $B_u^+ \to  l^+ \nu_l$ and $B^- \to \pi^0 \mu^- \bar \nu_\mu$  processes, where $l=e, \mu, \tau$. We then  estimate the branching ratios and forward-backward asymmetries of   $B_{(s)} \to P (V)l \bar{\nu}_l$  processes, where $P(=K, \pi, \eta^{(\prime)})$ denotes the pseudoscalar meson and $V(=K^*, \rho)$ is  the vector meson.  We also find out various lepton non-universality parameters in these processes in the presence of new physics.

\end{abstract}
\maketitle
\section{Introduction}
In recent times, flavour physics has become quite interesting as several deviations at the level of  $(2-4)\sigma$ have persistently  been observed in  semileptonic $B$ decays.  Specifically,  the LHCb experiment has observed several anomalies in the rare semileptonic
$B$ decays driven by the  flavour changing neutral current (FCNC) $b \to s$ transitions. The most leading ones are the observation of $3.7 \sigma$ deviation in the angular observables $P_5^\prime$ \cite{P5p-exp, P5p},  the  decay rate  of $\bar B \to \bar K^{(*)} \mu^+ \mu^-$ mode \cite{Kstar-decayrate} and also the $3\sigma$ \cite{phi-decayrate} discrepancy in the decay rate of $B_s \to \phi \mu^+ \mu^-$ process in the low $q^2$ region. Besides these anomalies, recently LHCb and $B$ factories have observed the violation of lepton flavour universality   in $B \to D^{(*)} l \bar{\nu}_l$ and $B \to K^{(*)} l^+ l^-$ processes, which comprises some additional tension. 
 The lepton non-universality (LNU) parameter ($R_K$),  defined as the ratio of the branching fractions of  $B^+ \to K^+ \mu^+ \mu^-$ over $B^+ \to K^+ e^+ e^-$ and its measured value in the low $q^2 \in [1,6]$ region \cite{RK-exp}
\bea
R_K^{\rm Expt} = \frac{{\rm BR}(B^+  \to K^+ \mu^+ \mu^-)}{{\rm BR}(B^+ \to K^+ e^+ e^-)} =0.745^{+0.090}_{-0.074} \pm 0.036 ,
 \eea
has $2.6 \sigma$ deviation from the corresponding SM result  $R_K^{\rm SM} = 1.0003 \pm 0.0001$ \cite{RK-SM}.
In addition, very recently LHCb Collaboration has also  reported discrepancy  of  $2.2 \sigma$  in $R_{K^*}$ \cite{RKstar-exp} 
\bea
R_{K^*}^{\rm Expt} =\frac{{\rm BR}(B \to K^* \mu^+ \mu^-)}{{\rm BR}(B \to K^* e^+ e^-)}=  0.660^{+0.110}_{-0.070} \pm 0.024 ,
\eea
from the corresponding SM prediction $R_{K^*}^{\rm SM}= 0.92 \pm 0.02$  \cite{RKstar-SM}  in the $q^2 \in [0.045, 1.1]~{\rm GeV}^2$ bin and   $2.4 \sigma $  discrepancy \cite{RKstar-exp}
\bea
R_{K^*}^{\rm Expt} = 0.685^{+0.113}_{-0.069} \pm 0.047, 
\eea
 has been found  in  $q^2 \in [1.1, 6]~{\rm GeV}^2$ region from its SM predicted value $R_{K^*}^{\rm SM}= 1.00 \pm 0.01$  \cite{RKstar-SM}.
 
Analogously, in the charged current transition processes mediated through $b \to c \tau \bar{\nu}_\tau$,  LHCb as well as both the $B$ factories Belle and BaBar have measured  the LNU parameter $R_{D^{(*)}}$  in   $B \to D^{(*)} l \bar{\nu}_l$ decay processes  and the measured values \cite{RD-BaBar,RD-exp, RDstar-LHCb} 
\bea
&& R_D^{\rm Expt}= \frac{{\rm BR}(B  \to D \tau \nu_l)}{{\rm BR}(B  \to D l \nu_l)} =0.397 \pm  0.040 \pm 0.028\;, \\
 && R_{D^*} ^{\rm Expt}=\frac{{\rm BR}(B  \to D^* \tau \nu_l)}{{\rm BR}(B  \to D^* l \nu_l)}= 0.316 \pm 0.016 \pm 0.010\;,
 \eea   
have respectively $1.9\sigma$ and $3.3\sigma$ deviation from the corresponding SM predictions \cite{RD-SM, RDstar-SM}
\bea
 R_D^{\rm SM} = 0.300 \pm 0.008\;, ~~~~~~~~~
 R_{D^*}^{\rm SM} = 0.252 \pm 0.003\;. 
\eea
In this context, we wish to explore the possibility of observing LNU parameters and other  asymmetries in the rare semileptonic $b \to u l \bar{\nu}_l$ decay processes,  in order to corroborate the observed results on  lepton non-universality.

In the SM, the  $V-A$ current structure of the weak interactions  describes various charged current interactions for  all three generation of quarks and leptons to a high precision. However, the recent experimental
data indicates that among all the leptonic and semileptonic decays of $B$ mesons, the  decay processes involving third generation of fermions in  the final state are comparatively  less precise than the first two generations.  The coupling of the third generation fermions to the electroweak gauge sector is relatively  stronger due to the heavier mass of the tau lepton  and thus, more sensitive to new physics (NP) which could modify the $V - A$ structure of the SM. The decays with third generation fermions in the final state are sensitive to non-SM contributions arising from the violation of LFU, hence, these processes could be ideally suited for  probing the  NP signature.   In this respect,  the study of $B \to (\pi, \rho, \eta^{(\prime)}) l \bar{\nu}_l$ and $B_s \to K^{(*)} l \bar{\nu}_l$ charged current processes, involving the  quark level transitions $b \to u$ would be quite interesting to test the lepton flavour non-universality.  In this paper, we adopt the model-independent approach to analyze the effect of NP in the rare semileptonic $b \to u l \bar{\nu}_l$ decay processes. For this purpose, we consider the  generalized effective Lagrangian, including  the possible new  parameters allowed by Lorentz invariance. We constrain the new  coefficients by using the experimental data on the branching fractions of $B_u^+ \to l^+ \nu_l$ processes. We then compute the branching ratios, forward-backward asymmetries and various LNU  parameters of   semileptonic $B \to (\pi, \rho, \eta^{(\prime)}) l \nu_l$ and $B_s \to K^{(*)}l \nu_l$ processes.   Although these processes have been extensively studied  in the literature \cite{Bourrely, Dutta1,  soni, Wang, Wang2, Feldmann, Bern,  etaprime, eta, eta-etaprime, Ruckl}, in the context of various new physics models and also in model-independent way, but the search for lepton nonuniversality parameters are not being explored.

The outline of the paper is as follows. In section II, we describe the most general effective Lagrangian responsible for the $b \to u l \bar{\nu}_l$ processes. We also show the constraints on the new parameters by using the branching ratios of $B_u^+ \to l^+ \bar{\nu}_l$ processes. The constraint on new physics couplings from the $B^- \to \pi^0 \mu^- \bar \mu_\nu$ process is presented in section III.   We  also estimate the branching ratios, forward-backward asymmetries and  the LNU   parameters of the $B \to P l \bar{\nu}_l$  processes, where $P(=K, \pi, \eta^{(\prime)})$ represents the pseudoscalar meson,  in section III. In section IV, we study the rare semileptonic $B \to V l \bar{\nu}_l$  processes, where $V(=K^*, \rho)$ denotes the vector meson. Our findings are summarized in section V. 
\section{General effective Lagrangian for $b \to u l \bar{\nu}_l$ transitions}
The most general effective Lagrangian for $b \to u l \bar{\nu}_l$ process is given by \cite{Lagrangian}
\begin{eqnarray}
\mathcal L_{\rm eff} &=&
-\frac{4\,G_F}{\sqrt{2}}\,V_{u b}\,\Bigg\{(1 + V_L)\,\bar{l}_L\,\gamma_{\mu}\,\nu_L\,\bar{u}_L\,\gamma^{\mu}\,b_L +
V_R\,\bar{l}_L\,\gamma_{\mu}\,\nu_L\,\bar{u}_R\,\gamma^{\mu}\,b_R 
\nn \\
&&+
S_L\,\bar{l}_R\,\nu_L\,\bar{u}_R\,b_L +
S_R\,\bar{l}_R\,\nu_L\,\bar{u}_L\,b_R + 
T_L\,\bar{l}_R\,\sigma_{\mu\nu}\,\nu_L\,\bar{u}_R\,\sigma^{\mu\nu}\,b_L  \Bigg\} + {\rm h.c.}\,,\label{ham}
\end{eqnarray}
where $G_F$ is the Fermi constant, $V_{ub}$ is the Cabibbo-Kobayashi-Maskawa (CKM)  matrix element and $q_{L(R)}=L(R) q$ are the chiral quark fields with $L(R)=(1\mp \gamma_5)/2$ as the projection operator.  Here  $V_{L, R}$, $S_{L, R}$ and $T_L$ are the    vector, scalar and tensor new physics  couplings associated with the left-handed neutrinos, which are zero in the SM. 
The constraint on the  new  coefficients obtained  from the leptonic $B_u^+ \to l^+ \nu_l$ processes  are discussed in the subsection below.
\subsection{Constraints on new couplings from rare leptonic  $B_u^+ \to l^+ \nu_l$ processes}
\begin{figure}[h] 
\centering
\includegraphics[scale=0.55]{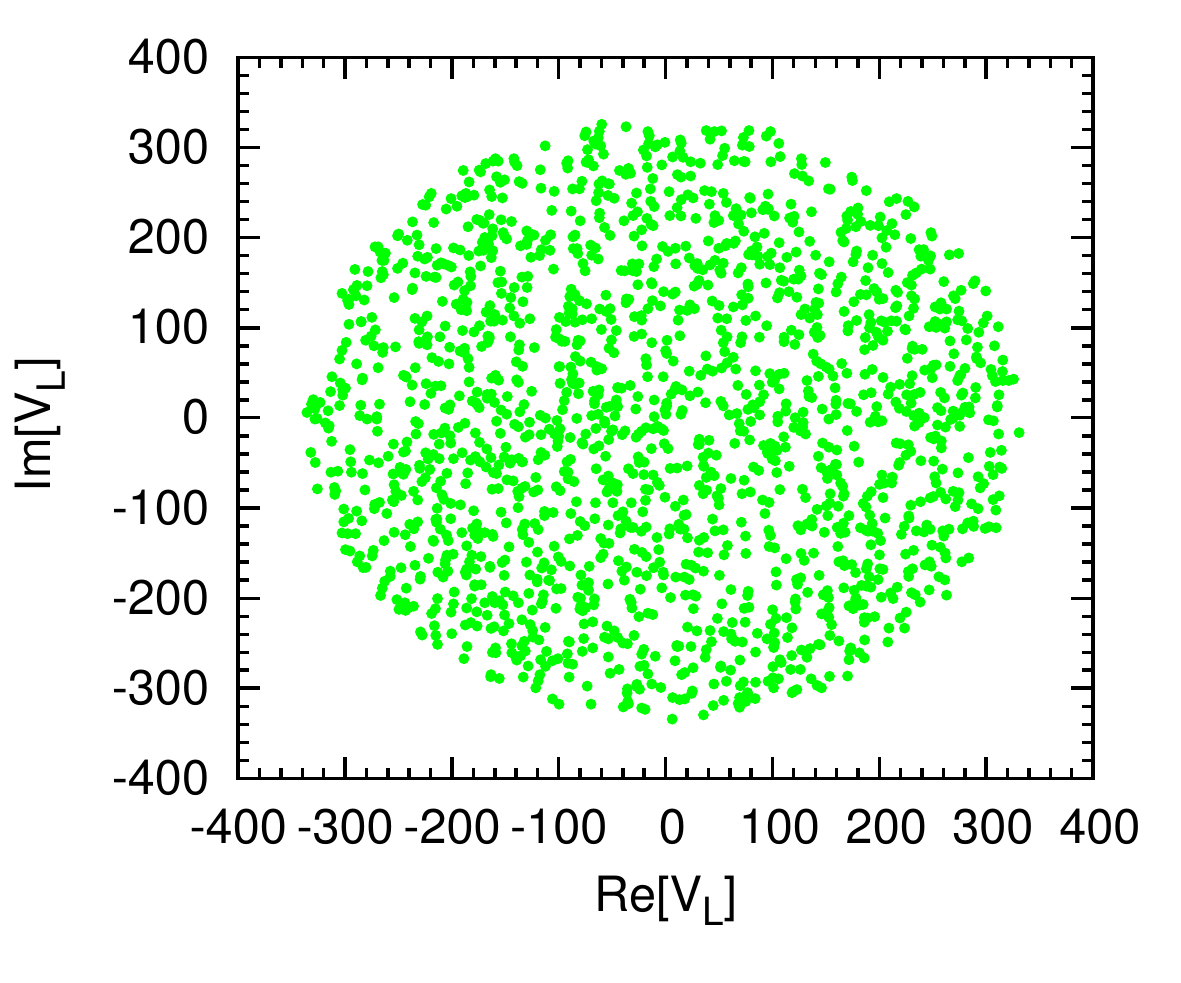}
\quad
\includegraphics[scale=0.55]{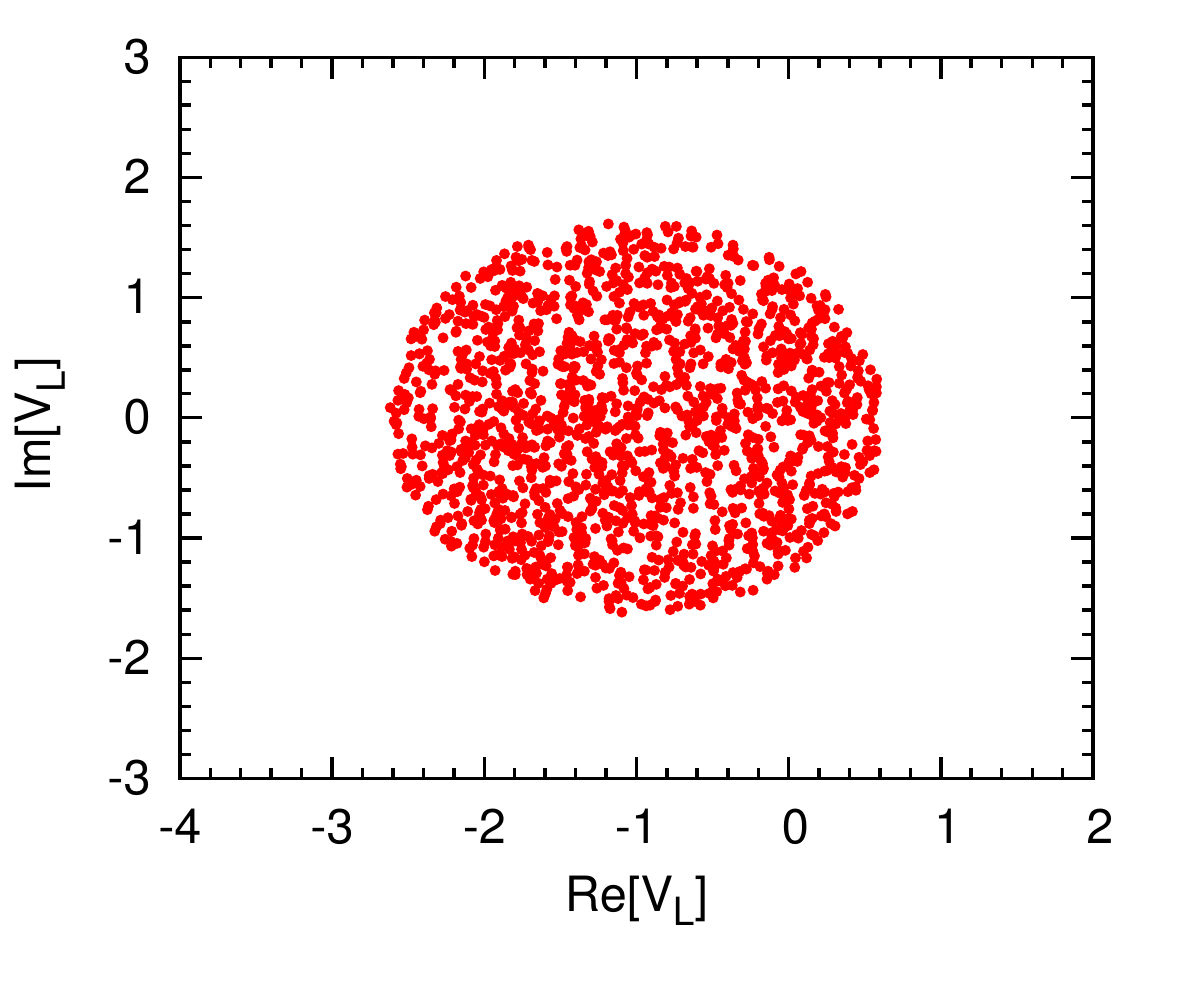}
\quad
\includegraphics[scale=0.55]{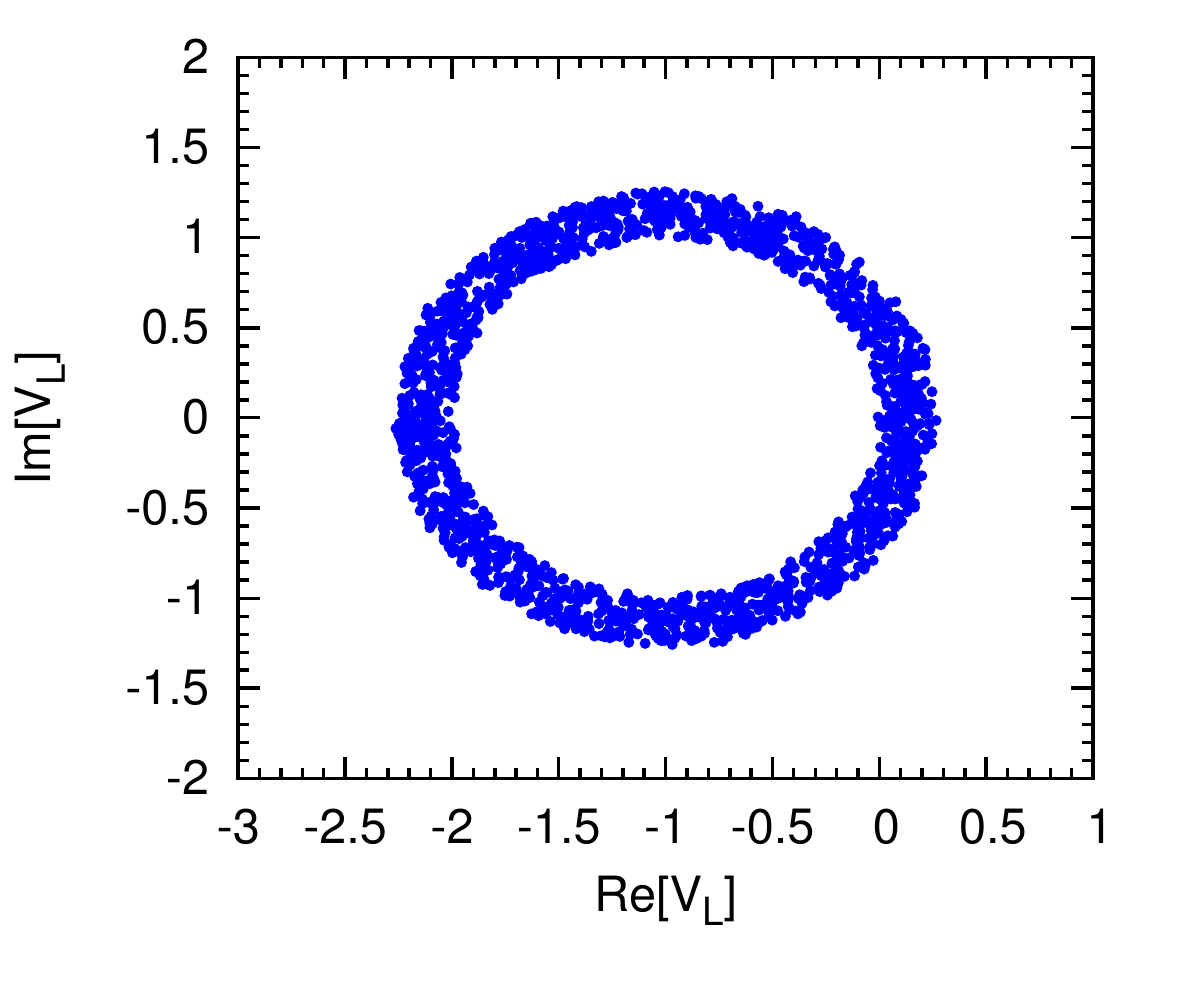}
\caption{Constraint on the  real and imaginary parts of   $V_L$  parameter obtained  from   $B_u^+ \to e^+ \nu_e$ (top-left panel), $B_u^+ \to \mu^+ \nu_\mu$ (top-right panel) and $B_u^+ \to \tau^+ \nu_\tau$ (bottom panel). } \label{Bulnu-VL}
\end{figure}

\begin{figure}[h]
\centering
\includegraphics[scale=0.55]{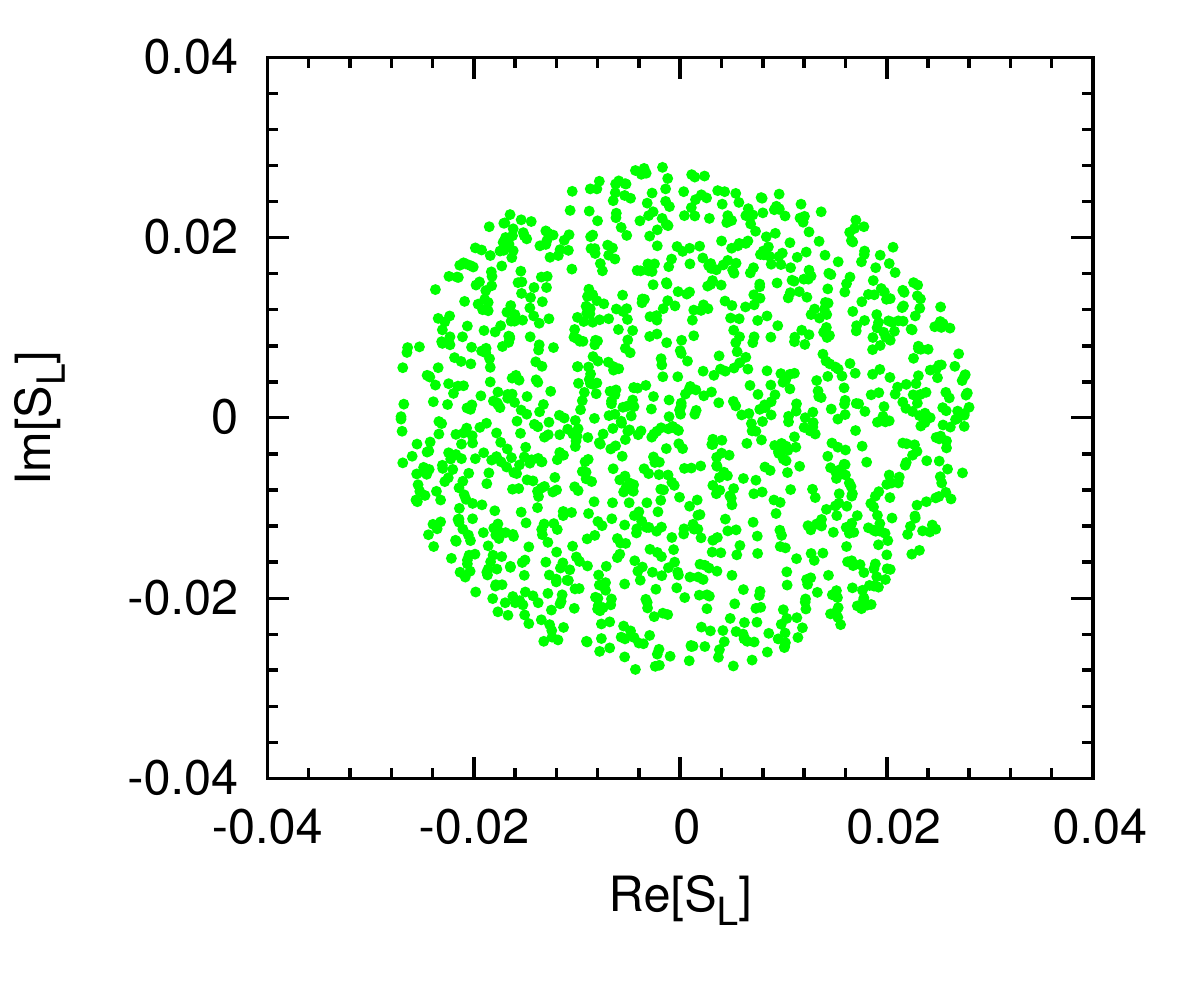}
\quad
\includegraphics[scale=0.55]{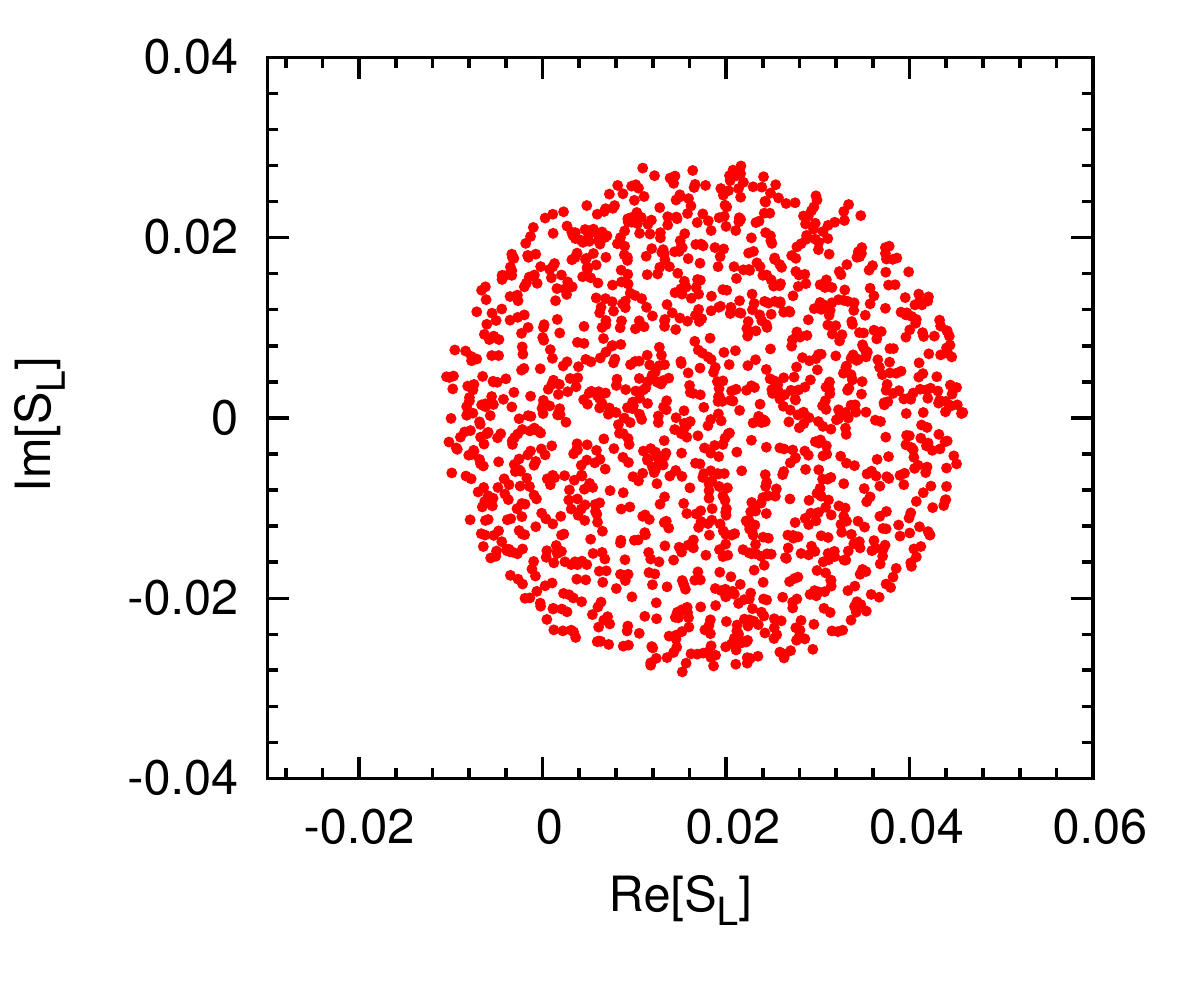}
\quad
\includegraphics[scale=0.55]{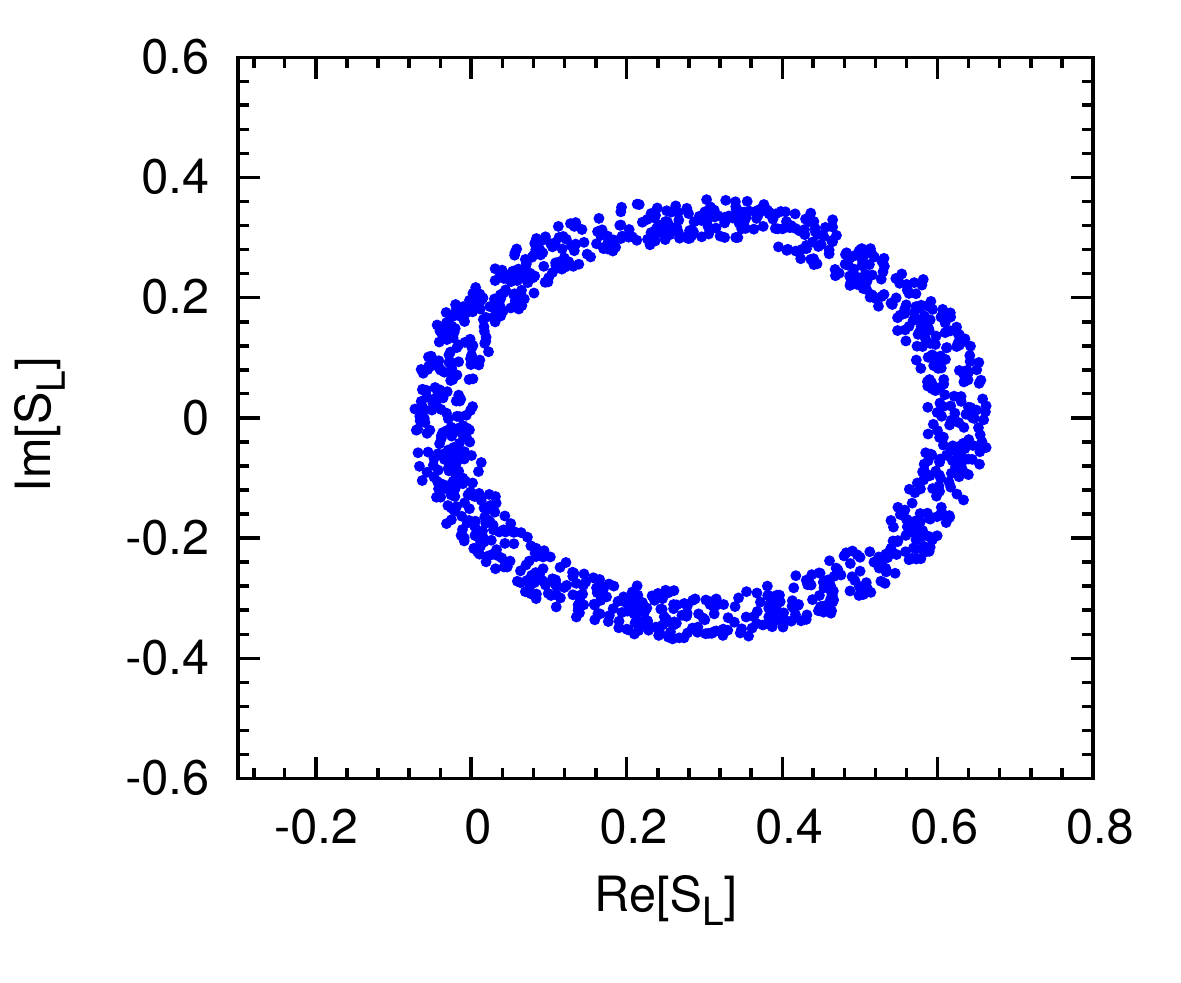}
\caption{Constraint on the real and imaginary parts of the $S_L$  parameter  obtained from   $B_u^+ \to e^+ \nu_e$ (top-left panel), $B_u^+ \to \mu^+ \nu_\mu$ (top-right panel) and $B_u^+ \to \tau^+ \nu_\tau$ (bottom panel).} \label{Bulnu-SL}
\end{figure}

The rare leptonic $B_u^+ \to l^+ \nu_l$ processes are mediated by the quark-level transitions $b \to u $  and are theoretically very clean. The only non-perturbative quantity involved in these processes is the decay constant of $B_u$ meson. Including  the new  coefficients from Eqn. (\ref{ham}), the branching ratios of  $B_u^+ \to l^+ \nu_l$ processes in the presence of NP are given by \cite{Bu-BR}
\bea \label{BR-Bulnu}
{\rm BR} (B_{u}^+ \to l^+  \nu_l)&=&\frac{G_F^2 M_{B_u} m_l^2}{8 \pi} \Big( 1-\frac{m_l^2}{M_{B_u}^2}\Big)^2 f_{B_u}^2 \left | V_{ub} \right |^2 \tau_{B_u^+} \nn \\ && \times \Big | \left(1 + V_L-V_R \right)-\frac{M_{B_u}^2}{m_l (m_b +m_u)}  \left(S_L - S_R \right) \Big |^2,
\eea
where $M_{B_u}~(f_{B_u})$ is the mass (decay constant) of $B_u$ meson and  $m_l$ is the lepton mass.
In our analysis, all the particle masses and the life time of $B_u^+$ meson are taken  from \cite{pdg}. The decay constant  of $B_u$ meson is taken  as $f_{B_u} = 190.5~(4.2)$ MeV \cite{Bu-decayconstant}, and for the CKM matrix element, we use the Wolfenstein parametrization with the values $A=0.811 \pm 0.026$, $\lambda=0.22506\pm 0.00050$,
$\bar{ \rho} =0.124^{+0.019}_{-0.018}$ and $\bar{ \eta} =0.356 \pm 0.011$ \cite{pdg}.  Using these   values, the obtained  branching fractions of $B_{u}^+ \to l^+  \nu_l$ processes  in the SM   are given as 
\bea
&&{\rm BR} (B_u^+ \to e^+ \nu_e)|^{\rm SM}= (8.9 \pm 0.23) \times 10^{-12}, \nn \\
&&{\rm BR} (B_u^+ \to \mu^+ \nu_\mu)|^{\rm SM}= (3.83 \pm 0.1)  \times 10^{-7}, \nn \\
&&{\rm BR} (B_u^+ \to \tau^+ \nu_\tau)|^{\rm SM}= (8.48 \pm 0.28) \times 10^{-5},
\label{Bu-SM}
\eea
and the corresponding experimental values are \cite{pdg}
\bea
&&{\rm BR} (B_u^+ \to e^+ \nu_e)|^{\rm Expt} ~\textless ~9.8 \times 10^{-7},\nn \\
&& {\rm BR} (B_u^+ \to \mu^+ \nu_\mu)|^{\rm Expt}~ \textless ~1.0 \times 10^{-6}, \nn  \\ 
&&{\rm BR} (B_u^+ \to \tau^+ \nu_\tau)|^{\rm Expt} =(1.09 \pm 0.24) \times 10^{-4}.
\label{Bu-Exp}
\eea
Since $B_u^+ \to l^+ \nu_l$ processes do not receive any contribution from tensor coupling, we ignore the effect of tensor operator in this work.  In our analysis, we consider the new  coefficients $V_{L,R},~S_{L,R}$ as complex.   For simplicity, we consider the presence of only one  coefficient at a time and constrain its real and imaginary parts by comparing the predicted SM branching fractions of $B_u^+ \to l^+ \nu_l$ processes with the   corresponding experimental results. For  $B_u^+ \to \tau^+ \nu_\tau$, we compare with the  $1\sigma$ range of observed data.  In Fig. \ref{Bulnu-VL}, we show the  constraints on the real and imaginary parts of the $V_L$  coefficient obtained from the  $B_u^+ \to  e^+ \nu_e$ (top-left panel), $B_u^+ \to  \mu^+ \nu_\mu$ (top-right panel) and $B_u^+  \to \tau^+ \nu_\tau$ (bottom panel) processes.
Analogously,   the  allowed ranges of the real and imaginary parts of $S_{L}$   coefficient  derived from   the  $B_u^+ \to e^+ \nu_e$ (top-left panel), $B_u^+ \to \mu^+ \nu_\mu$ (top-right panel) and $B_u^+  \to \tau^+ \nu_\tau$ (bottom panel) processes are shown in Fig. \ref{Bulnu-SL}. 
The constraint on the imaginary part of the $V_R~(S_R)$  coefficient is same as  $V_L~(S_L)$ coefficient and the corresponding real part is  related by ${\rm Re}[V_R]~({\rm Re}[S_R])= - {\rm Re}[V_L]~({\rm Re}[S_L])$.   It should be noted that the bounds obtained from   $B_u^+ \to  e^+\nu_e (\mu^+ \nu_\mu)$ process are comparatively weak as only the upper limits on  the branching ratios of these  processes exist.  Furthermore, the bounds on new coefficients obtained from $B_u^+ \to e^+ \nu_e$ process are too weak to make reasonable predictions for   the observables associated with $b \to u e^+ \nu_e$ decay modes.  Therefore, we only present the results for semileptonic $B$ decays with $\mu(\tau)$ in the final state. 

%
%
%
%
%
%
%
%
%
%
%
%
%
%
%
%
%

\section{$B \to P l \bar \nu_l$  processes}

In this section, we discuss the  rare $B \to P l \bar{\nu}_l$ processes, where  $P=\pi, K, \eta^{(\prime)}$.  The matrix elements of various hadronic currents
between the initial $B$ meson and the final pseudoscalar  meson $P$, can be parametrized in terms of two 
form factors $F_0$, $F_1$ \cite{Pformfactor-expr, Sakaki} as
\bea
\langle P(k) | \bar{u}\gamma_{\mu}\,b | B(p_B) \rangle &=& F_{1}(q^2)\Big[(p_B + k)_{\mu} - \frac{M_B^2 - M_P^2}{q^2}q_{\mu}\Big] + F_0(q^2)\frac{M_B^2 - M_P^2}{q^2}q_{\mu}\;, \
\eea
where $p_B$ and $k$ are respectively the four momenta of the $B$ and $P$ mesons  and $q = p_B -k$ is
the momentum transfer. Now using the above form factors, the double differential decay distribution of $B \to P l \nu_l$ processes  in terms of the helicity amplitudes $H_0$, $H_t$ and $H_S$ are given by \cite{Sakaki}
\bea
\frac{d\Gamma(B \to P l \bar \nu_l)}{dq^2} &=& {G_F^2 |V_{ub}|^2 \over 192\pi^3 M_B^3} q^2 \sqrt{\lambda_P(q^2)} \left( 1 - {m_l^2 \over q^2} \right)^2  \nn \\         &\times& \Bigg \lbrace \Big | 1 + V_L + V_R \Big |^2 \left[ \left( 1 + {m_l^2 \over 2q^2} \right) H_{0}^{2} + {3 \over 2}{m_l^2 \over q^2} \, H_{t}^{2} \right] + {3 \over 2} |S_L + S_R|^2 \, H_S^{2} \nn \\ &+& 3{\rm Re}\left[ ( 1 + V_L + V_R ) (S_L^* + S_R^* ) \right] {m_l \over \sqrt{q^2}} \, H_S H_{t}  \Bigg \rbrace,  \label{br-exp}
\eea
where 
\bea
\lambda_P=\lambda (M_B^2, M_P^2, q^2)  = M_B^4+M_P^4+q^4-2(M_B^2M_P^2+M_P^2q^2+M_B^2q^2)\;, 
\eea
and the  helicity amplitudes $(H_{0,t,S})$ in terms of the form factors $(F_{0,1})$ are given as 
\bea
&&H_0(q^2) = \sqrt{\frac{\lambda_P \left(q^2 \right)}{q^2}} F_1 \left(q^2 \right), \nn \\
&&H_t(q^2)  = \frac{M_B^2-M_P^2}{\sqrt{ q^2}} F_0 \left(q^2 \right), \nn \\
&&H_S (q^2) = \frac{M_B^2-M_P^2}{m_b-m_u} F_0 \left(q^2 \right). 
\eea
Here $M_P$ is the mass of the $P$ meson and  $m_b~(m_u)$ is the mass of the $b~(u)$ quark. 

The  lepton forward-backward asymmetry, which is an interesting observable to look for NP, defined as
\bea
A_{FB} (q^2) = \frac{\int_0^1 \frac{d\Gamma}{dq^2 d\cos \theta}d\cos \theta-\int_{-1}^0 \frac{d\Gamma}{dq^2 d\cos \theta}d\cos \theta}{d\Gamma /dq^2}\;.
\eea
Besides the branching ratio and forward-backward asymmetry, another important observable  is the LNU ratio. Similar to  $R_{D^{(*)}}$ observables, we   define the LNU parameter for $B \to P l \nu_l$ processes as
\bea
R_P^{\tau \mu}=\frac{{\rm BR}(B \to P \tau \bar{\nu}_{\tau}) }{{\rm BR}(B \to P \mu \bar{\nu}_{\mu})}, 
\eea
in order to scrutinize the  violation of lepton universality effect in $b \to u l \nu_l$ decays. In  Ref. \cite{RDstar-SM}, the authors have studied the  lepton  universality violating  ratio ${\rm BR}(B \to P \tau \bar{\nu}_{\tau}) / {\rm BR}(B \to P l \bar{\nu}_{l})$, where $l=e, \mu$.  Since the constraints on new coefficients obtained from $B_u^+ \to e^+ \nu_e$ process are too weak, it would not be possible to  predict  reasonably constrained result for  the ${\rm BR}(B \to P \tau \bar{\nu}_{\tau}) / {\rm BR}(B \to P e \bar{\nu}_{e})$ ratio. Therefore, we only consider the ${\rm BR}(B \to P \tau \bar{\nu}_{\tau}) / {\rm BR}(B \to P \mu \bar{\nu}_{\mu})$ parameter in our analysis. 

In order to explore few other  observables which are sensitive to NP in  the $b \to u l \bar{\nu}_l$ processes, we define the  parameter $R_{P P^\prime}^{l}$ as ratio of branching fractions of  $B \to P l^- \bar \nu_l$ to $B \to P^\prime l^- \bar \nu_l$ processes
\bea
R_{P P^\prime}^{l} = \frac{{\rm BR}(B \to P l^- \bar \nu_l)}{{\rm BR}(B \to P^\prime l^- \bar \nu_l)}.
\eea
These processes differ only in the spectator quark content and hence, any deviation from SM prediction, if observed would hint towards the existence of NP.

After setting the stage, we now proceed  for numerical analysis. We consider all the particle masses and the life time of $B$ meson from the Ref. \cite{pdg}. To make predictions for the various observables or to extract information about potentially new short distance
physics, one should have sufficient  knowledge on the associated hadronic form factors. 
  For the form factors of $\bar B_s \to K^+ l^- \bar \nu_l$ processes, we consider  the perturbative QCD (PQCD) calculation \cite{Wang, Wang2}  based on the $k_T$ factorization \cite{keum} at
next-to-leading order (NLO) in $\alpha_s$ \cite{Wang3}, which gives
\bea
F_1^{B_s\to K}(q^2) &=&  {F_1^{B_s\to K}(0)} \left(\frac{1}{(1-q^2/M_{B_s}^2)} + \frac{a_1 q^2/M_{B_s}^2}{(1-q^2/M_{B_s}^2)(1-b_1 q^2/M_{B_s}^2)}\right),\nn \\
F_0^{B_s\to K}(q^2) &=& \frac{F_0^{B_s\to K}(0)} {(1-a_0 q^2/M_{B_s}^2+ b_0 q^4/M_{B_s}^4)}~,
\eea
where $M_{B_s}$  is the mass of $B_s$ meson and the values of the parameters $a_{0,1}$, $b_{0,1}$ and $F_{0,1}^{B_s \to K}$ are listed in Table I. 
 \begin{table}[htb]
\begin{center}
\caption{Numerical values of the $B_s \to K$ form factors in the PQCD approach \cite{Wang}. }
\begin{tabular}{| c |c | c | c| }
\hline
Parameters~ & PQCD~  \\
 \hline
 \hline
 
$F_0(0)$ &~ $0.26^{+0.04}_{-0.03}\pm 0.02$~ \\
  
$a_0$   ~&~ $0.54 \pm 0.00 \pm 0.05$~ \\

$b_0$   ~&~ $-0.15 \pm 0.00 \pm 0.00$~ \\

\hline

$F_1(0)$ &~ $0.26\pm 0.035\pm 0.02$~ \\
  
$a_1$   ~&~ $0.57 \pm 0.01 \pm 0.02$~ \\

$b_1$   ~&~ $0.50 \pm 0.01 \pm 0.05$~ \\

\hline  
\end{tabular}
\end{center}
\end{table}


For $B \to \pi$ form factors, we use the light cone sum rule (LCSR) results as input for a $z$-series parametrization which yield the $q^2$ shape in the whole semileptonic region of $B \to \pi l \nu_l$ processes. The $q^2$ dependence of the form factors is parametrized as \cite{pi-formfactor}
\bea
&&F_{1}(q^2) = \frac{F_{1}(0)}{\Big(1-\frac{q^2}{M_{B^*}{2}}\Big)}\Bigg\{1 + \sum_{k=1}^{N-1}\,b_k\,\Big(z(q^2,t_0)^k - z(0,t_0)^k -
(-1)^{N-k}\,\frac{k}{N}\,\Big[z(q^2,t_0)^N - z(0,t_0)^N\Big]\Big)\Bigg\}, \nonumber \\
&&F_0(q^2) = F_0(0)\,\Bigg\{1 + \sum_{k=1}^N\,b_k^0\,\Big(z(q^2,t_0)^k - z(0,t_0)^k\Big)\Bigg\}, 
\eea
where $N=2$ for $F_1(q^2)$ form factor and for $F_0(q^2)$ form factor, $N=1$. Here the function $z(q^2, t_0)$ is defined as  \cite{zseries}
\bea \label{zq2}
&&z(q^2,t_0) = \frac{\sqrt{(M_B + M_{\pi})^2 - q^2} - \sqrt{(M_B + M_{\pi})^2 - t_0}}{\sqrt{(M_B + M_{\pi})^2 - q^2} + \sqrt{(M_B + M_{\pi})^2 - t_0}}\;, 
\eea
where  $t_0 = (M_B + M_{\pi})^2 - 2\sqrt{M_B\,M_{\pi}}\,\sqrt{(M_B + M_{\pi})^2 - q^2_{\rm min}}$ is the auxiliary parameter.
Here the  values of  various   parameters involved are  $F_{1}(0) = F_0(0)= 0.281 \pm 0.028$,  $b_1 =-1.62 \pm 0.70$ and $b^0_1= -3.98 \pm 0.97$ \cite{pi-formfactor}. 

The $B^- \to \eta^{(\prime)} l^- \bar \nu_l$ processes are also mediated by the flavour changing charged current (FCCC) transitions $b \to u $. 
For the  study of  these  processes, we use $SU(3)_F$ flavour symmetry  to relate the form factors of $F_1^{B \to \eta^{(\prime)}}$ to $F_1^{B \to \pi}$.  We choose the scheme as  discussed in \cite{eta-etaprime-mixing, eta-etaprime-mixing2}, and consider
\bea
 &&|\eta \rangle = \cos \phi |\eta_q \rangle  - \sin \phi |\eta_s \rangle, \nn \\
 && |\eta^\prime \rangle = \sin \phi |\eta_q \rangle  + \cos \phi |\eta_s \rangle,
\eea
for the $\eta-\eta^\prime$ mixing, where $| \eta_q \rangle = (u\bar u + d\bar d)/ \sqrt{2}$, $\eta_s = s \bar s$ and $\phi$ is the fitted mixing angle $(\phi = 39.3^\circ)$ \cite{eta-etaprime-mixing2}. With these input parameters in hand, we now proceed  to discuss four different new physics scenarios and their  effect  on $b \to u l \nu_l$ processes.
 \begin{figure}[h]
\centering
\includegraphics[scale=0.55]{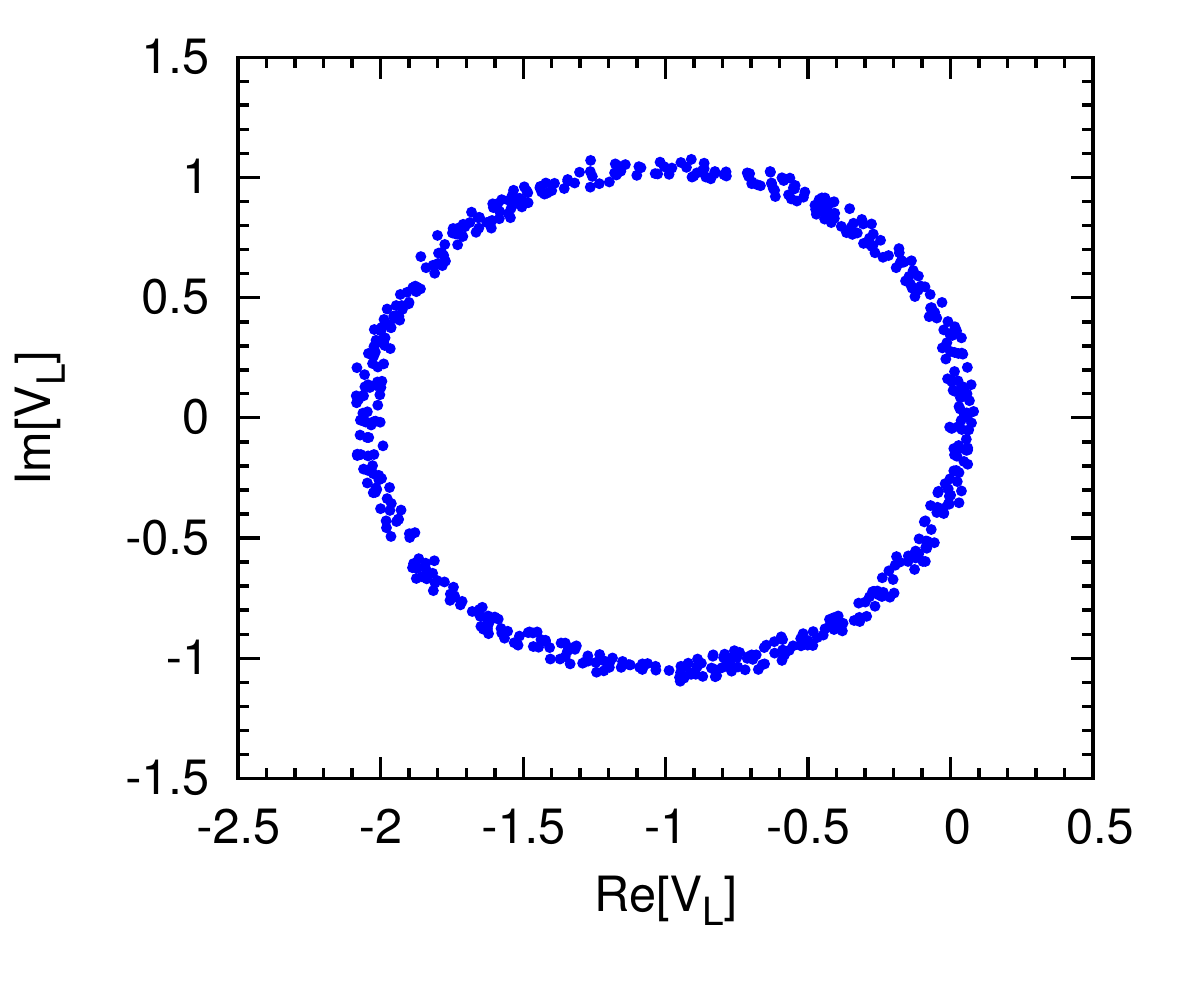}
\quad
\includegraphics[scale=0.55]{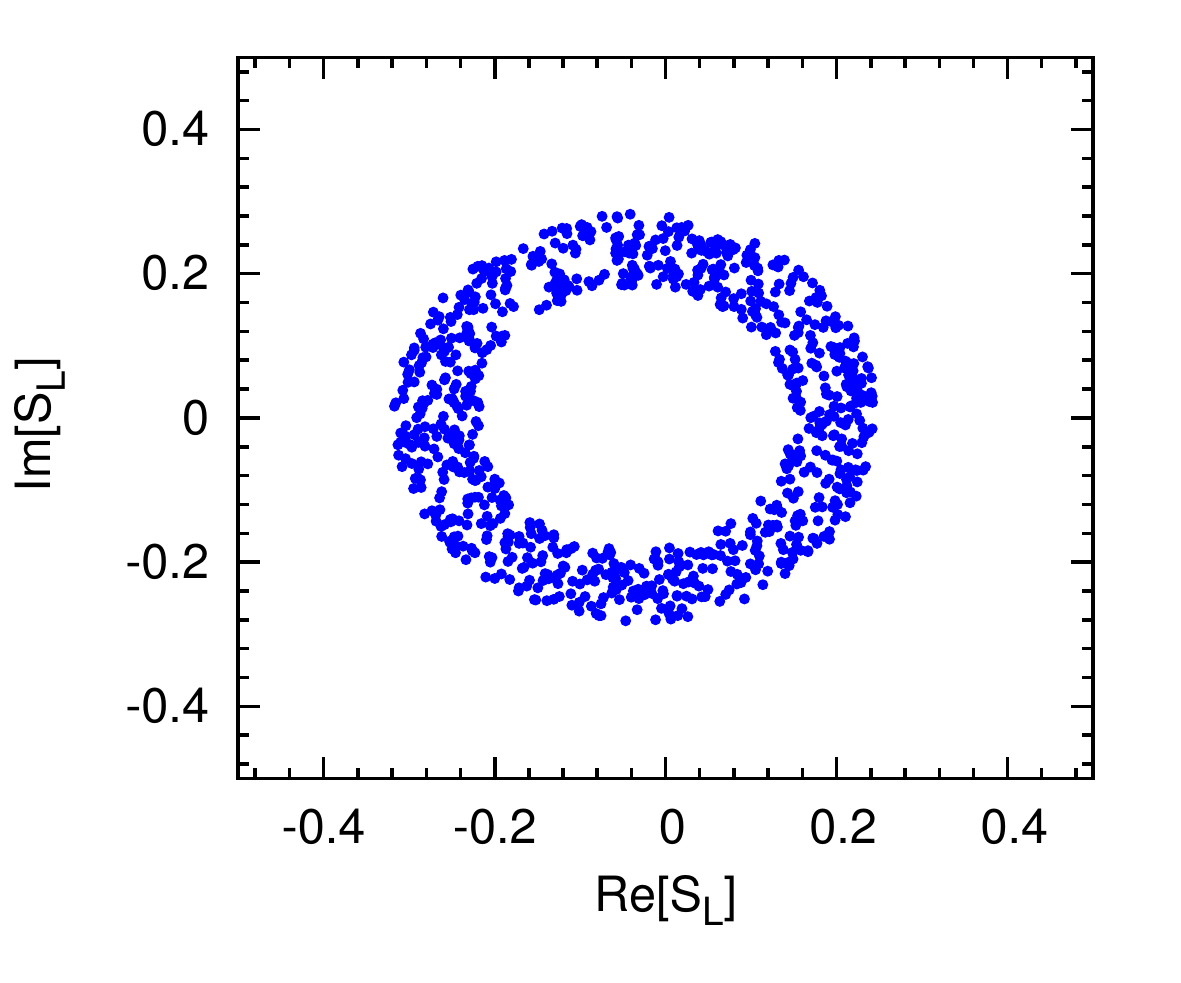}
\caption{Constraint on the real and imaginary parts of the $V_L$  (left panel) and $S_L$ (right panel) parameters  obtained from   $B_u^- \to \pi^0  \mu \bar\nu_\mu$ process.} \label{B2Pi}
\end{figure}

\subsection{Case A: Effect of $V_L$ only}

In this case, we assume that only the new  $V_L$  coefficient is present in addition to SM contribution, in the   effective Lagrangian (\ref{ham}). From Eqn. (\ref{br-exp}) it should be noted that  as the NP has the same structure as the SM, the SM decay rate gets modified by the factor $|1 +V_L|^2$.
 The constraints on  the real and imaginary parts of  $V_L$ coefficient for $b \to u \tau \bar \nu_\tau$ are   obtained from the branching ratio of $B_u^+ \to \tau^+\nu_\tau$ process as discussed  in section II. 
From the bottom panel of Fig. \ref{Bulnu-VL}, one can notice that  the constraint on $V_L$ is $|V_L| \leq 2.5$, obtained from  $B_u \to \tau \bar \nu_\tau$ process. In our analysis, we consider the   values for real and imaginary parts of $V_L$, which give the maximum and minimum values of the branching ratio within the $1\sigma$ limit.   Thus, imposing the extrema conditions,  the allowed parameters  are found as $({\rm Re}[V_L], {\rm Im}[V_L])^{\rm max}= (0.130,  0.761)$ and  $({\rm Re}[V_L], {\rm Im}[V_L])^{\rm min}= (-0.929, 0.841)$. For $b \to u \mu \bar \nu_\mu$ transition as only the upper limit of $B_u \to \mu \bar \nu_\mu$ is known, it will not provide any strict bound on the NP coefficient $V_L$.  Therefore, to avoid overestimation of the predicted  values of various physical observables, we consider the branching ratio of $B^- \to \pi^0 \mu^-  \bar \nu_\mu $ process. Comparing the SM predicted value ${\rm BR}(B^- \to \pi^0 \mu^- \bar \nu_\mu)^{\rm SM}=(7.15 \pm 0.55) \times 10^{-5}$ with the $1\sigma$ range of corresponding measured value    ${\rm BR}(B^- \to \pi^0 \mu^- \bar \nu_\mu)^{\rm Expt}=(7.80 \pm 0.27) \times 10^{-5}$, we obtain the maximum and minimum values of the $V_L$ parameter as  $({\rm Re}[V_L], {\rm Im}[V_L])^{\rm max}= (-0.233,  0.769)$ and  $({\rm Re}[V_L], {\rm Im}[V_L])^{\rm min}= (-0.833, 0.968)$. The corresponding allowed parameter space is shown in the left panel of Fig. \ref{B2Pi}.

\begin{figure}[h]
\centering
\includegraphics[scale=0.4]{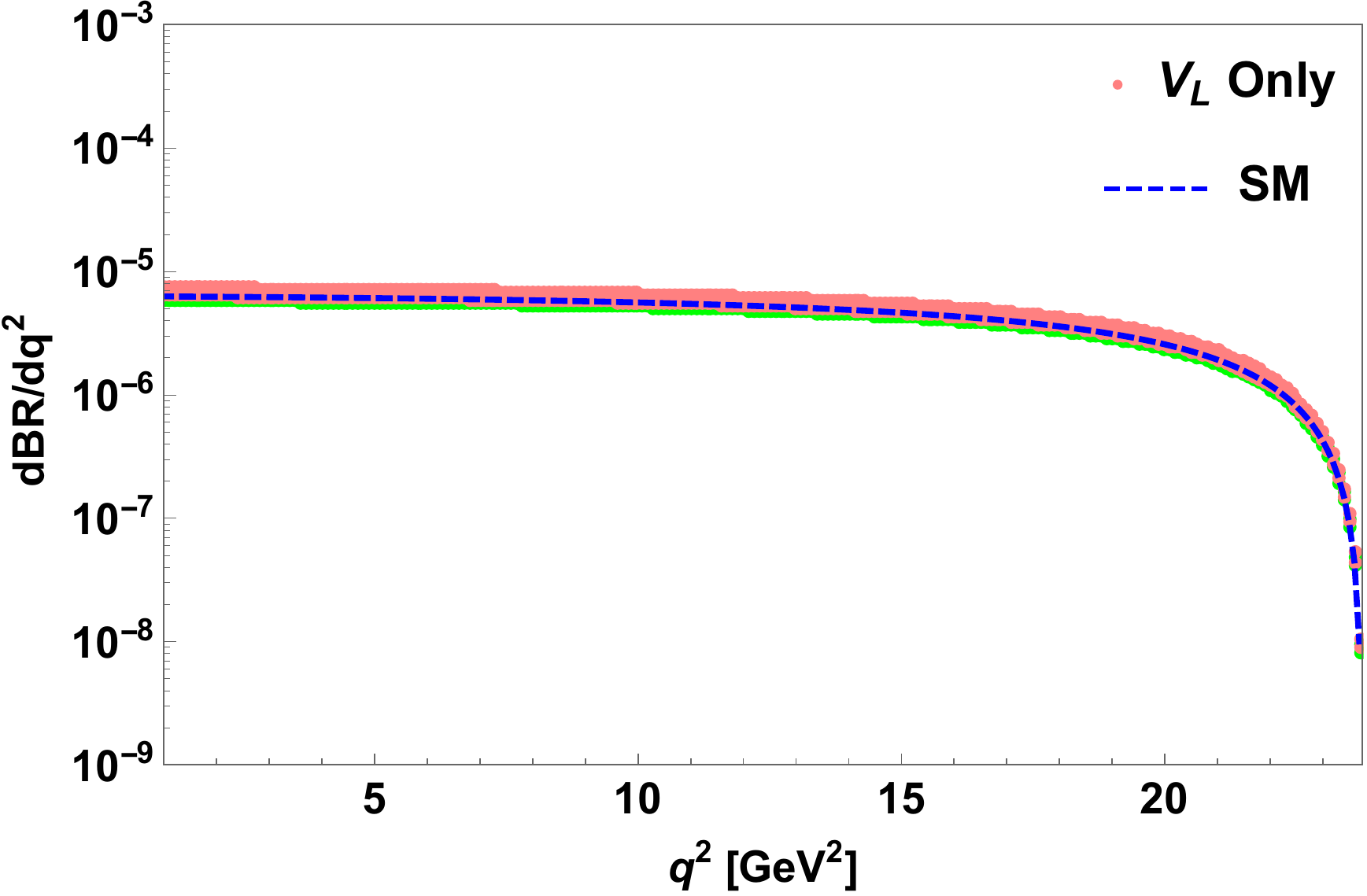}
\quad
\includegraphics[scale=0.4]{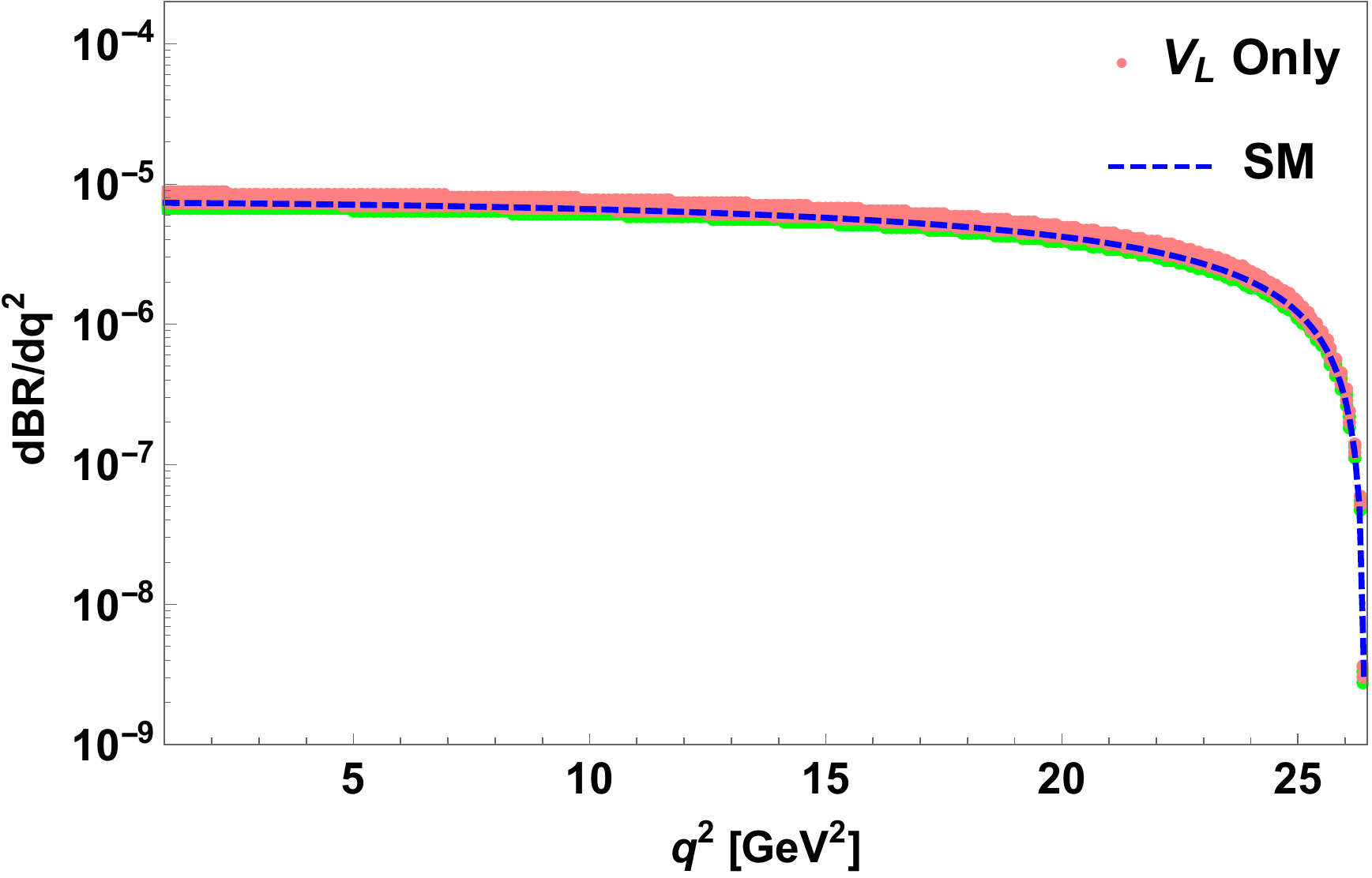}
\quad
\includegraphics[scale=0.4]{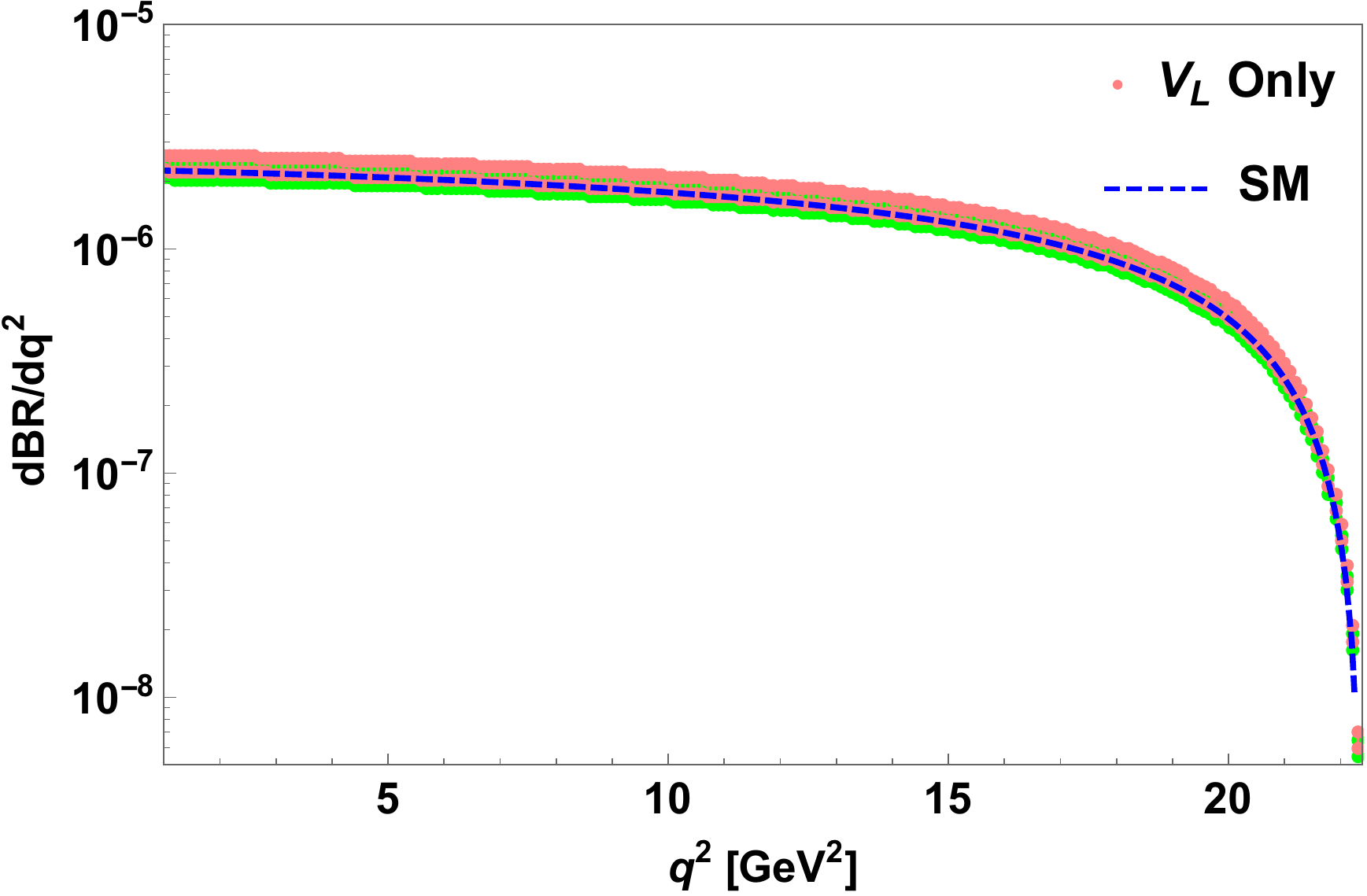}
\quad
\includegraphics[scale=0.4]{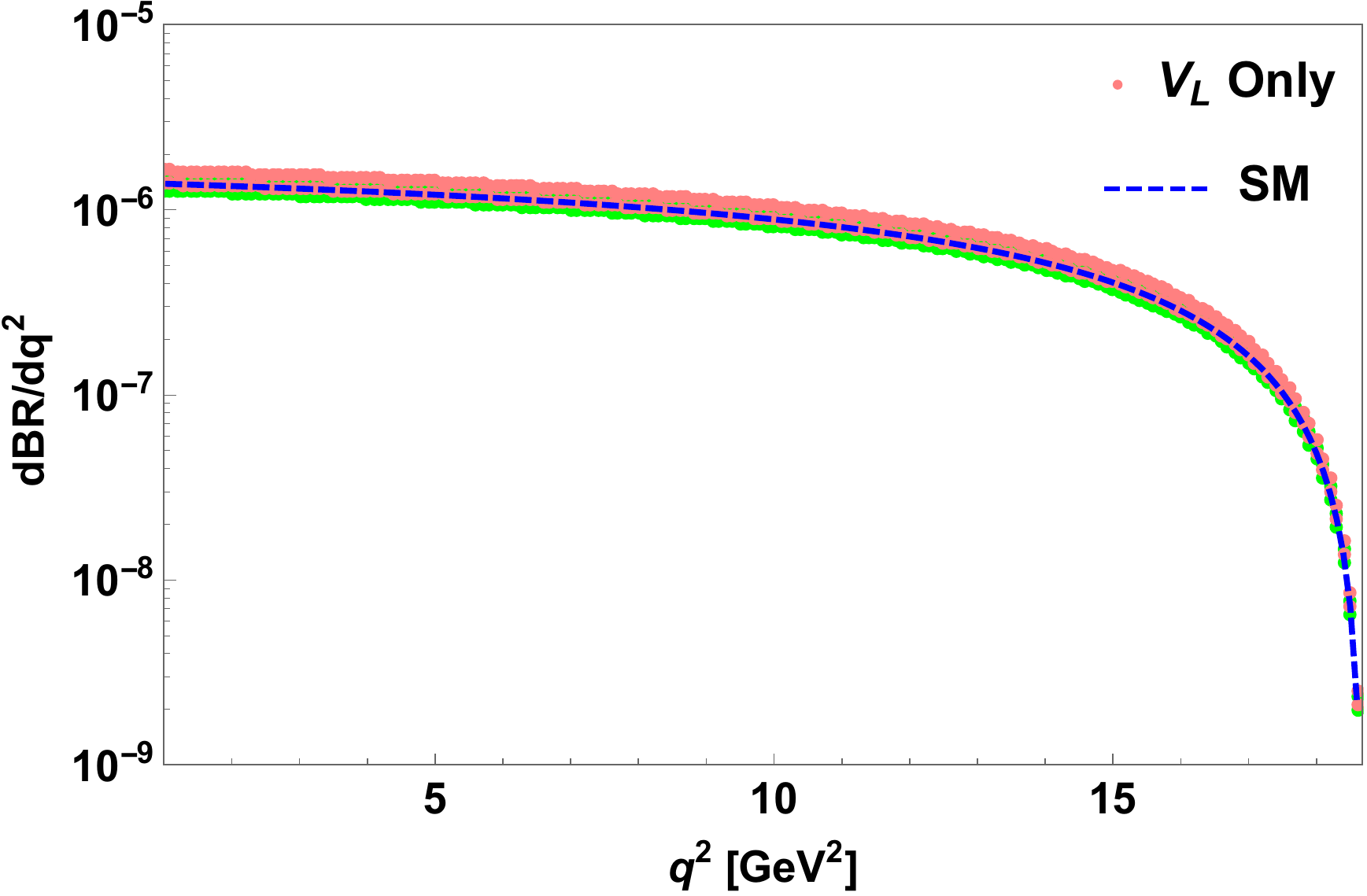}
\caption{The plots for  the $q^2$ variation of the branching ratios of $\bar B_s \to K^+ \mu^- \bar \nu_\mu$  (top-left panel), $ \bar B^0 \to \pi^+ \mu^-  \nu_\mu$  (top-right panel), $ B^- \to \eta \mu^- \bar \nu_\mu$  (bottom-left panel) and $B^- \to \eta^{\prime } \mu^- \bar \nu_\mu$  (bottom-right panel) processes for the NP contribution coming from  only $V_L$  coupling.   Here the red bands represent the   contributions due to the $V_L$ coupling. The blue dashed lines are for the SM contribution  and the green bands are due to the contributions coming from the  theoretical uncertainties.} \label{brmu-VL}
\end{figure}

 Using the allowed constrained values, we show the plots for the variation of branching fractions of various $B \to P \mu^-  \bar \nu_\mu $ processes   with respect to $q^2$ in  Fig. \ref{brmu-VL}, both in the SM and in NP scenario.  Here the plot for $\bar B_s \to K^+ \mu^- \bar \nu_\mu$ process is represented in the top-left panel,  the top-right panel is for the branching ratio of $ \bar B^0 \to \pi^+ \mu^-  \bar  \nu_\mu$,  the bottom-left plot is for $ B^- \to \eta \mu^- \bar \nu_\mu$ process and the branching ratio of $B^- \to \eta^{\prime }  \mu^- \bar \nu_\mu$ process is presented in the bottom-right panel.  
\begin{figure}[h]
\centering
\includegraphics[scale=0.4]{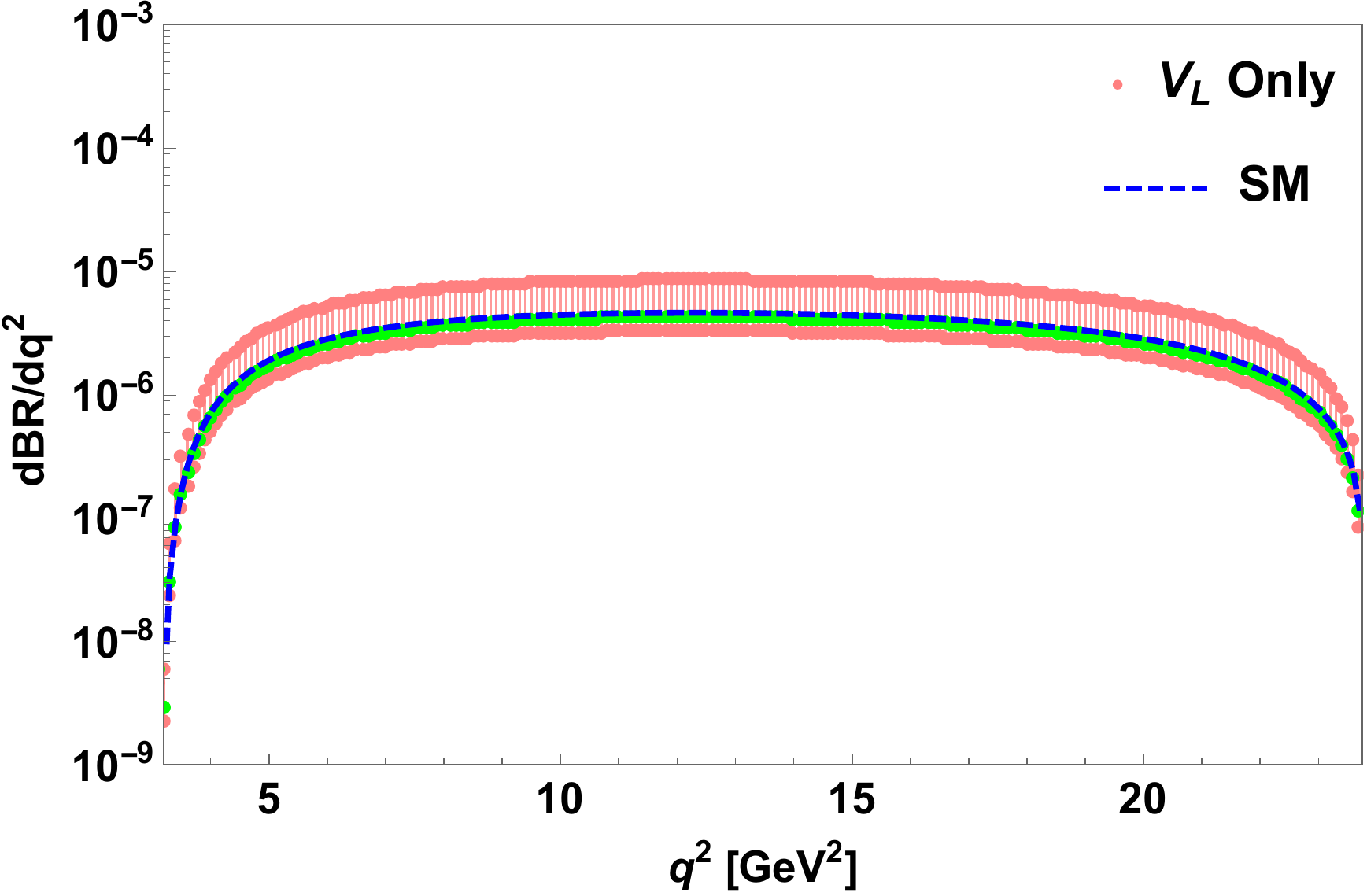}
\quad
\includegraphics[scale=0.4]{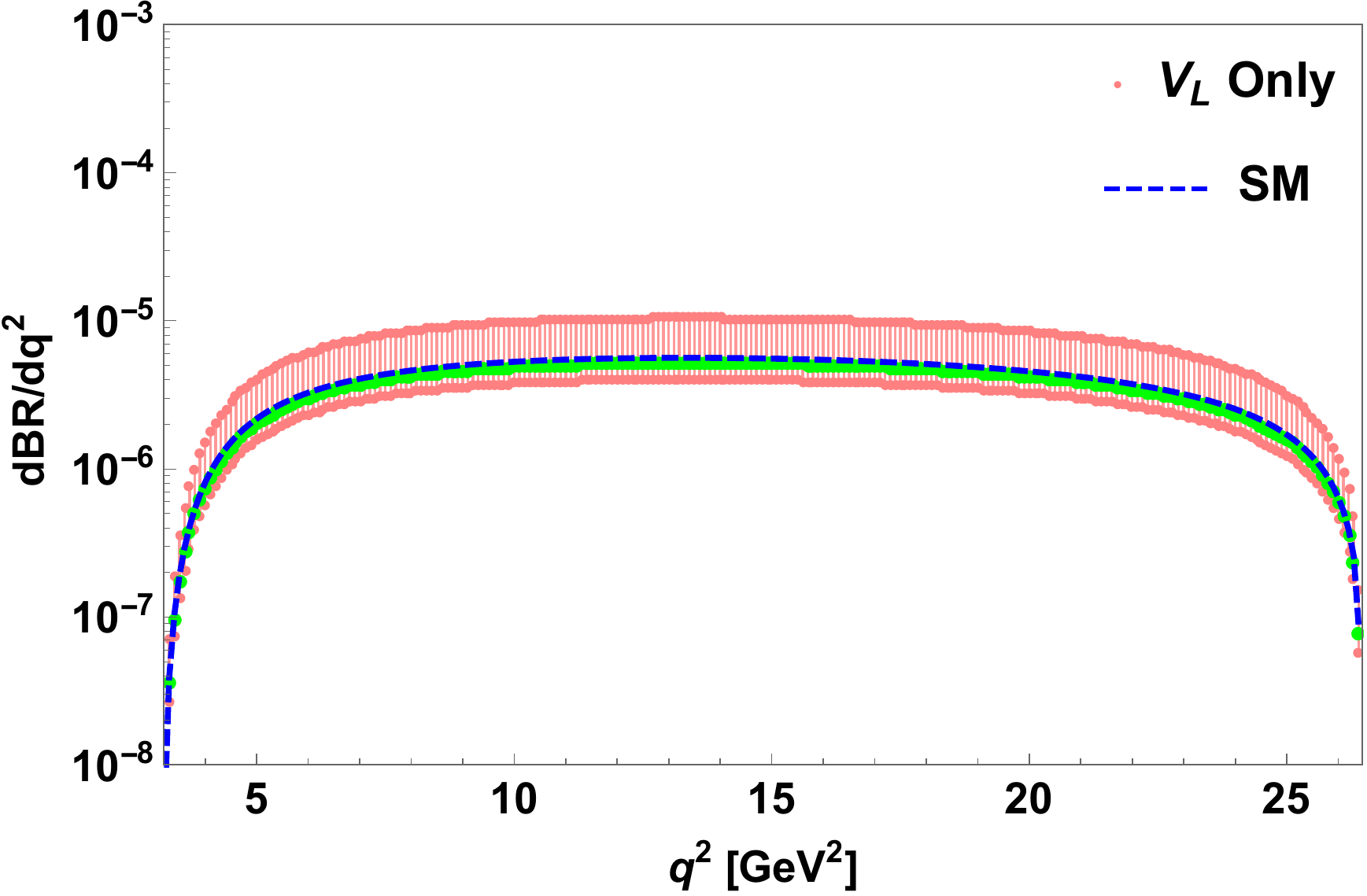}
\quad
\includegraphics[scale=0.4]{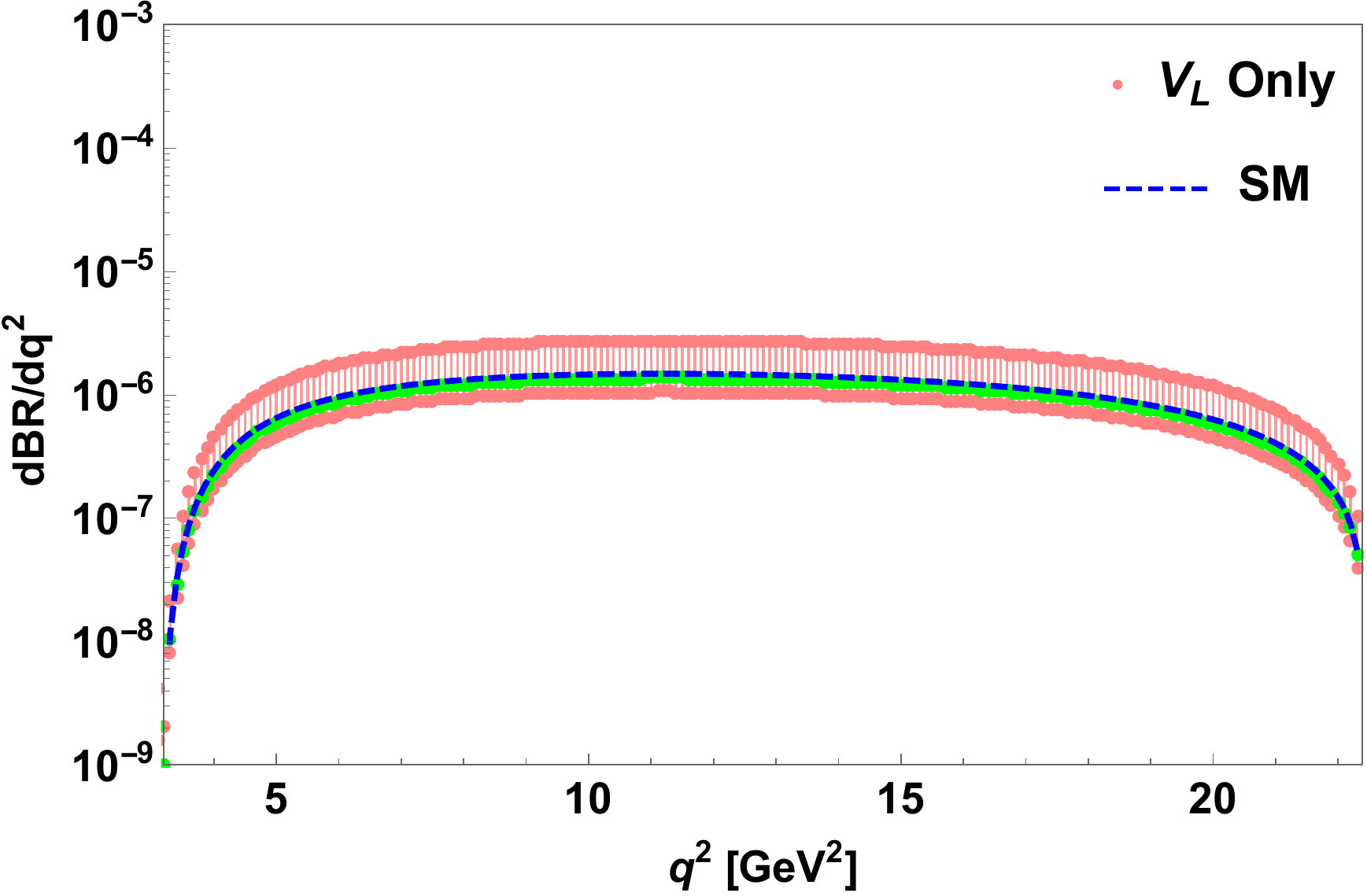}
\quad
\includegraphics[scale=0.4]{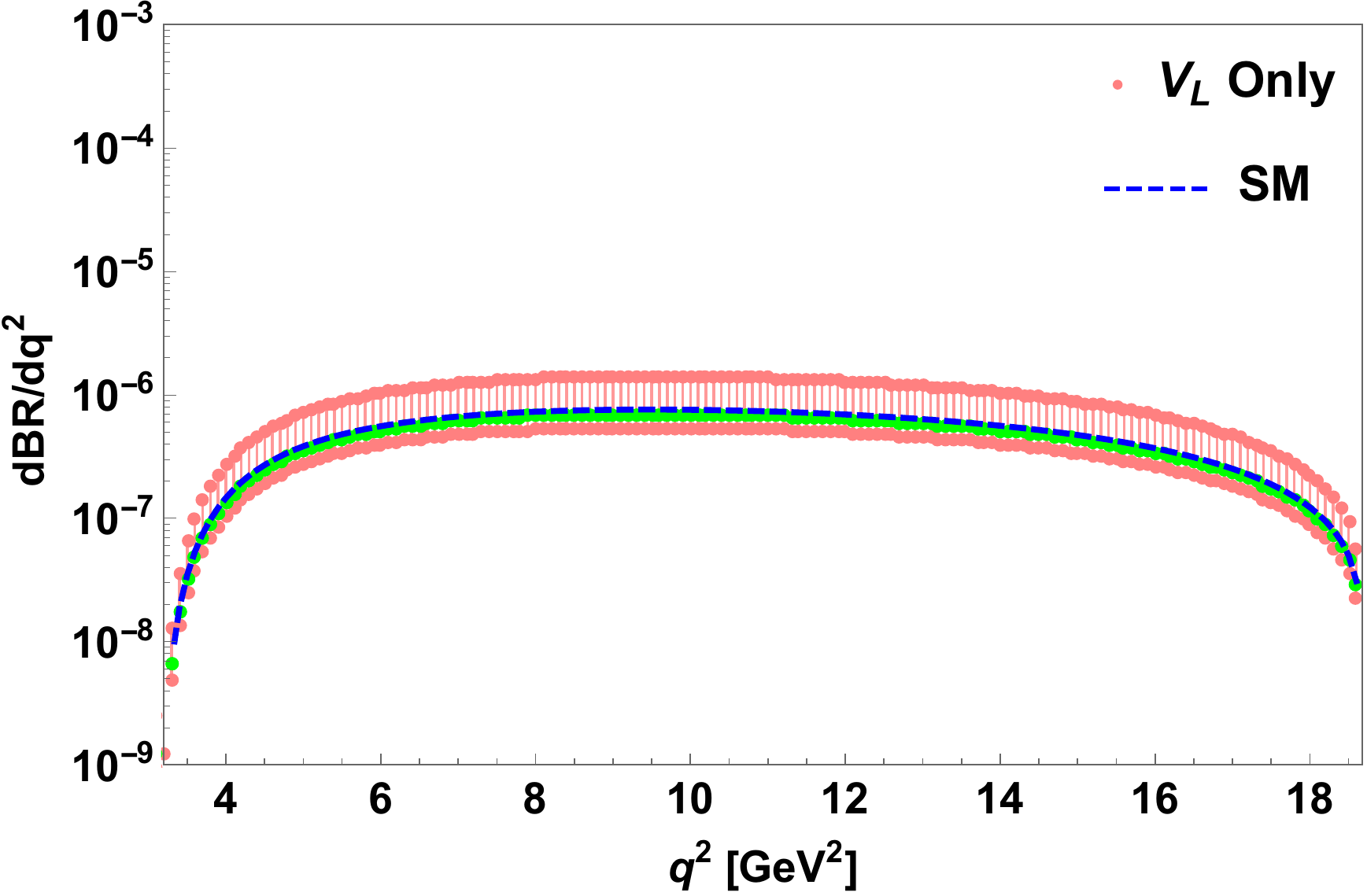}
\caption{The plots for  the $q^2$ variation of the branching ratios of $\bar B_s \to K^+ \tau^- \bar \nu_\tau$  (top-left panel), $  \bar B^0 \to \pi^+ \tau^-   \nu_\tau$  (top-right panel), $ B^- \to \eta \tau^- \bar \nu_\tau$  (bottom-left panel) and $ B^- \to \eta^{\prime } \tau^- \bar \nu_\tau$  (bottom-right panel) processes for the NP contribution due to $V_L$  coupling. } \label{Brtau-VL}
\end{figure}
 In these figures, the red bands are due to the contribution coming from  $V_L$ new physics parameter in addition to  SM and the  blue dashed lines are due to   SM. The green bands are  the corresponding SM theoretical uncertainties, which arise due to the uncertainties in the SM input parameters such as CKM elements and form factors.  Analogous plots for the  variation of the branching ratios of $\bar B_s \to K^+ \tau^- \bar \nu_\tau$ (top-left panel), $\bar B^0 \to \pi^+ \tau^- \bar \nu_\tau$ (top-right panel), $B^- \to \eta \tau^- \bar \nu_\tau$ (bottom-left panel) and $B^- \to \eta^{\prime } \tau^- \bar \nu_\tau$ (bottom-right panel) processes  are shown in  Fig. \ref{Brtau-VL}.  The integrated values of the  branching ratios for these processes are given in Table II.    Due to the inclusion of new $V_L$  coefficient, we found certain deviation in the   branching ratios of $B \to P \tau \bar \nu_\tau$ processes from the SM values, whereas the deviation in the branching ratios of $B \to P \mu \bar \nu_\mu$ processes are relatively small. Our predicted results for $ B \to (\pi, \eta^{(\prime)}) l \nu_l$  processes are consistent with the existing  experimental data \cite{pdg} 
  \bea
 &&  {\rm BR}(B^+ \to \eta l^+ \nu_l)^{\rm Expt} = (3.8 \pm 0.6) \times 10^{-5}, ~~~
 {\rm BR}(B^0 \to \pi^- l^+ \nu_l)^{\rm Expt} = (1.45 \pm 0.05) \times 10^{-4},\nn\\
  &&{\rm BR}(B^+ \to \eta^\prime l^+ \nu_l)^{\rm Expt} = (2.3 \pm 0.8) \times 10^{-5}, ~~~
 {\rm BR}(B^0 \to \pi^- \tau^+ \nu_\tau)^{\rm Expt} <  2.5 \times 10^{-4}.  
  \eea   
\begin{figure}[h]
\centering
\includegraphics[scale=0.4]{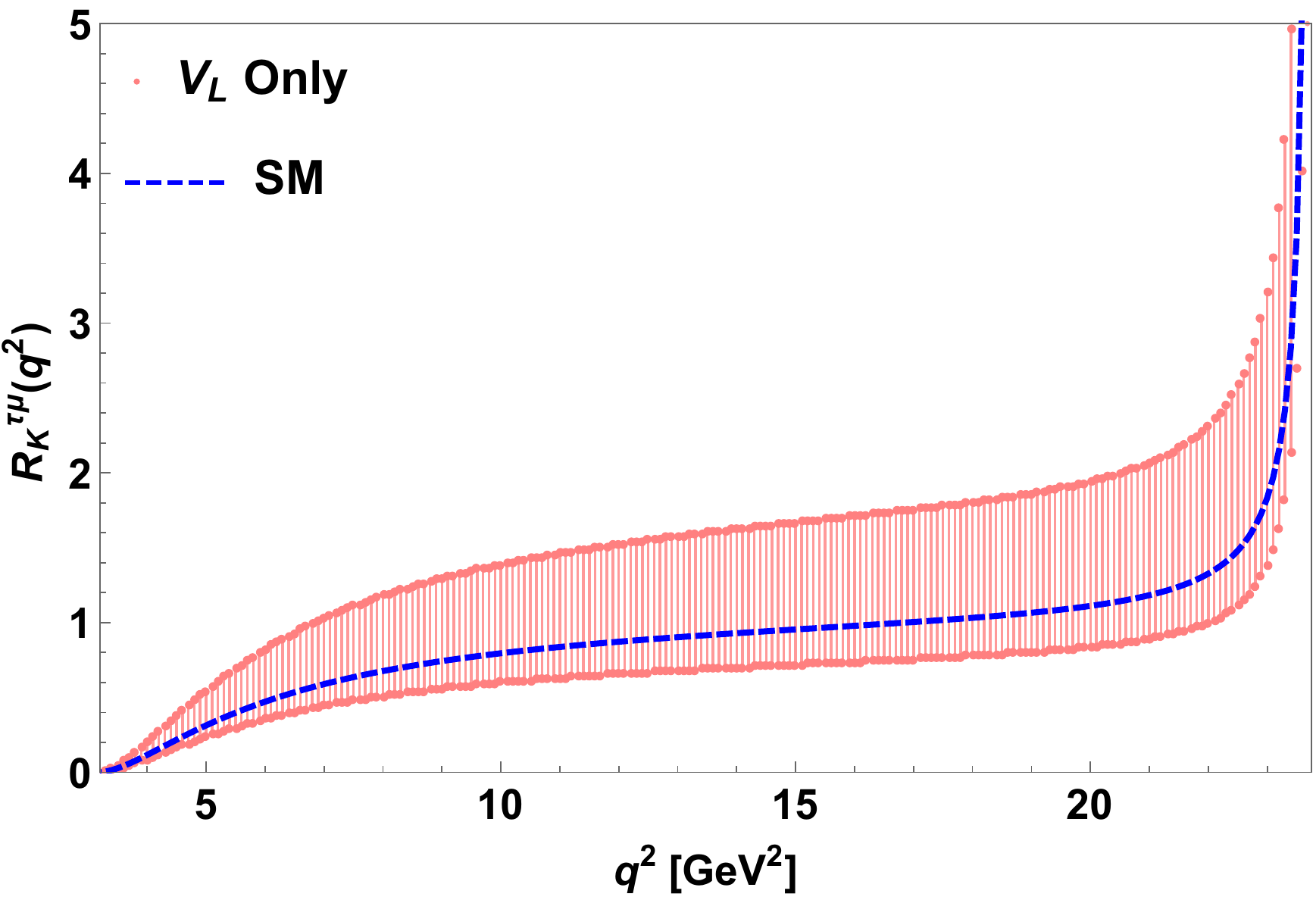}
\quad
\includegraphics[scale=0.4]{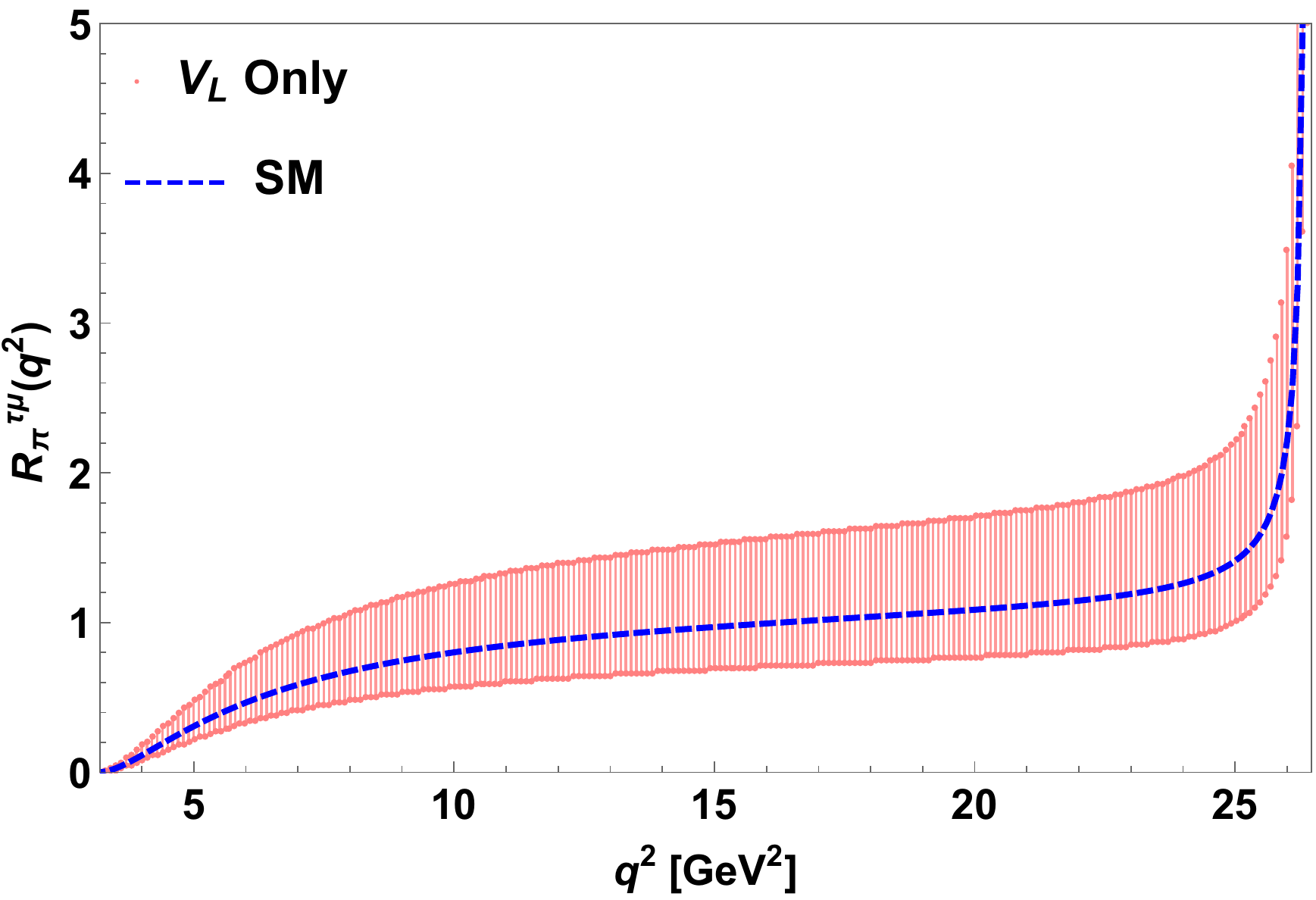}
\quad
\includegraphics[scale=0.4]{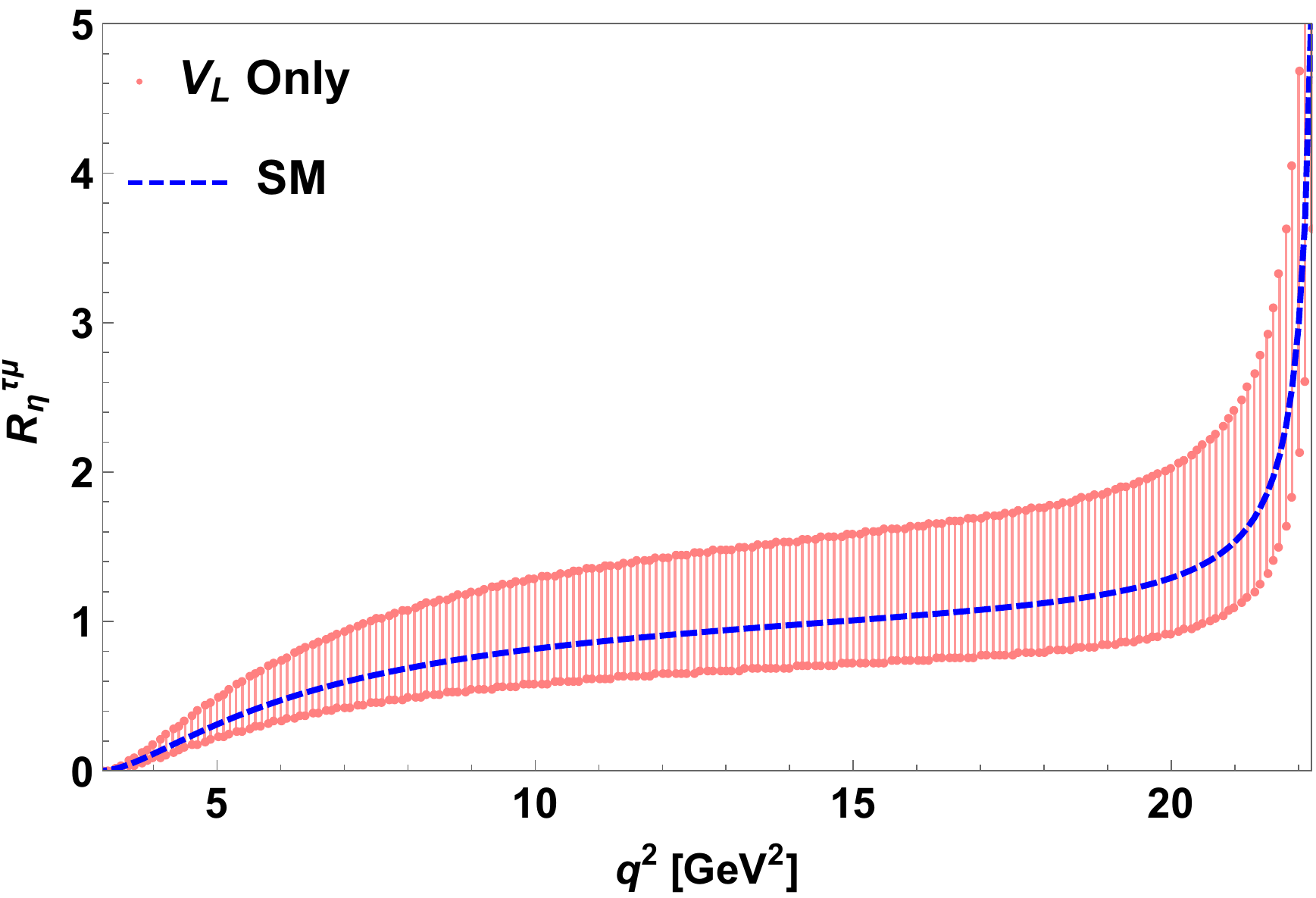}
\quad
\includegraphics[scale=0.4]{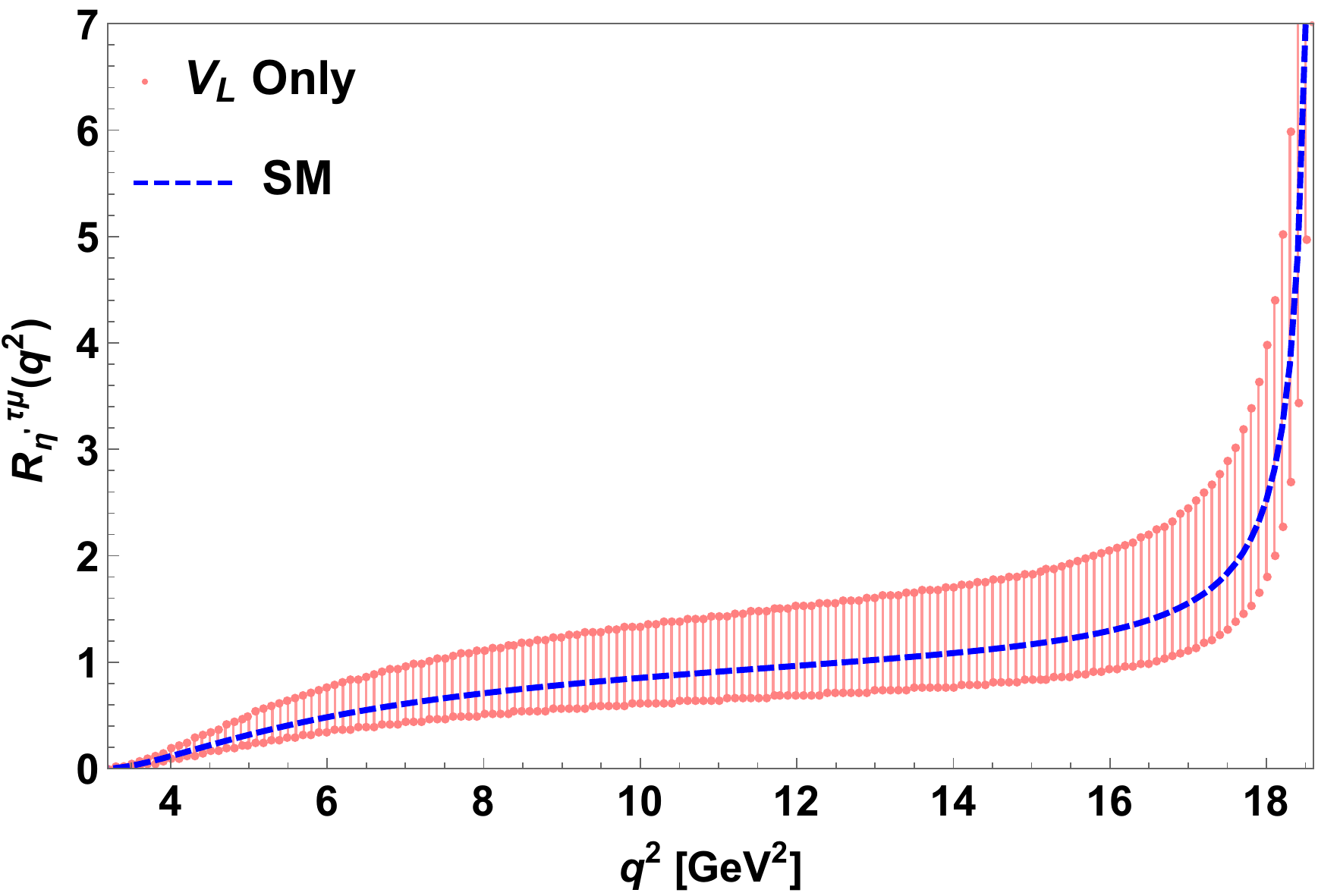}
\caption{The plots for the LNU parameters  $R_K^{\tau \mu}(q^2)$ (top-left panel), $R_\pi^{\tau \mu}(q^2)$ (top-right panel), $R_\eta^{\tau \mu}(q^2)$ (bottom-left panel) and $R_{\eta^\prime}^{\tau \mu }(q^2)$ (bottom-right panel) for the NP contribution due to  $V_L$ coupling.  } \label{RK-VL}
\end{figure}
Since the  $V_L$   contribution has the same structure as SM, the forward-backward asymmetry parameter of  $B \to P \mu^- \bar \nu_\mu ~(\tau^- \bar \nu_\tau)$ processes do not deviate from their  SM values, and  the  corresponding integrated values (integrated over the whole $q^2$ range) are presented in  Table II.  In Fig. \ref{RK-VL}, we show the plots for the  LNU parameters of  $\bar B_{(s)} \to P l \bar \nu_l$ processes,   $R_K^{\tau \mu }$ (top-left panel), $R_{\pi}^{\tau \mu } $ (top-right panel), $R_\eta^{\tau \mu}$ (bottom-left panel) and $R_{\eta^\prime}^{\tau \mu }$ (bottom-right  panel).  Including only $V_L$ coupling, we also compute the  $R_{\pi K}^l$, $R_{\pi \eta}^l$  and $R_{\pi \eta^\prime}^l$  parameters, however, no deviation has been found from their corresponding SM result.  The numerical values of these  parameters  are listed in Table III. 
 
\subsection{Case B: Effect of  $V_R$ only}
\begin{figure}[h]
\centering
\includegraphics[scale=0.4]{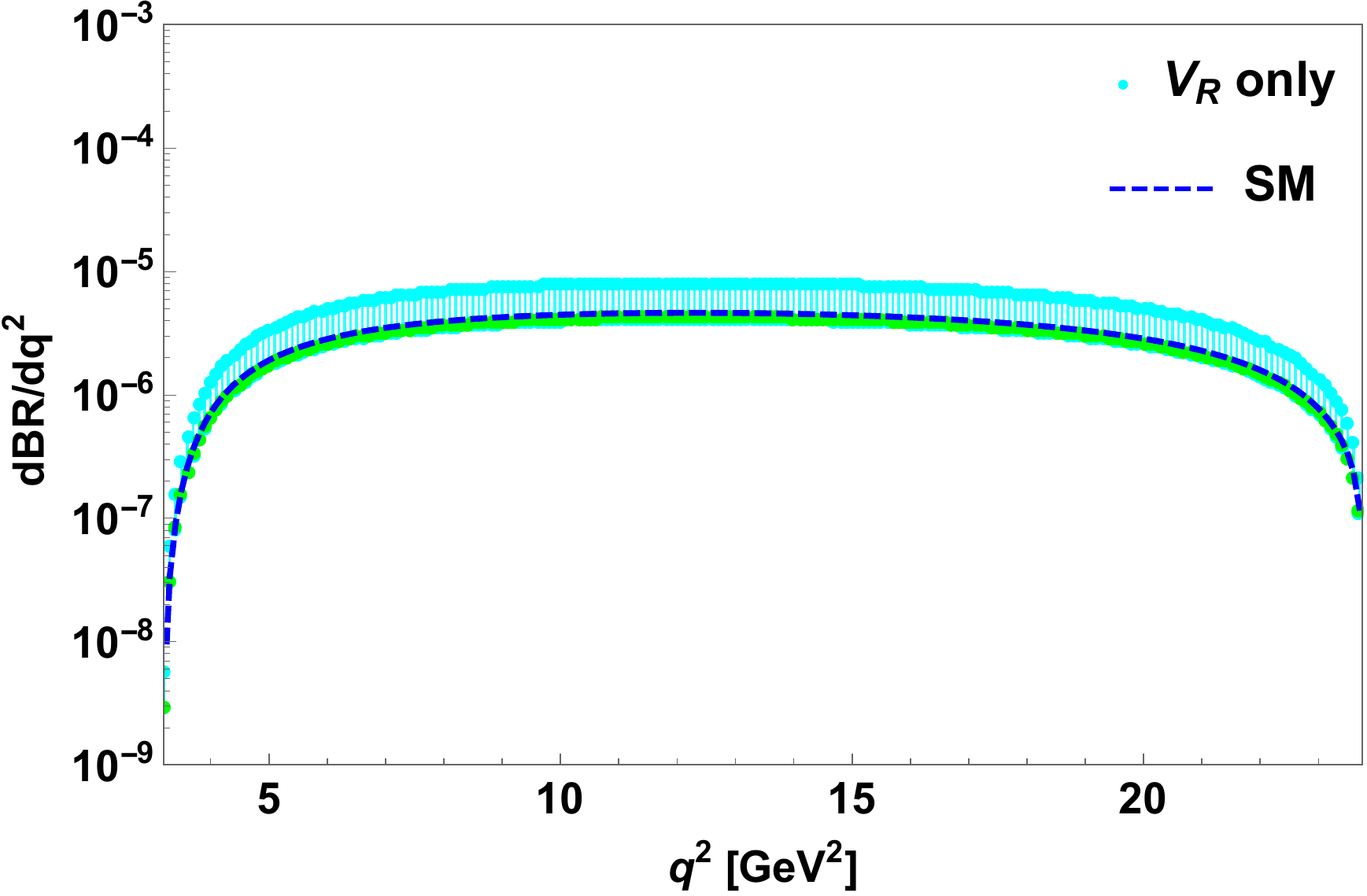}
\quad
\includegraphics[scale=0.4]{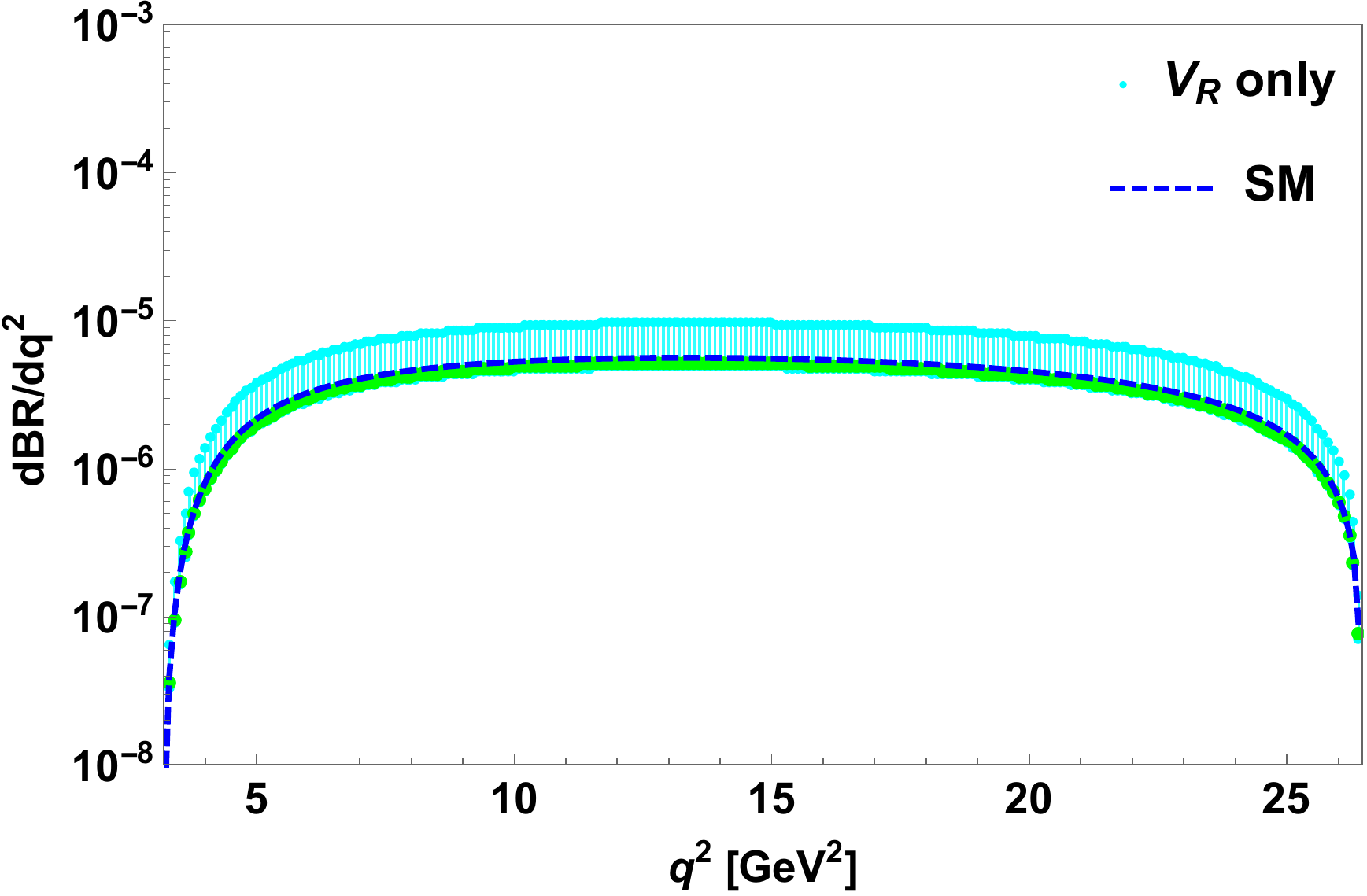}
\quad
\includegraphics[scale=0.4]{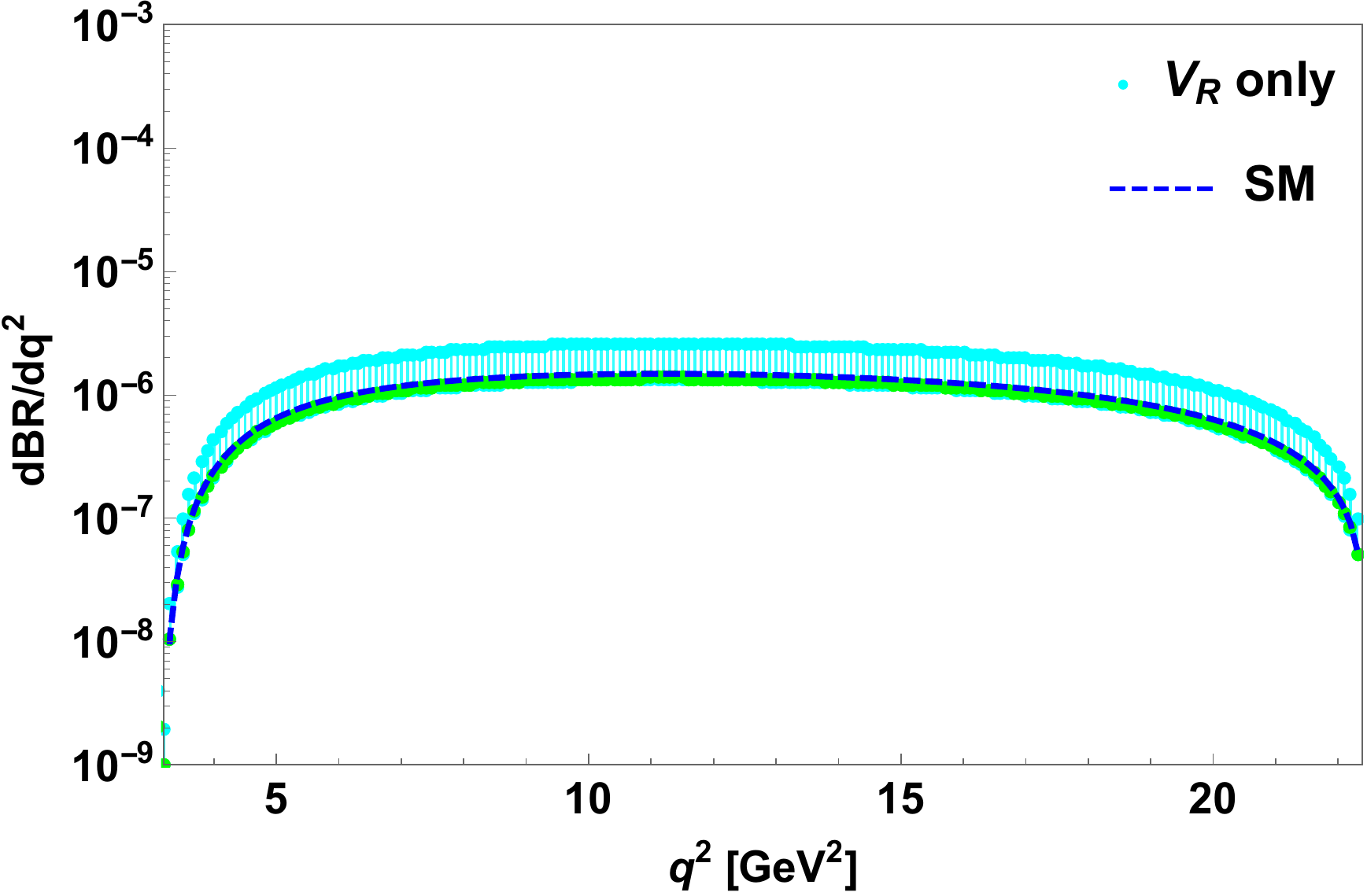}
\quad
\includegraphics[scale=0.4]{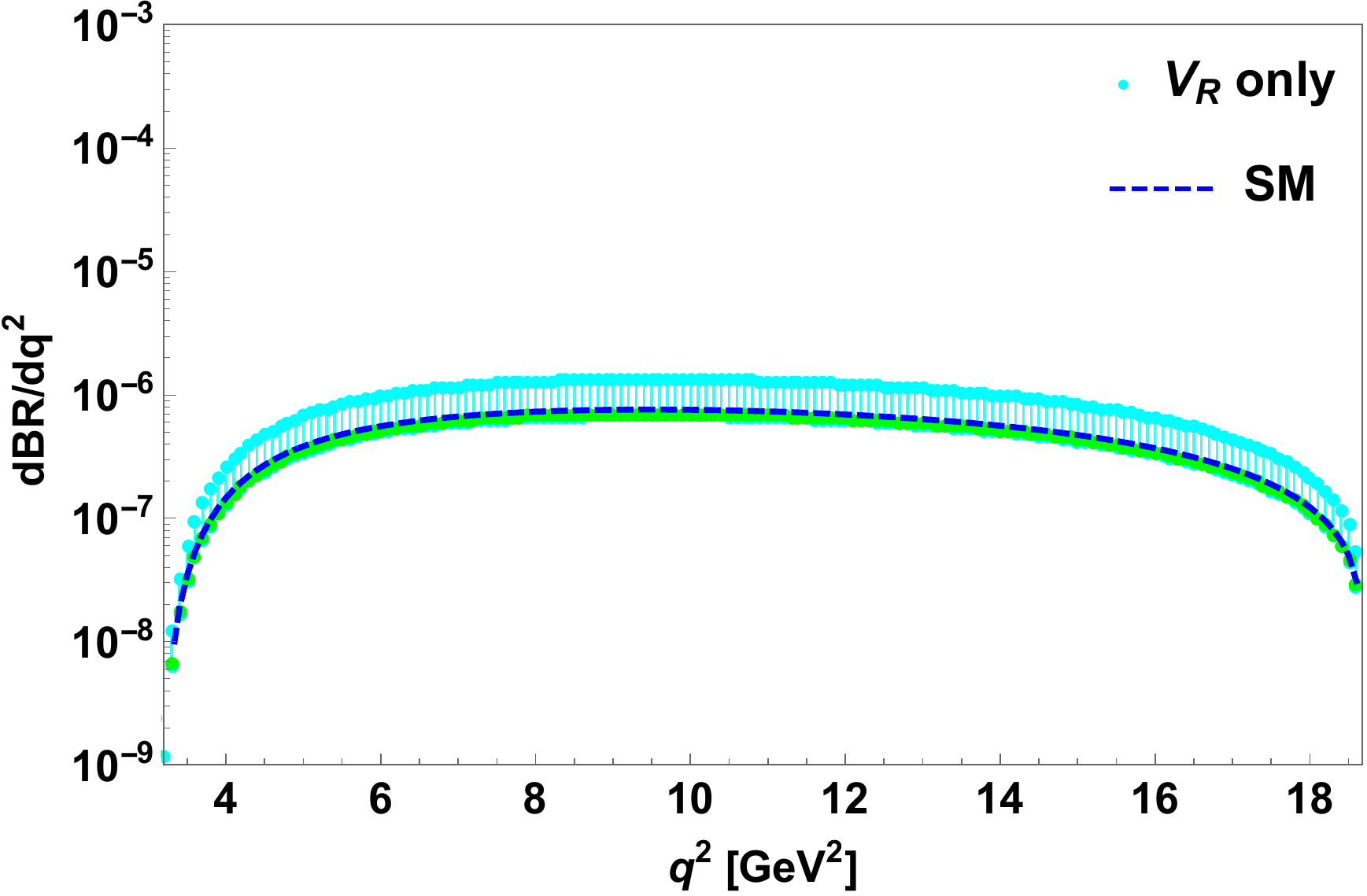}
\caption{The plots for the branching ratios of  $B_s \to K^+ \tau^- \bar \nu_\tau$  (top-left panel),  $\bar B^0  \to \pi^+ \tau^- \bar \nu_\tau$  (top-right panel), $ B^- \to \eta \tau^- \bar \nu_\tau$  (bottom-left panel) and $B^- \to \eta^{\prime } \tau^- \bar \nu_\tau$  (bottom-right panel) processes for the NP contribution of only $V_R$  coupling. Here the cyan bands are for the  $V_R$ NP coupling contributions.  } \label{brtau-VR}
\end{figure}
\begin{figure}[h]
\centering
\includegraphics[scale=0.4]{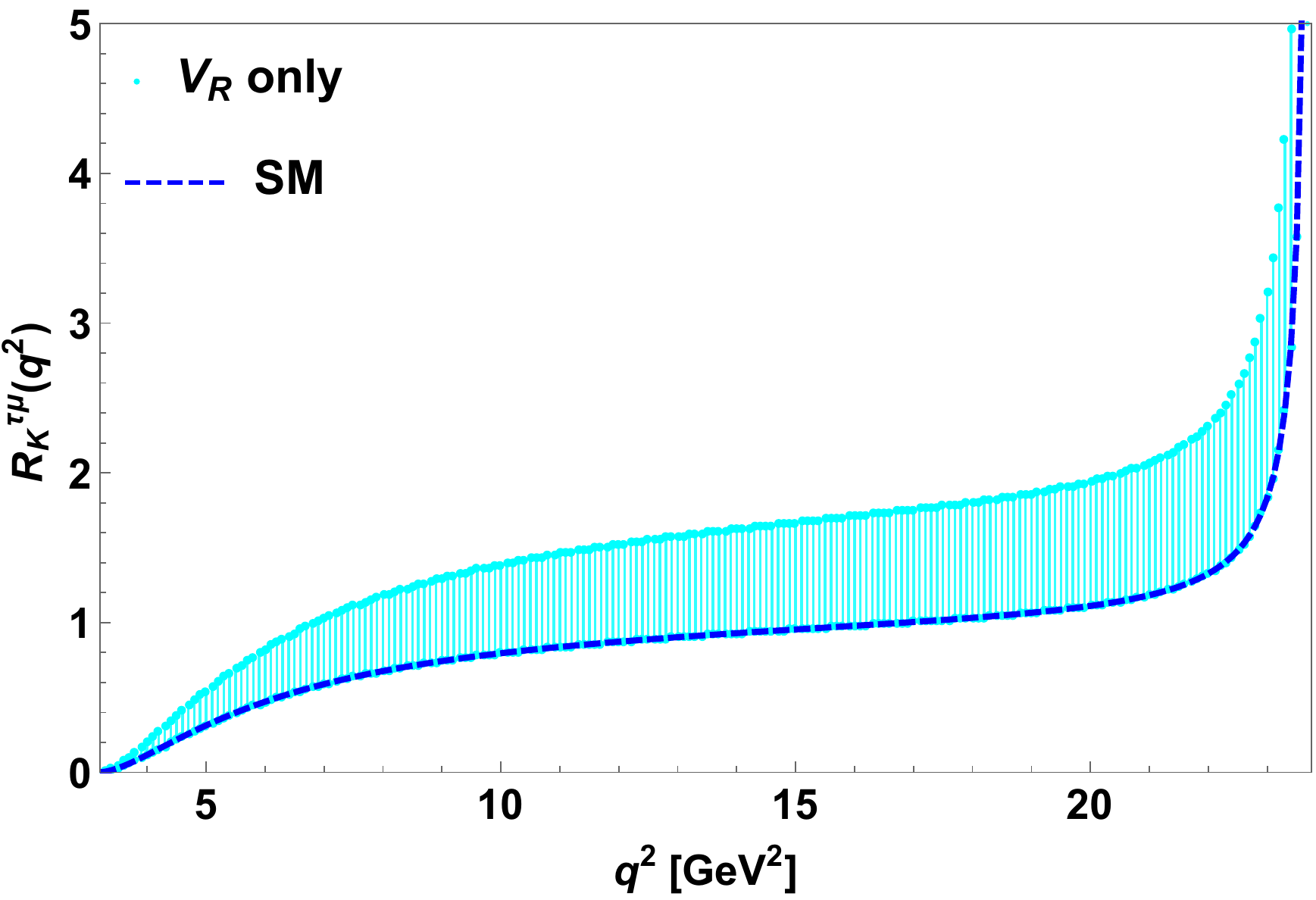}
\quad
\includegraphics[scale=0.4]{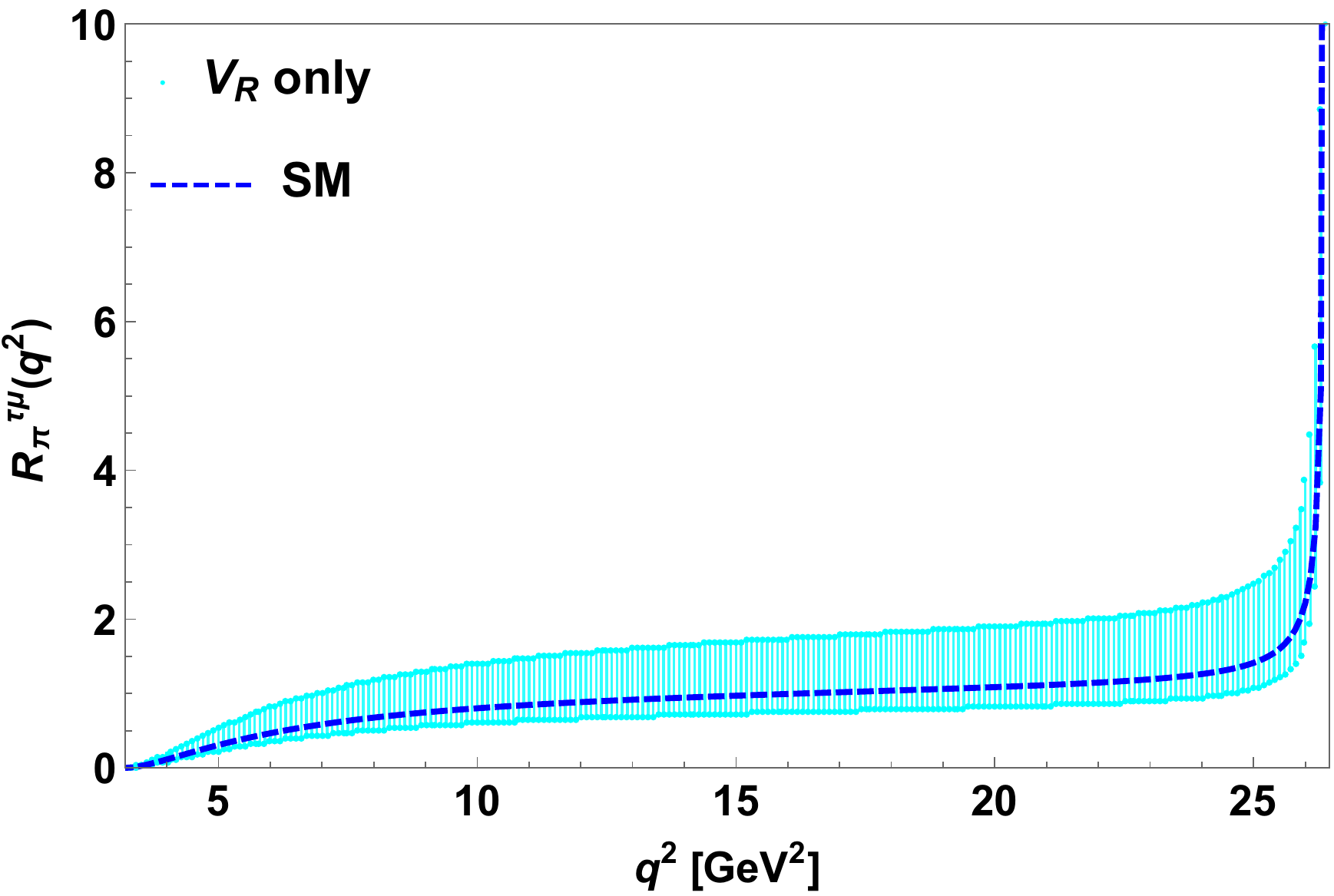}
\quad
\includegraphics[scale=0.4]{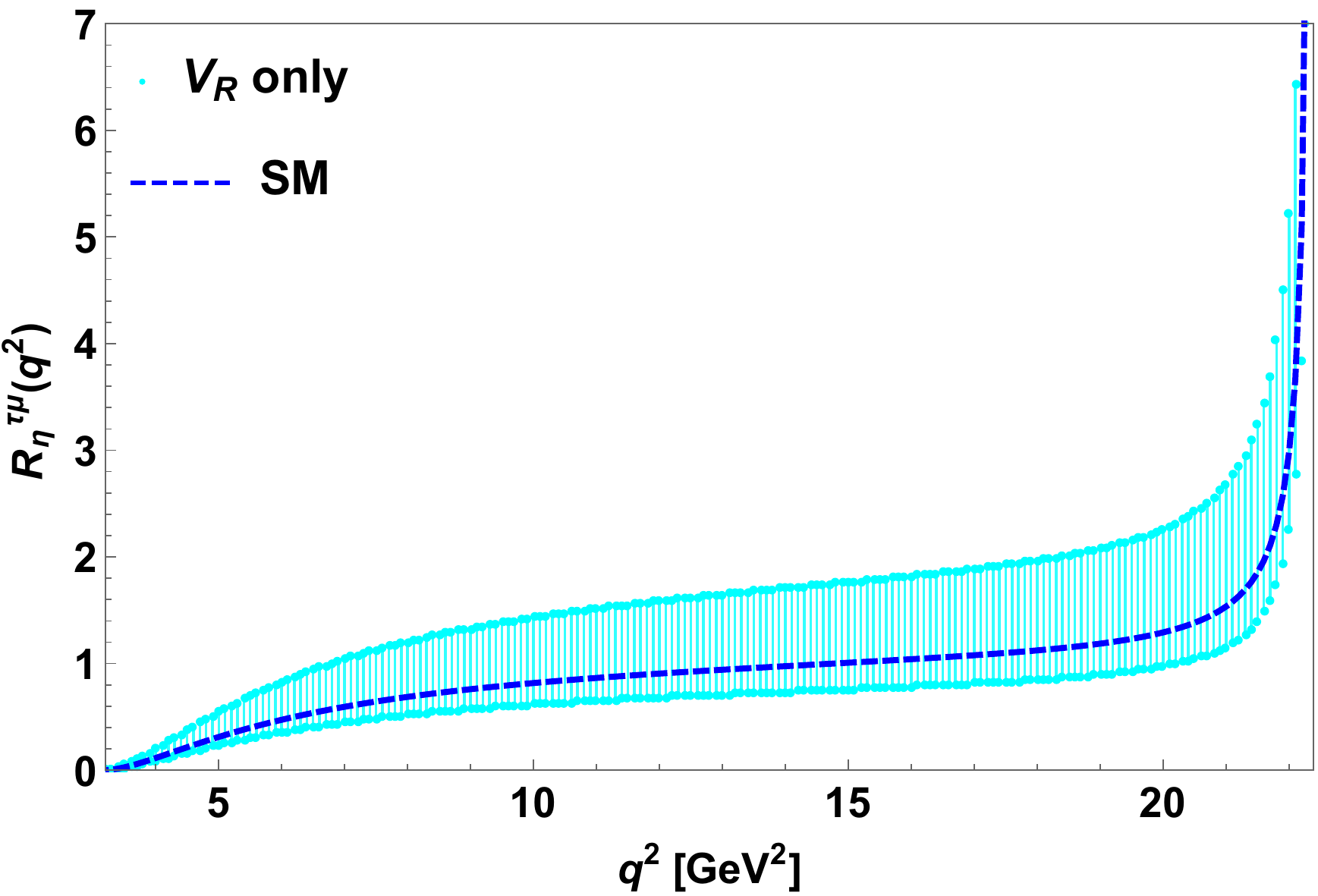}
\quad
\includegraphics[scale=0.4]{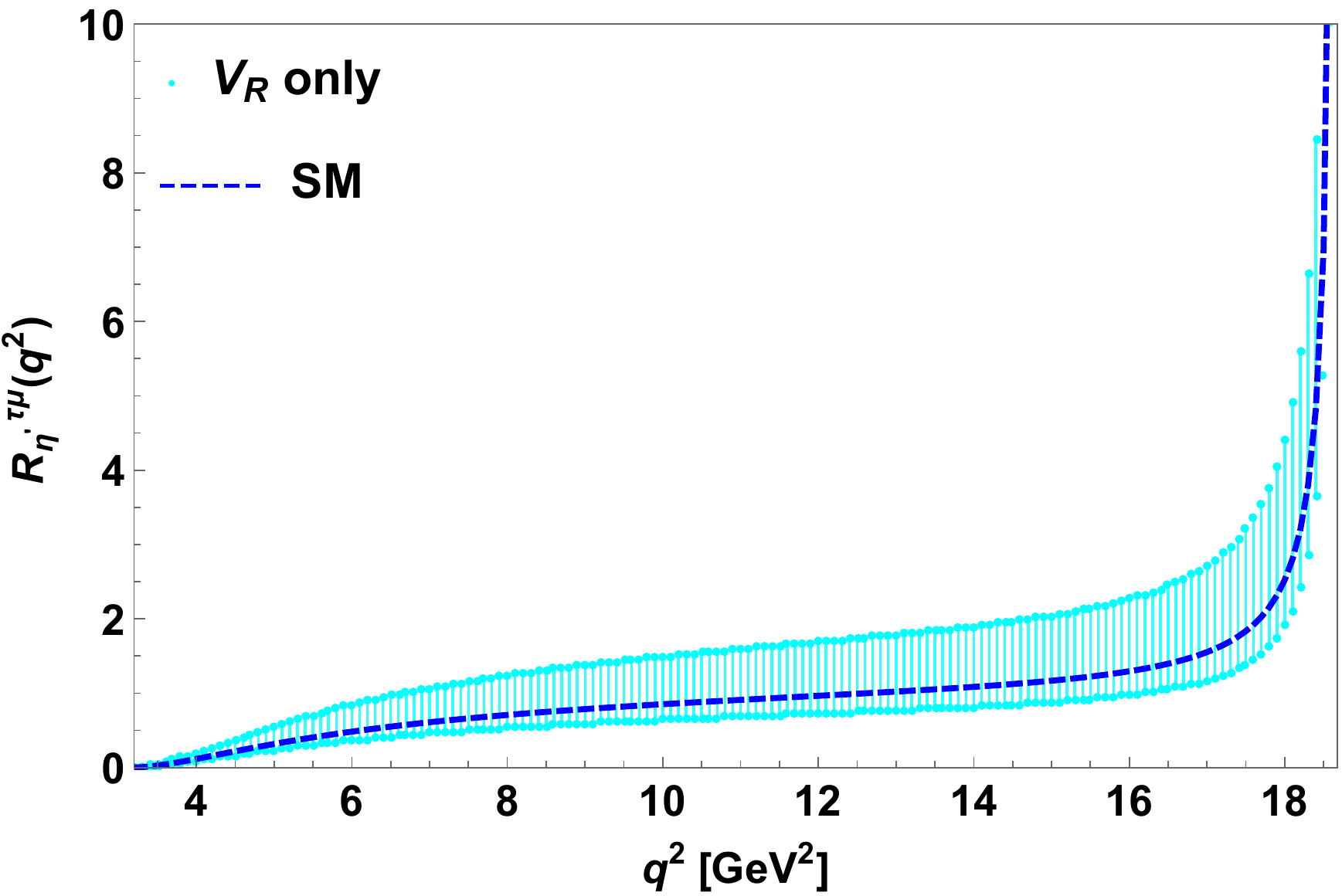}
\caption{The plots for the LNU parameters  $R_K^{\tau \mu }(q^2)$ (top-left panel), $R_\pi^{\tau \mu}(q^2)$ (top-right panel), $R_\eta^{\tau \mu}(q^2)$ (bottom-left panel) and $R_{\eta^\prime}^{\tau \mu}(q^2)$ (bottom-right panel). } \label{RK-VR}
\end{figure}

 Here we consider the effect of only $V_R$ coefficient in addition to the SM contribution.  The constraints obtained on real and imaginary parts of $V_R$ coupling  from $B_u \to \tau \nu$ process  are related to that of $V_L$ as ${\rm Re}[V_R]=-{\rm Re}[V_L]$ and ${\rm Im}[V_R]={\rm Im}[V_L]$, and thus, allowed parameter space for $V_R$ is same as that of $V_L$ with a sign flip for the real parts. The minimum and maximum values of the $V_R$ parameters are obtained using the  extrema conditions as    $({\rm Re}[V_R], {\rm Im}[V_R])^{\rm max}= (-0.242, -0.561 )$ and  $({\rm Re}[V_R], {\rm Im}[V_R])^{\rm min}= (0.259,  -0.406)$. 
 However, the constraints on $V_R$  obtained from $B^-\to \pi^0 \mu^- \bar  \nu_\mu$ for  
$b \to u \mu \bar \nu_\mu$ transition are same as $V_L$. Thus, the predicted branching ratios for $B \to P \mu \bar \nu_\mu$ processes in the presence of   $V_R$ coupling are same as those with $V_L$ coupling.
Using  the allowed values of the couplings,   the plots for the branching ratios of $\bar B_s \to K^+ \tau^-  \bar \nu_\tau$ (top-left panel), $\bar B^0 \to \pi^+ \tau^-  \bar \nu_\tau$ (top-right panel),   $B^- \to \eta \tau^- \bar \nu_\tau$ (bottom-left panel) and $B^- \to \eta^{\prime } \tau^-\bar  \nu_\tau$ (bottom-right panel) processes in the presence of  $V_R$ coupling are shown in  Fig. \ref{brtau-VR}. In these plots, the cyan bands are obtained by using the allowed parameter space of $V_R$. 
The predicted integrated values of branching ratios of these processes are listed in Table II. Like the previous case,  the forward-backward asymmetry parameters are also not affected due to  $V_R$ coupling. In Fig. \ref{RK-VR}, we present the plots for  the LNU parameters $R_K^{\tau \mu}(q^2)$ (top-left panel), $R_\pi^{\tau \mu} (q^2)$ (top-right panel), $R_\eta^{\tau \mu} (q^2) $ (bottom-left panel) and $R_{\eta^\prime}^{\tau \mu} (q^2)$ (bottom-right panel).  In the presence of $V_R$ coupling, the  parameters $R_{\pi K}^l$, $R_{\pi \eta^{(\prime)}}^l$  don't have  any deviation from their corresponding SM predictions.  In Table III, we present the   numerical values of these  parameters. 

 \begin{table}[htb]
\begin{center}
\caption{The predicted branching ratios and  forward-backward asymmetries  of  $\bar B_{(s)} \to P l \bar \nu_l$ processes, where $P=K, \pi, \eta^{(\prime)}$ and  $l=\mu, \tau$ in the SM and in the presence of   $V_{L, R}$ NP couplings. }
\begin{tabular}{| c |c | c | c| c | }
\hline
Observables~ & Values in the SM~ &~ Values for $V_L$ coupling~&~ Values for $V_R$ coupling \\
 \hline
 \hline
 
  ${\rm BR}(\bar B_s \to K^+  \mu^- \bar \nu_\mu)$   ~&~ $(1.03 \pm 0.082) \times 10^{-4}$~ &~ $(1.03-1.22) \times 10^{-4}$ ~&~ $(1.03-1.22) \times 10^{-4}$ ~\\
 
${\rm BR}(\bar B_s \to K^+  \tau^- \bar \nu_\tau)$ ~ &~ $(6.7 \pm 0.536) \times 10^{-5}$~ &~ $(0.477-1.24) \times 10^{-4}$ ~&~ $(0.6-1.17) \times 10^{-4}$ ~ \\

$\langle A_{FB}^\mu \rangle$ ~&~$(2.98 \pm 0.238) \times 10^{-3}$ ~ &~ $2.98  \times 10^{-3}$~&~$2.98 \times 10^{-3}$~ \\

$\langle A_{FB}^\tau \rangle$ ~&~$0.275 \pm 0.022 $ ~ &~ $0.275$~&~$0.275$~ \\
\hline

${\rm BR}(\bar B \to \pi^+  \mu^-\bar  \nu_\mu)$   ~&~ $(1.35 \pm 0.1) \times 10^{-4}$~ &~ $(1.35-1.59) \times 10^{-4}$ ~&~ $(1.35-1.59) \times 10^{-4}$ ~\\
 
${\rm BR}(\bar B \to \pi^+  \tau^- \bar \nu_\tau)$ ~ &~ $(9.4 \pm 0.752) \times 10^{-5}$~ &~ $(0.67-1.75) \times 10^{-4}$ ~ &~ $(0.824-1.62) \times 10^{-4}$ ~\\

 $\langle A_{FB}^\mu \rangle$ ~&~$(2.94 \pm 0.235) \times 10^{-3}$ ~ &~ $2.94 \times 10^{-3}$~&~$2.94  \times 10^{-3}$~ \\

$\langle A_{FB}^\tau \rangle$ ~&~$(0.27 \pm 0.021) $ ~ &~ $0.27$~&~$0.27$~ \\
\hline

${\rm BR}(B^- \to \eta  \mu^-\bar  \nu_\mu)$   ~&~ $(3.143 \pm 0.25) \times 10^{-5}$~ &~ $(3.143-3.7) \times 10^{-5}$ ~&~ $(3.143-3.7) \times 10^{-5}$ ~\\
 
${\rm BR}(B^- \to \eta  \tau^- \bar \nu_\tau)$ ~ &~ $(1.96 \pm 0.16) \times 10^{-5}$~ &~ $(1.4-3.64) \times 10^{-5}$ ~&~ $(1.75-3.43) \times 10^{-5}$ ~ \\

$\langle A_{FB}^\mu \rangle$ ~&~$(3.45 \pm 0.276) \times 10^{-3}$ ~ &~ $3.45  \times 10^{-3}$~&~$3.45  \times 10^{-3}$~ \\

$\langle A_{FB}^\tau \rangle$ ~&~$(0.292 \pm 0.023) $ ~ &~ $0.292$~&~$0.292$~ \\
\hline

${\rm BR}(B^- \to \eta^{\prime }  \mu^- \bar  \nu_\mu)$   ~&~ $(1.45 \pm 0.116) \times 10^{-5}$~ &~ $(1.45-1.7) \times 10^{-5}$ ~&~ $(1.45-1.7) \times 10^{-5}$ ~\\
 
${\rm BR}(B^- \to \eta^{\prime }  \tau^- \bar \nu_\tau)$ ~ &~ $(7.81 \pm 0.06) \times 10^{-6}$~ &~ $(0.56-1.45) \times 10^{-5}$ ~&~ $(0.695-1.37) \times 10^{-5}$ ~ \\

$\langle A_{FB}^\mu \rangle$ ~&~$(4.1 \pm 0.328) \times 10^{-3}$ ~ &~ $4.1  \times 10^{-3}$~&~$4.1 \times 10^{-3}$~ \\

$\langle A_{FB}^\tau \rangle$ ~&~$(0.317 \pm 0.026) $ ~ &~ $0.317$~&~$0.317$~ \\

 \hline
\end{tabular}
\end{center}
\end{table} 

\begin{table}[htb]
\begin{center}
\caption{The predicted values of various  parameters ($R_P^{\tau \mu}$ and $R_{PP'}^l$) of  $\bar B_{(s)} \to P l \bar \nu_l$ processes in the SM and in the presence of   $V_{L, R}$ NP couplings. }
\begin{tabular}{| c |c | c | c| c | }
\hline
Observables~ & Values in the SM~ &~ Values for $V_L$ coupling~&~ Values for $V_R$ coupling \\
 \hline
 \hline

$R_K^{\tau \mu}$ &~ $0.649$~ &~ $0.46-1.02$ ~&~ $0.489-1.13$ ~ \\
 
 $R_\pi^{\tau \mu}$ &~ $0.7$~ &~ $0.497-1.1$ ~&~ $0.528 - 1.22$ ~  ~\\

$R_\eta^{\tau \mu}$ &~ $0.624$~ &~ $0.45-0.982$~&~ $0.47-1.09$ ~ ~ \\

 $R_{\eta^\prime}^{\tau \mu}$ &~ $0.54$~ &~ $0.385-0.85$ ~ &~ $0.408-0.946$ ~ ~\\
 
 \hline
 
 $R_{\pi K}^{\mu }$ &~ $1.31$~ &~ $1.3-1.31$ ~ &~ $1.3-1.31$ ~ ~\\
  
$R_{\pi \eta}^{\mu}$   ~&~ $4.3$~ &~ $4.3$ ~&~ $4.3$ ~ ~\\
 
$R_{\pi \eta^\prime}^{\mu }$ ~ &~ $9.3$ ~&~ $9.3-9.35$ ~&~ $9.3-9.35$ ~  ~\\
 
\hline

$R_{\pi K}^{\tau}$   ~&~ $1.4$~ &~ $1.4-1.41$ ~&~ $1.373-1.39$ ~ ~\\

$R_{\pi \eta}^{\tau}$   ~&~ $4.8$~ &~ $4.785-4.808$ ~&~ $4.709-4.723$ ~~\\
 
$R_{\pi \eta^\prime}^{\tau }$ ~ &~ $12.0$ ~&~ $11.96-12.1$ ~ &~ $11.82-11.86$ ~~\\
 
 \hline
\end{tabular}
\end{center}
\end{table}


\subsection{Case C: Effect of $S_L$ only}
In this subsection, we wish to see the effect of only $S_L$ coupling on various observables associated with $B \to P l  \bar \nu_l$ processes. For $b \to u \tau \nu$ transition, using the extrema conditions,  we obtain the maxima and minima  of    $S_L$ parameter as  $({\rm Re}[S_L], {\rm Im}[S_L])^{\rm max}=( -0.1063,  -0.0063)$ and 
$({\rm Re}[S_L], {\rm Im}[S_L])^{\rm min}=( 0.5397, 0.0244)$, 
from the allowed parameter space in the bottom panel of  Fig. \ref{Bulnu-SL}. 
Analogously,  for $b \to u \mu \bar \nu_\mu$, the extrema values of $S_L$ are found to be
$({\rm Re}[S_L], {\rm Im}[S_L])^{\rm max}=( -0.163,  0.252)$ and 
$({\rm Re}[S_L], {\rm Im}[S_L])^{\rm min}=( 0.017, 0.176)$ and the corresponding $1\sigma$ range of allowed parameter space is shown in the right panel of Fig. \ref{B2Pi}.
 Including the additional  contributions from $S_L$ coupling, the obtained branching ratios for various processes are listed in Table IV. It is observed that  
  the branching ratios of  $\bar B_{(s)} \to P^+ \tau^- \bar  \nu_\tau$ processes comparatively  deviate more  than the corresponding processes with muon in the final state.  
\begin{figure}[h]
\centering
\includegraphics[scale=0.4]{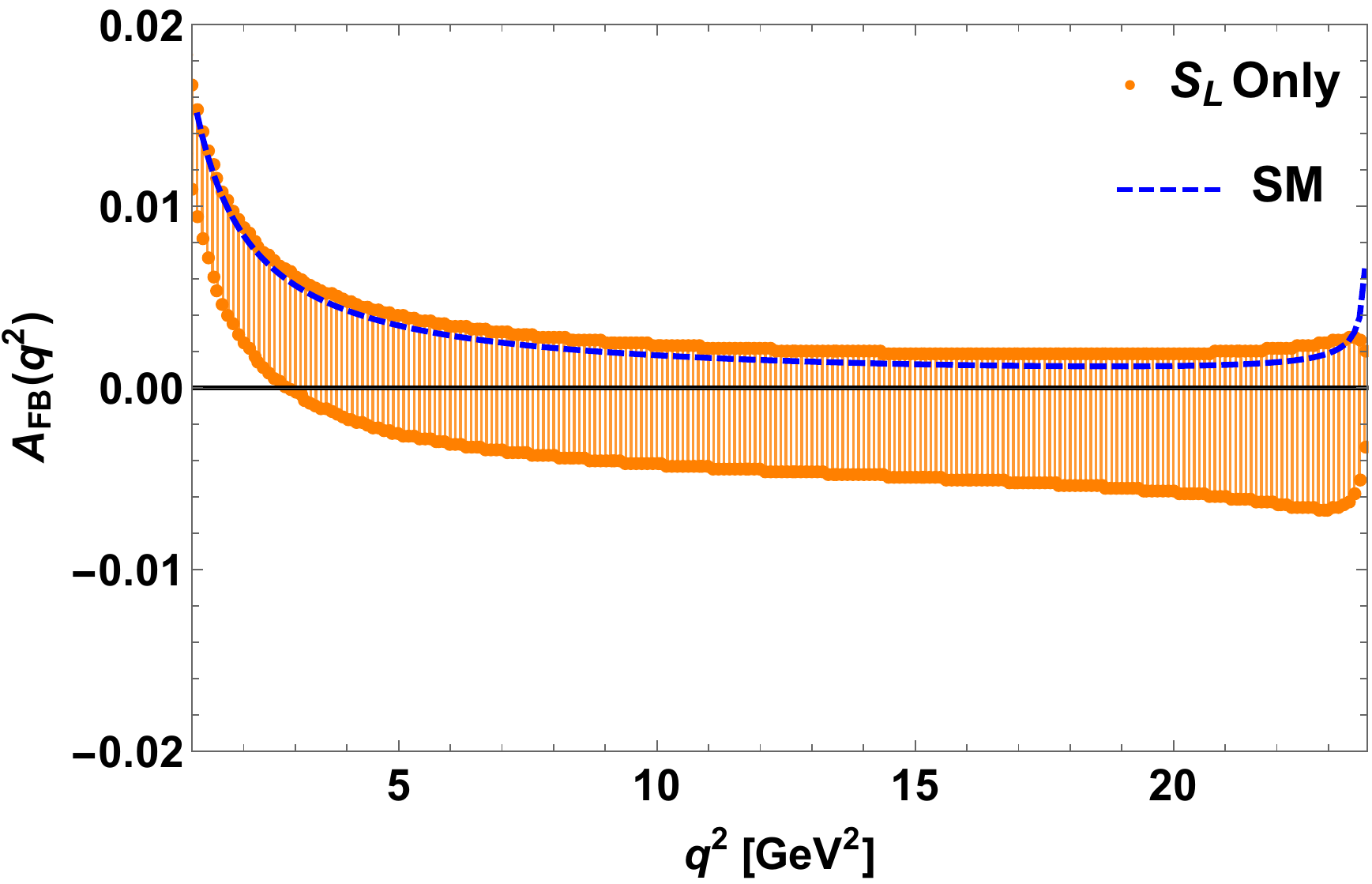}
\quad
\includegraphics[scale=0.4]{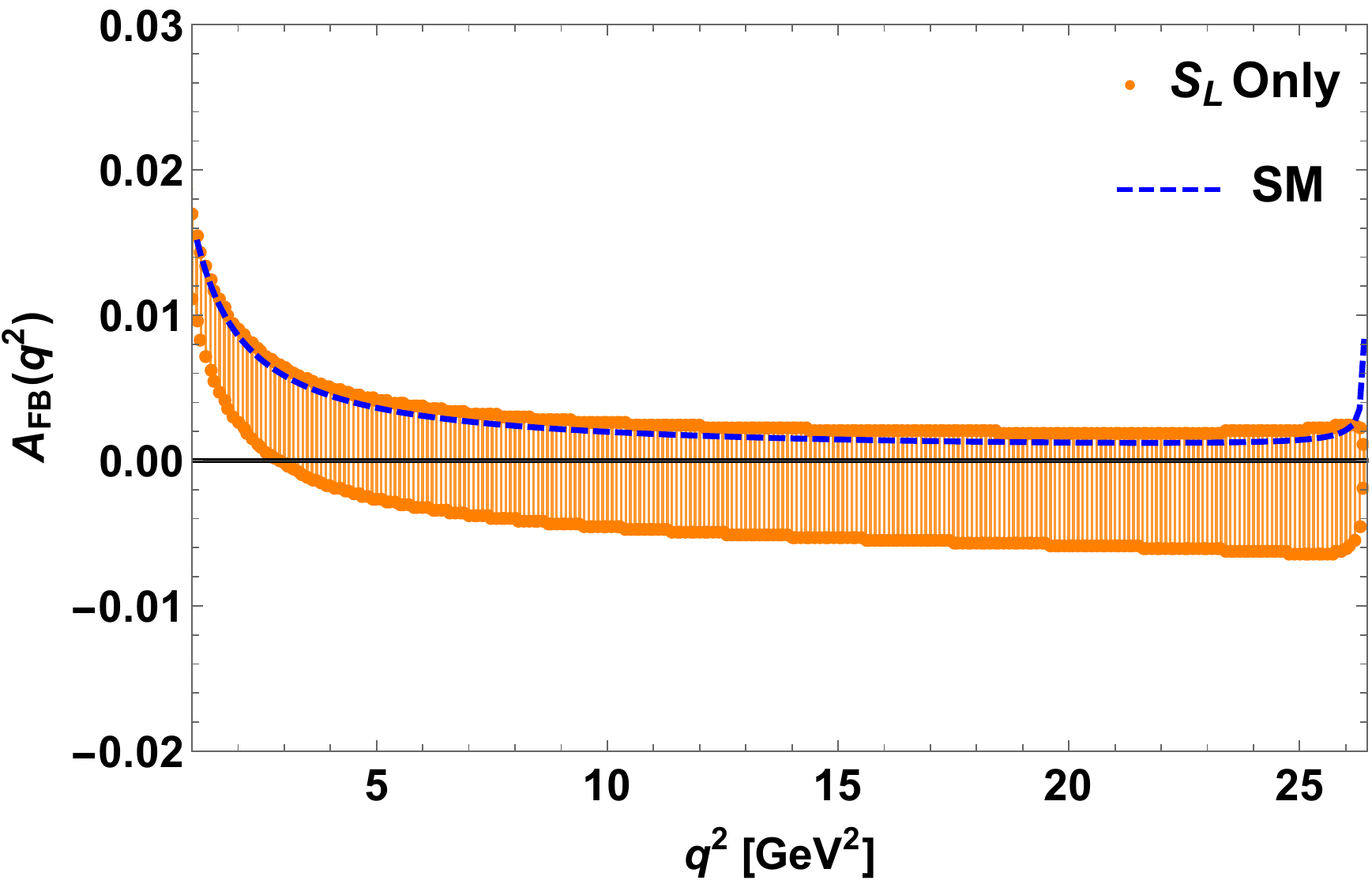}
\quad
\includegraphics[scale=0.4]{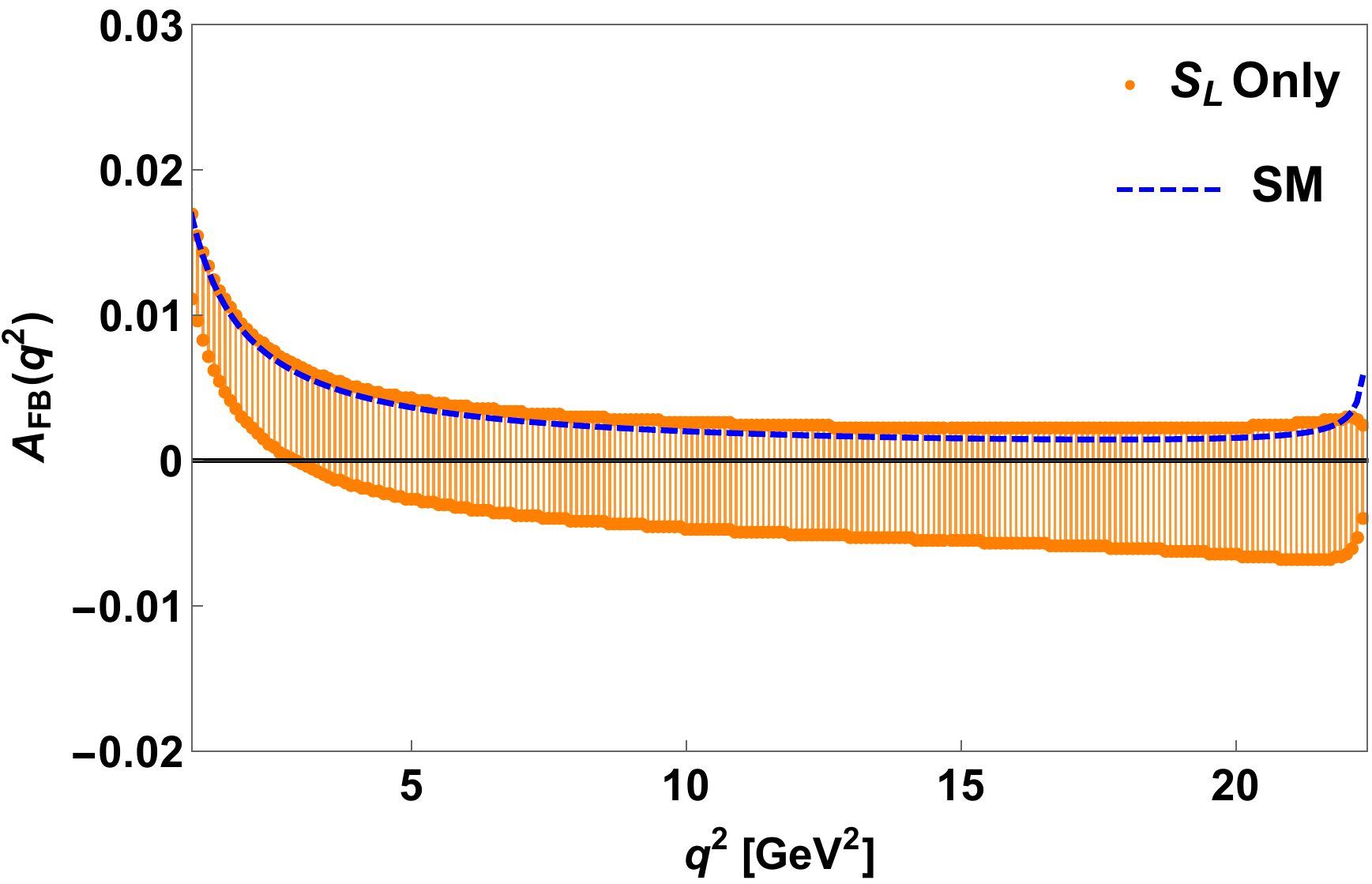}
\quad
\includegraphics[scale=0.4]{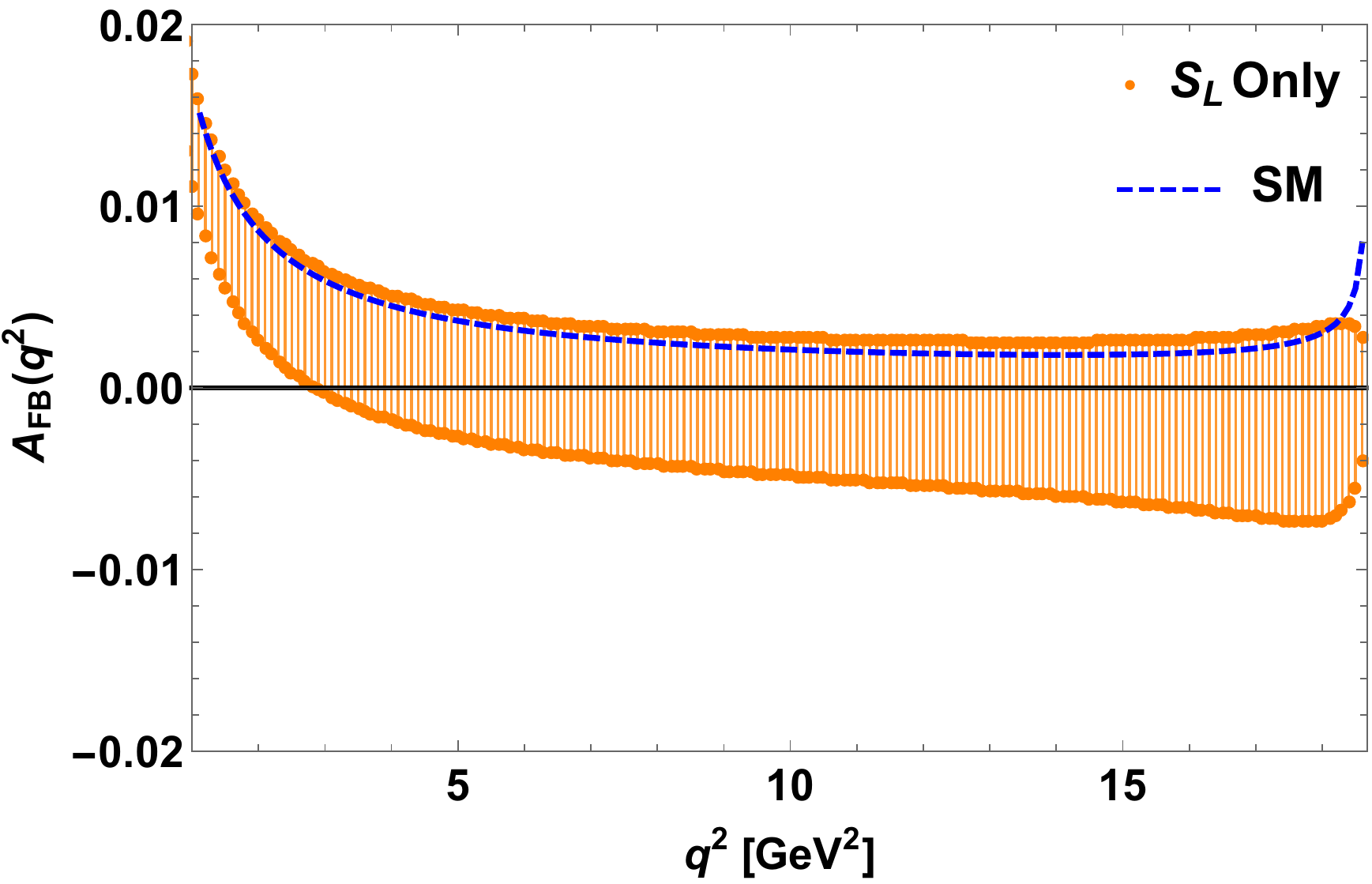}
\caption{The plots for the $q^2$ variation of  forward-backward asymmetry  of  $\bar B_s \to K^+ \mu^- \bar \nu_\mu$  (top-left panel),  $\bar B^0 \to \pi^+  \mu^- \bar  \nu_\mu$  (top-right panel), $B^- \to \eta \mu^- \bar \nu_\mu$  (bottom-left panel) and $B^- \to \eta^{\prime} \mu^- \bar \nu_\mu$  (bottom-right panel) processes. } \label{fbmu-SL}
\end{figure}
\begin{figure}[h]
\centering
\includegraphics[scale=0.4]{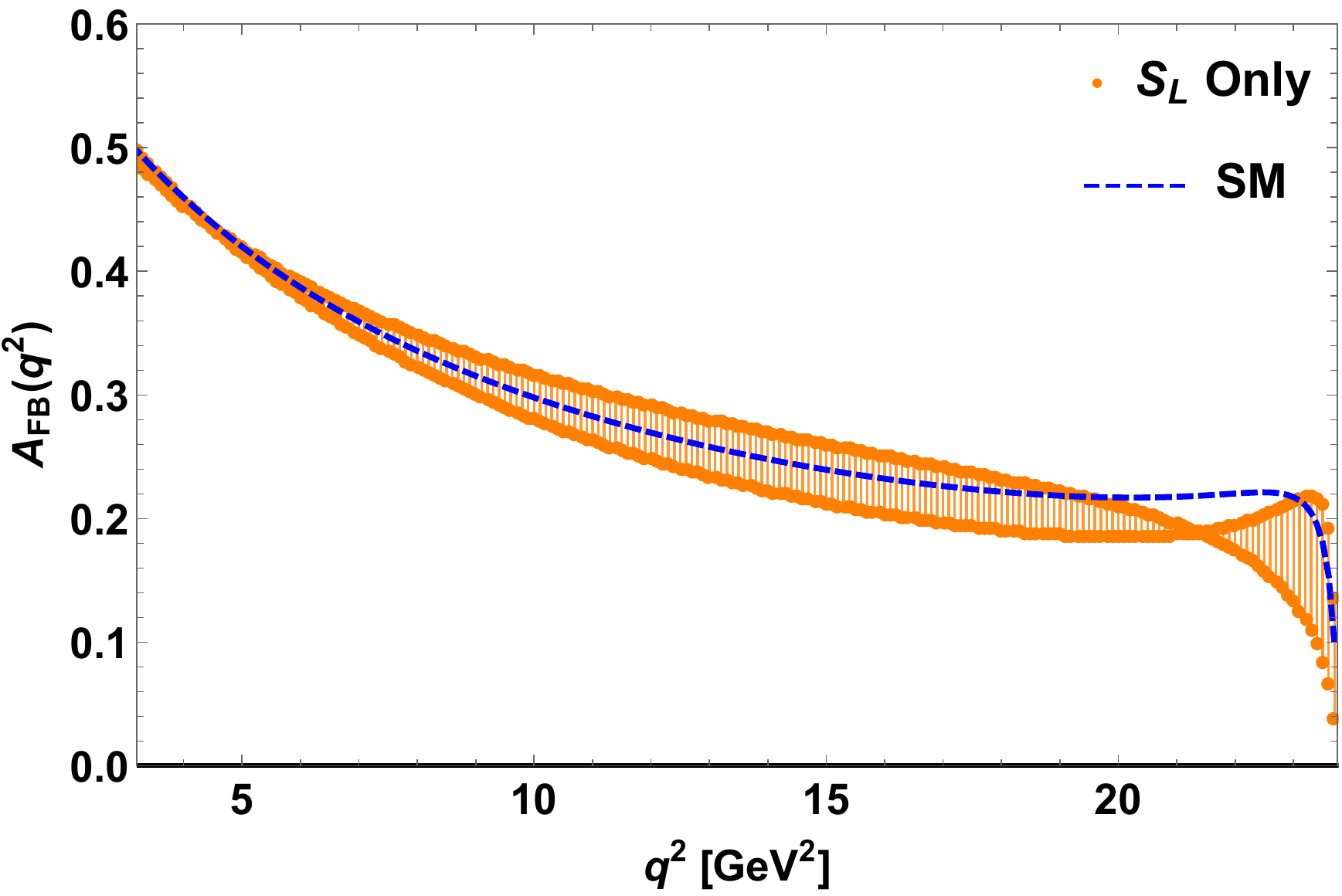}
\quad
\includegraphics[scale=0.4]{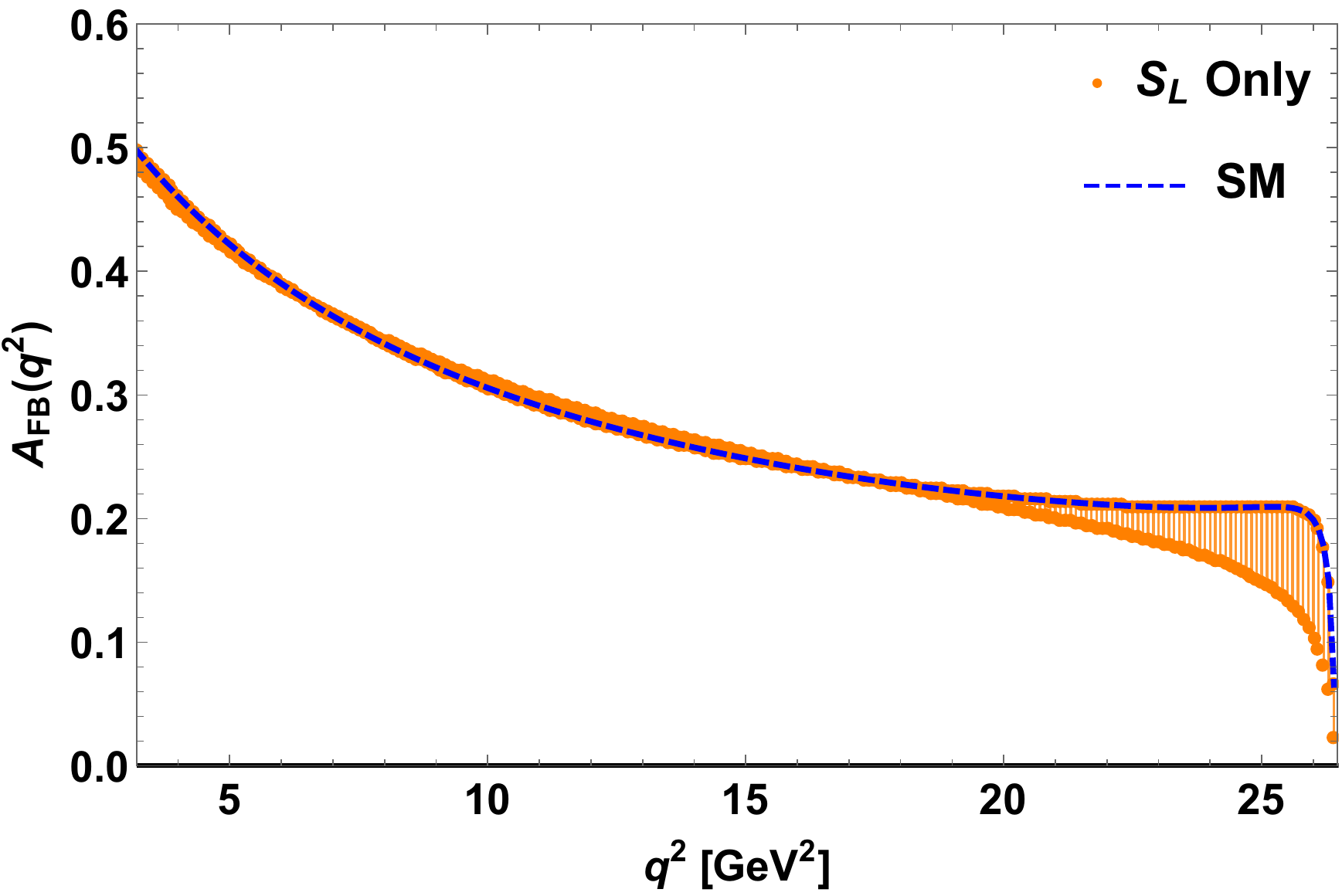}
\quad
\includegraphics[scale=0.4]{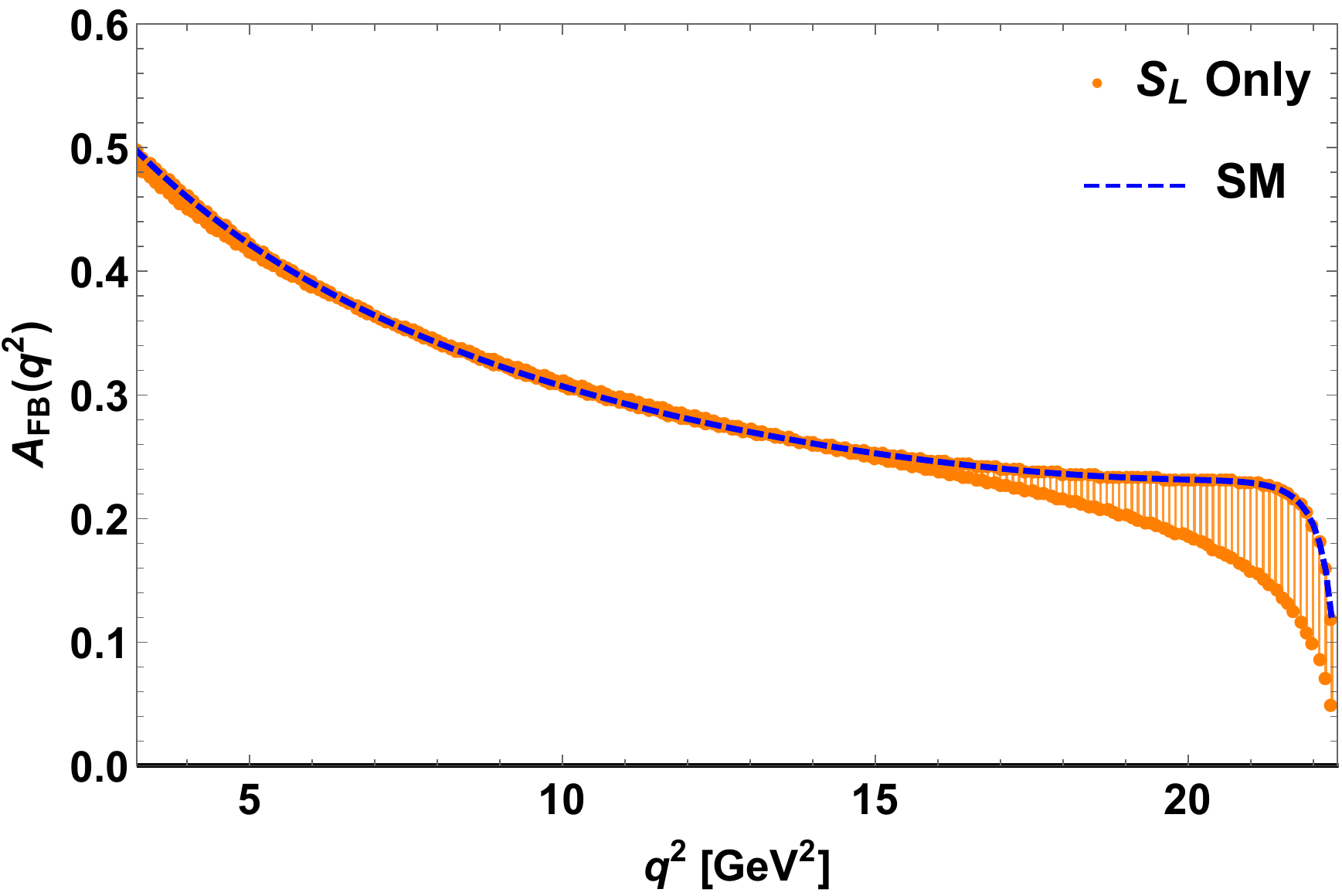}
\quad
\includegraphics[scale=0.4]{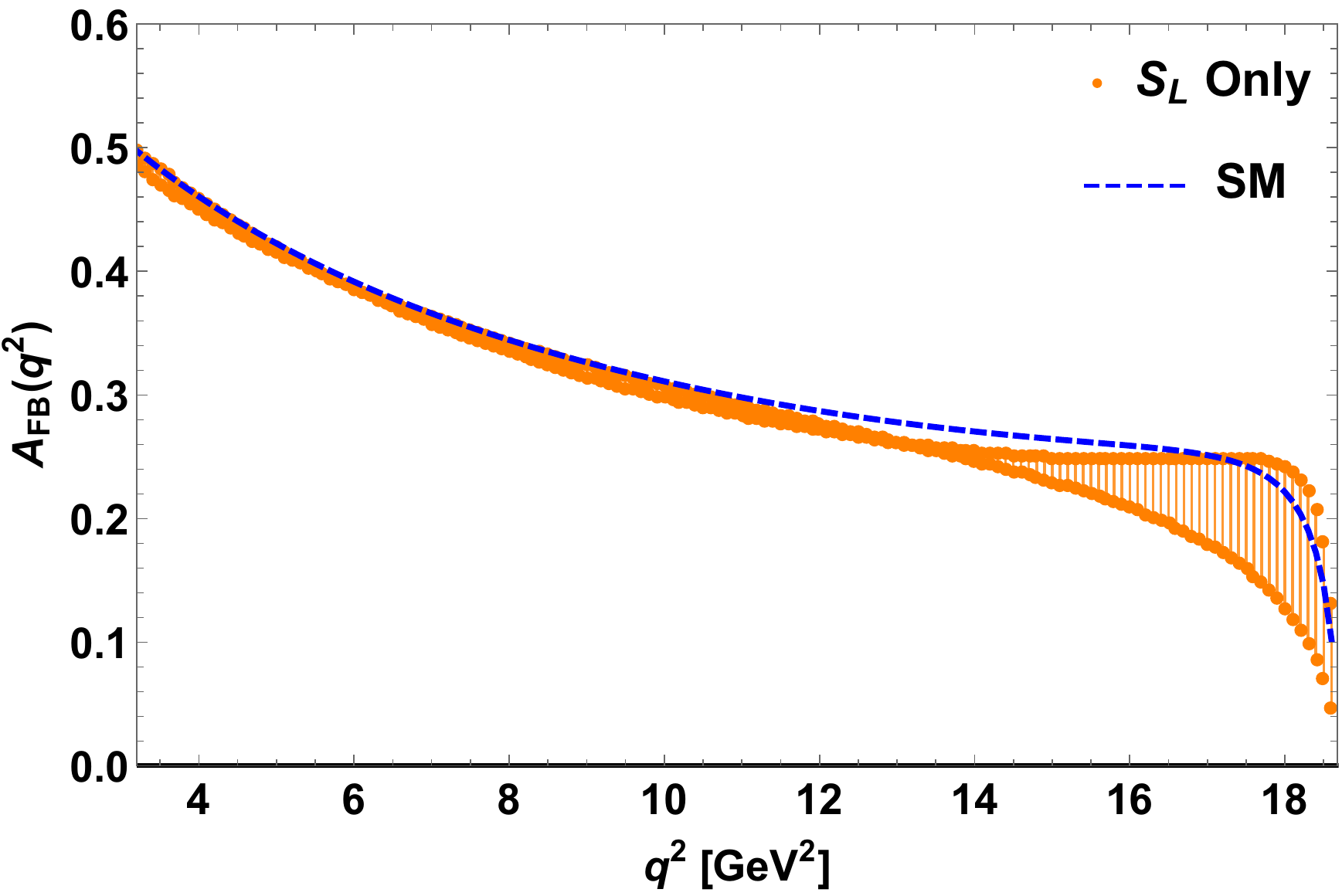}
\caption{The plots for the $q^2$ variation of  forward-backward asymmetry  of  $\bar B_s \to K^+ \tau^- \bar \nu_\tau$  (top-left panel),  $\bar B^0 \to \pi^+ \tau^- \bar \nu_\tau$  (top-right panel), $B^- \to \eta \tau^- \bar \nu_\tau$  (bottom-left panel) and $B^- \to \eta^{\prime } \tau^- \bar \nu_\tau$  (bottom-right panel) processes.} \label{fbtau-SL}
\end{figure}

\begin{figure}[h]
\centering
\includegraphics[scale=0.4]{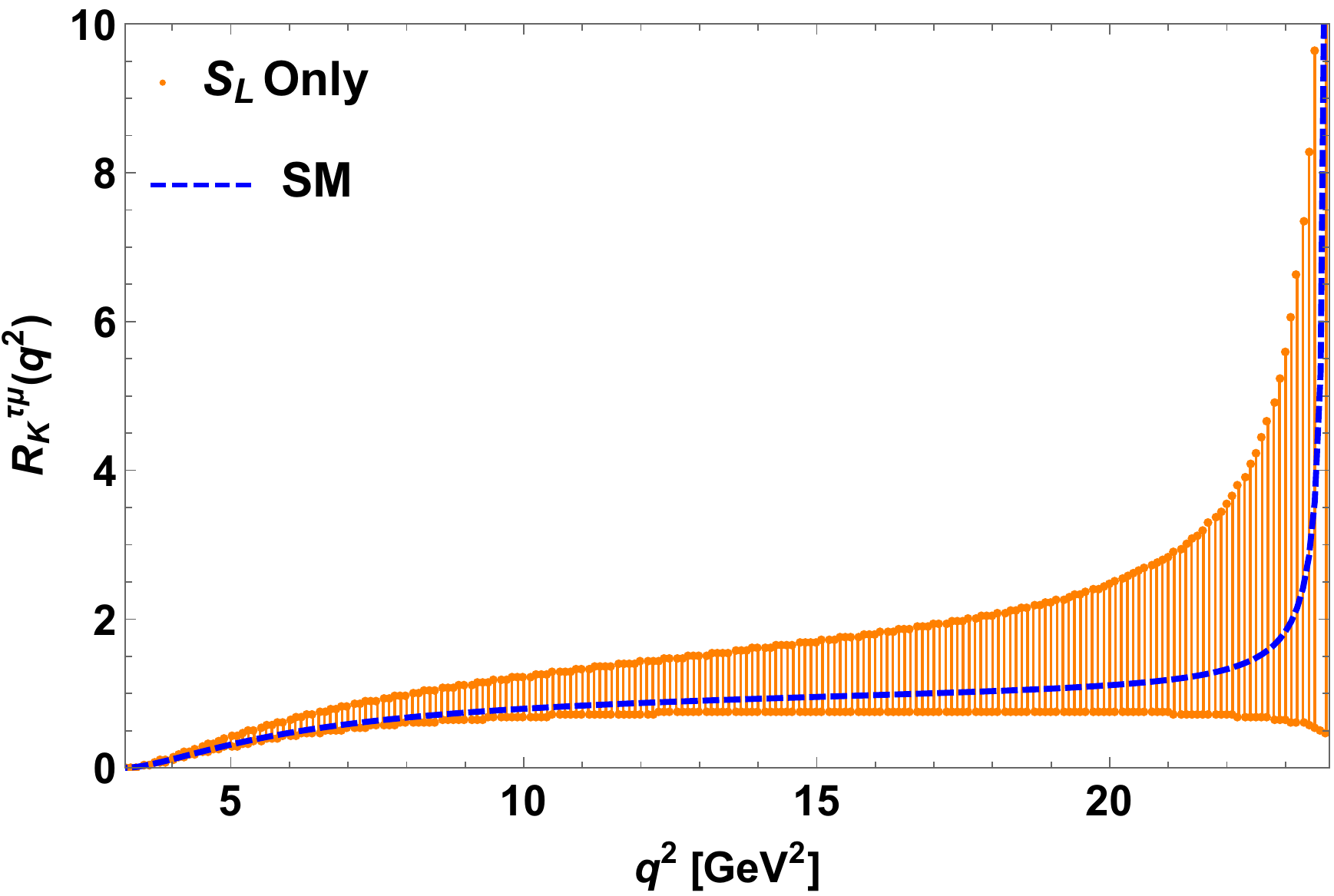}
\quad
\includegraphics[scale=0.4]{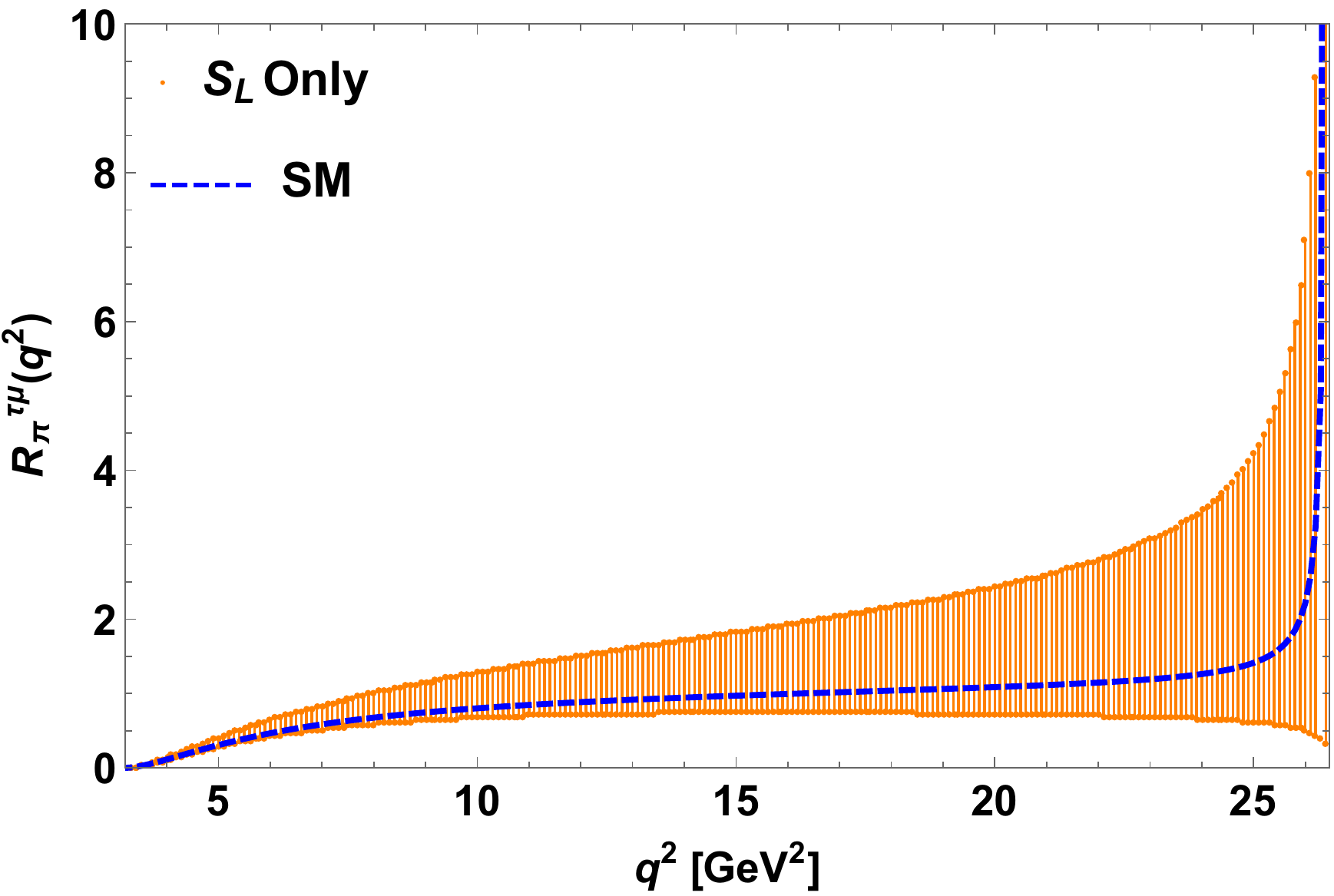}
\quad
\includegraphics[scale=0.4]{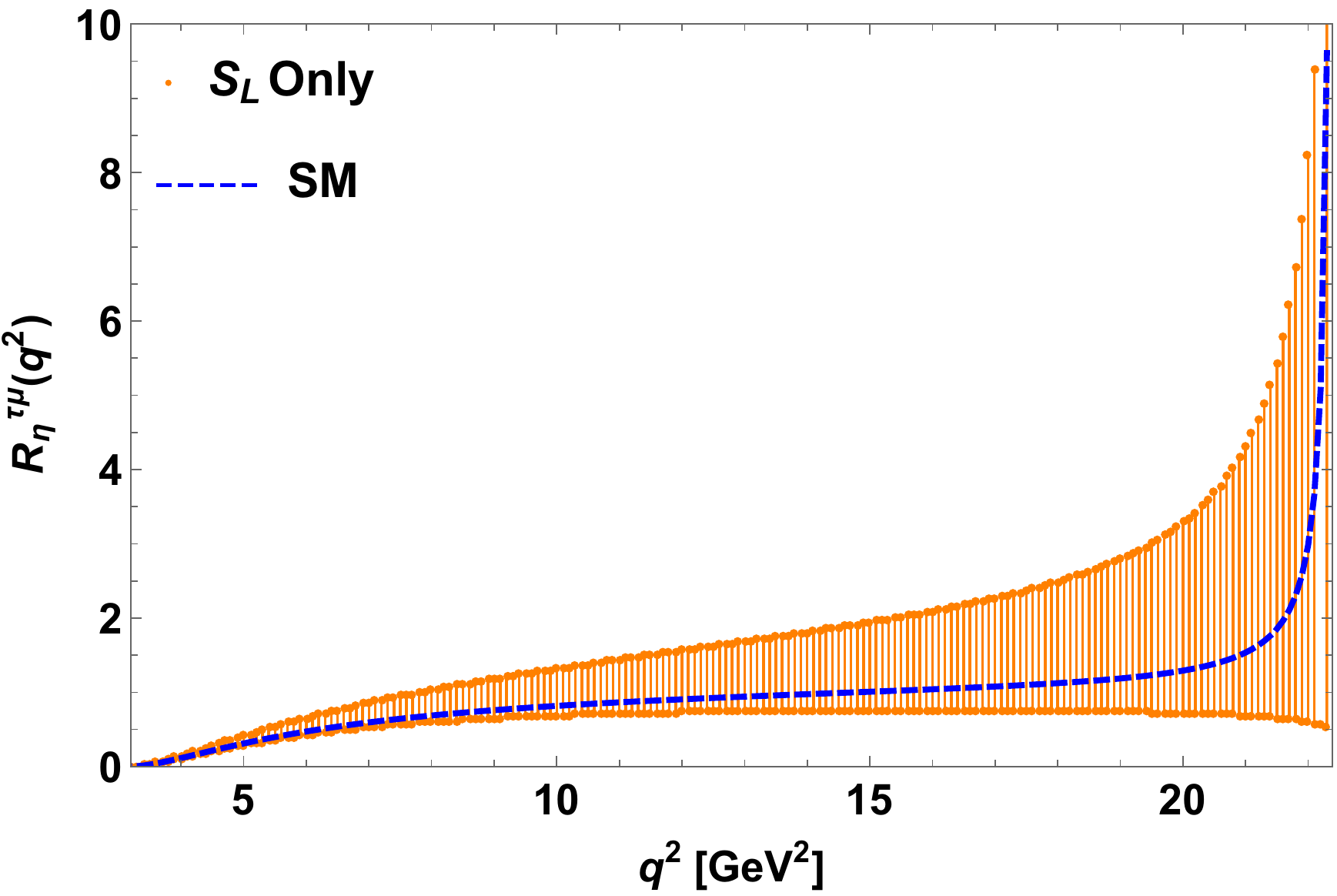}
\quad
\includegraphics[scale=0.4]{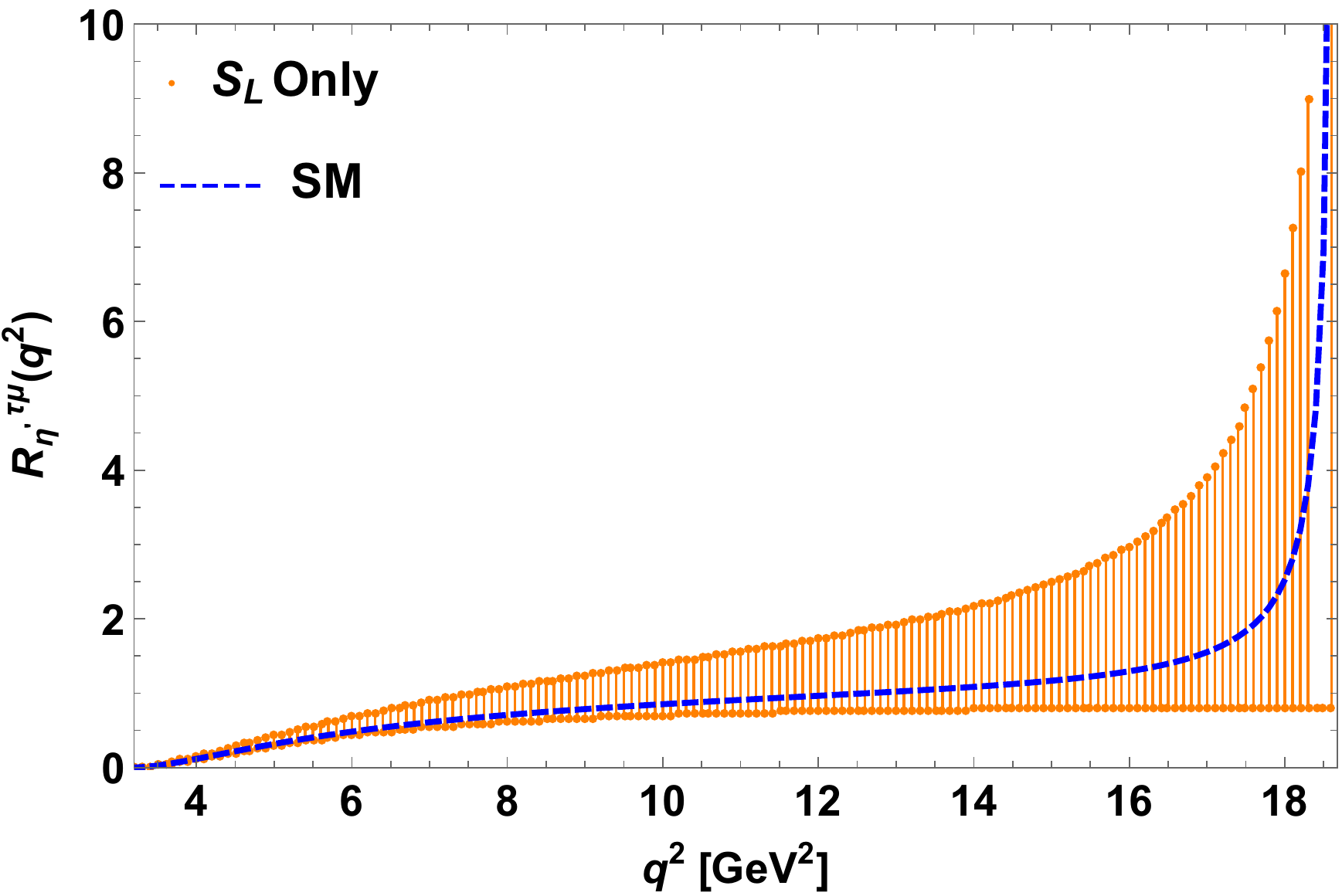}
\caption{The plots for the LNU parameters  $R_K^{\tau \mu }(q^2)$ (top-left panel), $R_\pi^{\tau \mu}(q^2)$ (top-right panel), $R_\eta^{\tau \mu}(q^2)$ (bottom-left panel) and $R_{\eta^\prime}^{\tau \mu}(q^2)$ (bottom-right panel) due to $S_L$ coupling. } \label{RK-SL}
\end{figure}

\begin{figure}[h]
\centering
\includegraphics[scale=0.4]{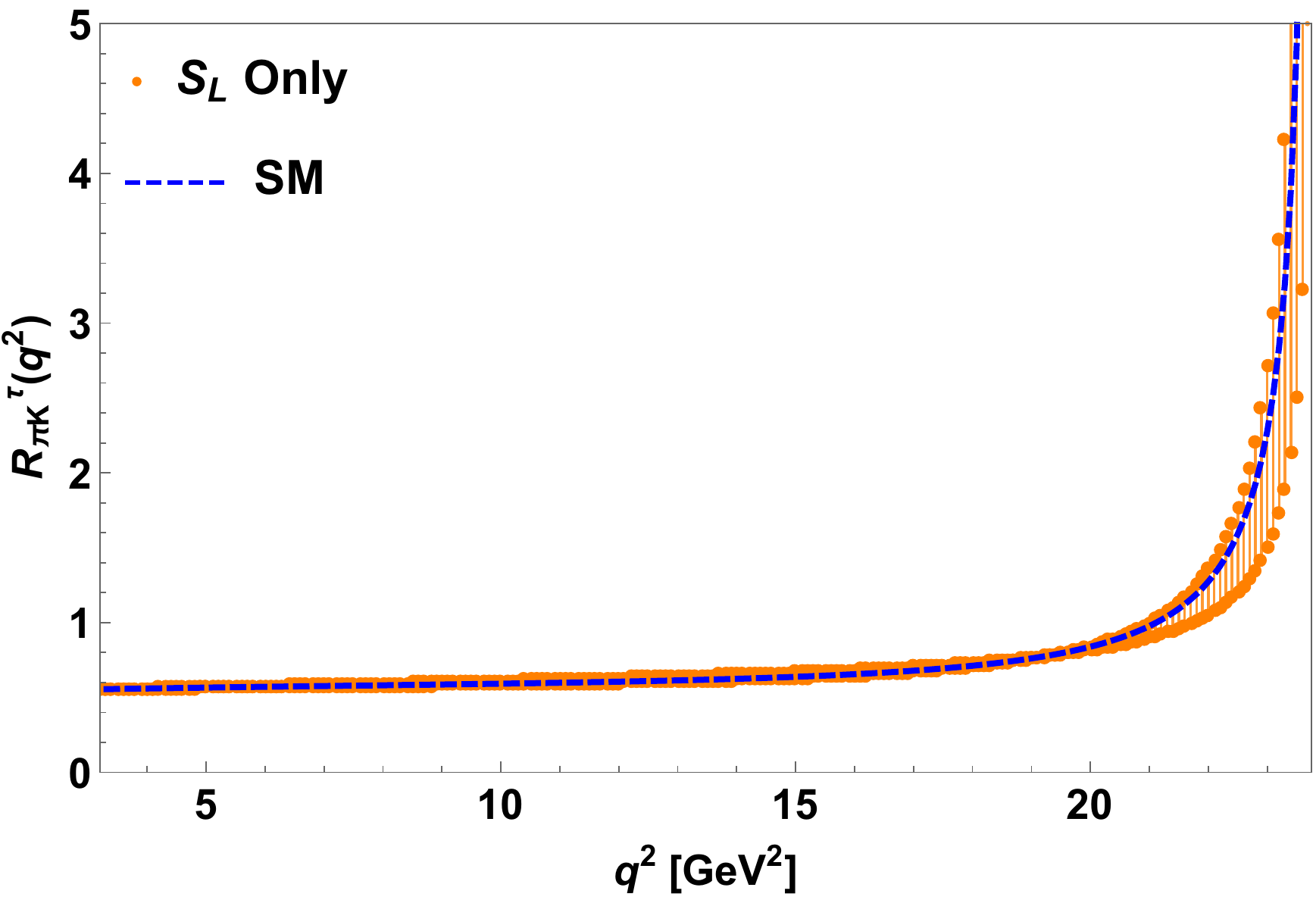}
\quad
\includegraphics[scale=0.4]{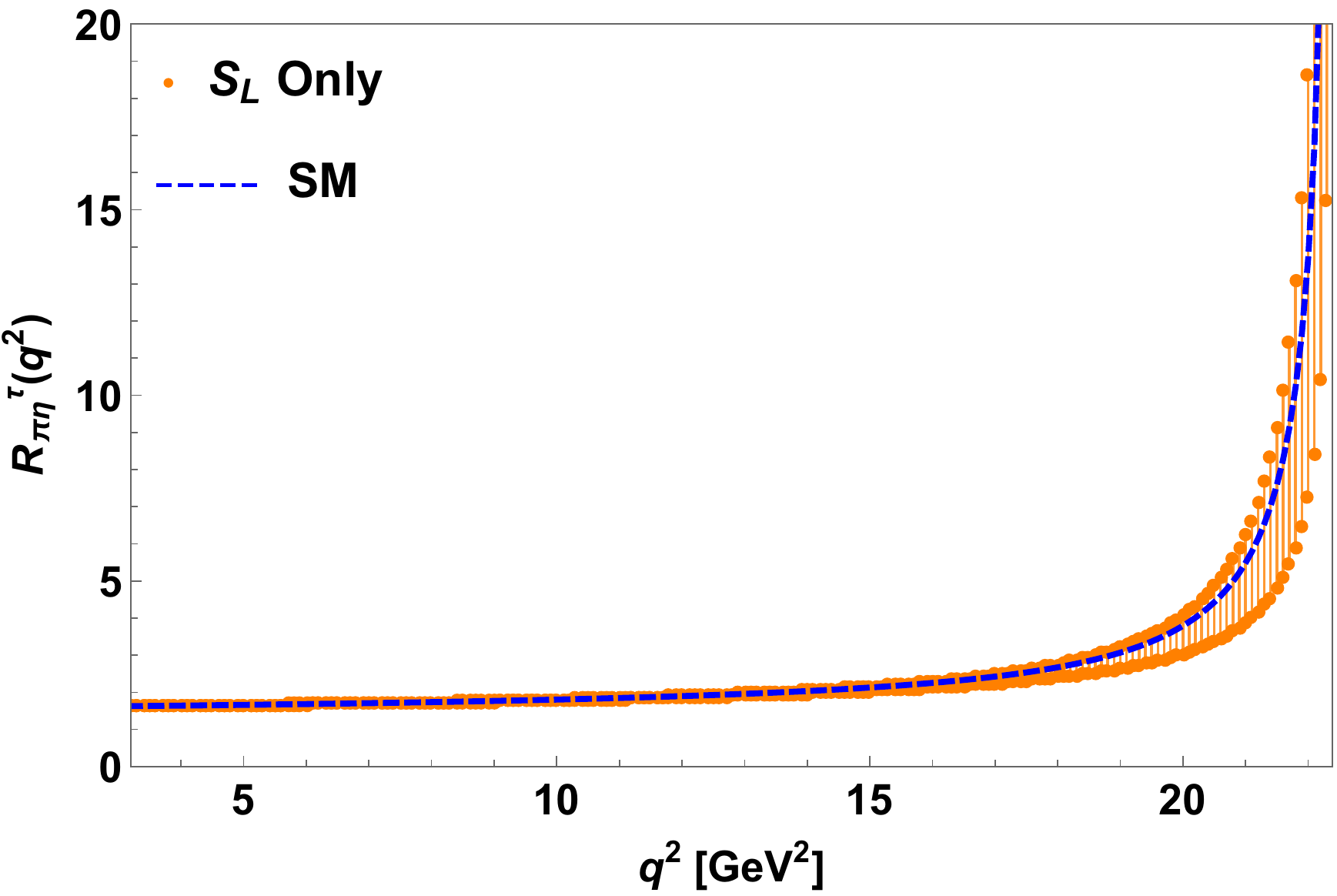}
\quad
\includegraphics[scale=0.4]{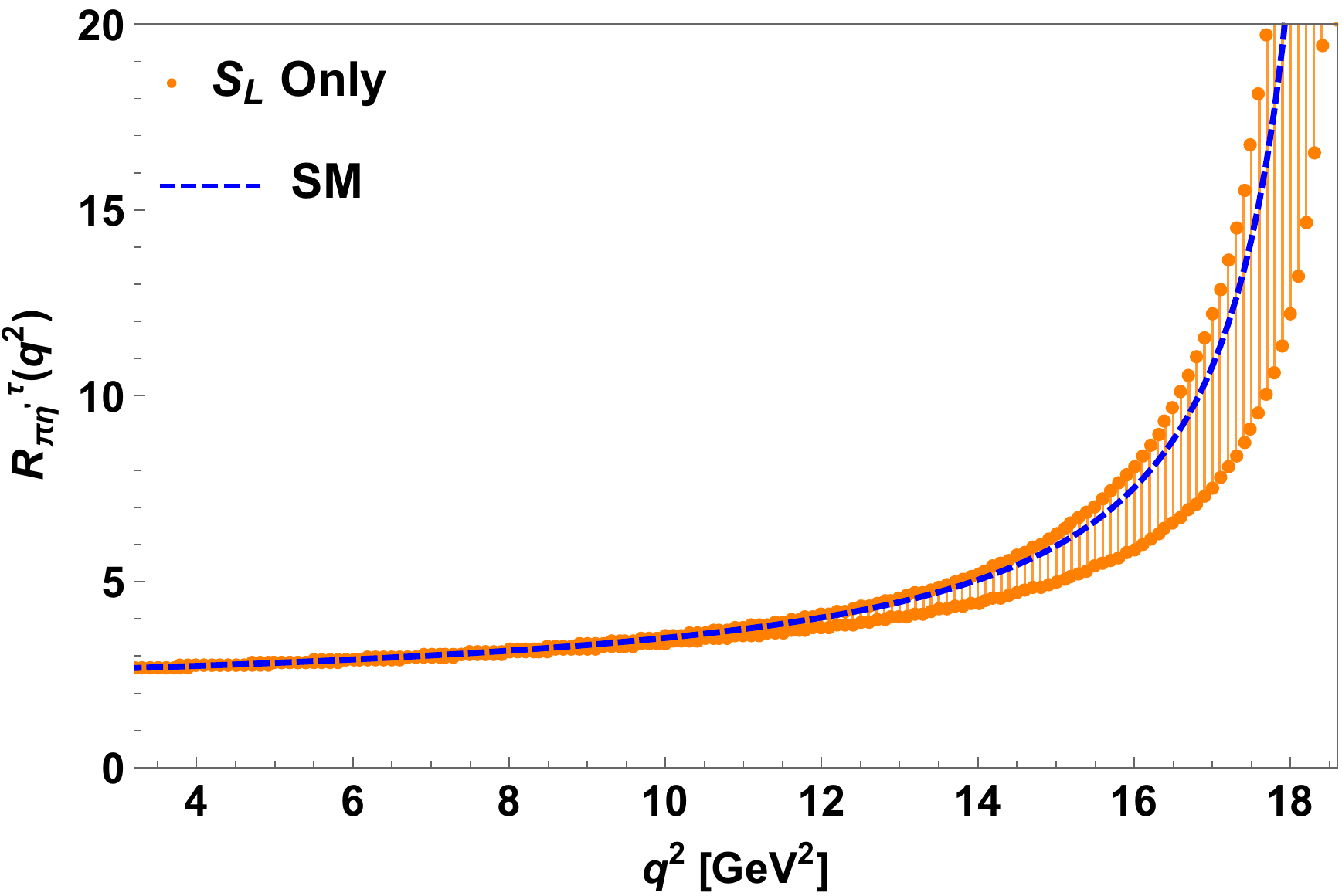}
\caption{The plots for  $R_{\pi K}^{ \tau}(q^2)$ (top-left panel), $R_{\pi \eta}^{\tau}(q^2)$ (top-right  panel) and $R_{\pi \eta^\prime}^{\tau }(q^2)$ (bottom panel) parameters. } \label{Rpik-SL}
\end{figure}

Fig. \ref{fbmu-SL} represents the $q^2$ variation of the forward-backwad asymmetry of $\bar B_s \to K^+ \mu^- \bar \nu_\mu$ (top-left panel), $\bar B^0\to \pi^+ \mu^- \bar \nu_\mu$ (top-right panel),   $B^- \to \eta \mu^- \bar  \nu_\mu$ (bottom-left panel) and $B^- \to \eta^{\prime } \mu^-\bar  \nu_\mu$ (bottom-right panel) processes for only $S_L$ coupling. The corresponding plots for $\bar B_{(s)} \to P \tau \bar \nu_\tau$ processes are given in Fig. \ref{fbtau-SL}. Due to the additional $S_L$ contribution, the forward-backward asymmetry parameters of these processes  deviate significantly from  SM. The corresponding integrated values  are presented in Table IV. Fig. \ref{RK-SL} represents the plots for the LNU parameters $R_K^{\tau \mu}(q^2)$ (top-left panel), $R_\pi^{\tau \mu} (q^2)$ (top-right panel), $R_\eta^{\tau \mu} (q^2) $ (bottom-left panel) and $R_{\eta^\prime}^{\tau \mu} (q^2)$ (bottom-right panel) verses $q^2$. The variation of  $R_{\pi K}^\tau$, $R_{\pi \eta^{(\prime)}}^\tau$ parameters with respect to $q^2$ are shown in Fig. \ref{Rpik-SL}. In Table V, we give the numerical values of these  parameters. 

 \begin{table}[htb]
\begin{center}
\caption{ Same as Table II  in the presence of   $S_{L, R}$ NP couplings. }
\begin{tabular}{| c |c | c |  }
\hline
Observables~ &~ Values for $S_L$ coupling~&~ Values for $S_R$ coupling \\
 \hline
 \hline
 
  ${\rm BR}( B_s \to K^+  \mu^- \bar \nu_\mu)$   ~ &~ $(1.1-1.15) \times 10^{-4}$ ~&~ $(1.1-1.15) \times 10^{-4}$ ~\\
 
${\rm BR}( B_s \to K^+  \tau^- \bar \nu_\tau)$ ~  &~ $(0.62-1.29) \times 10^{-4}$ ~&~ $(4.97-7.4) \times 10^{-5}$ ~ \\

$\langle A_{FB}^\mu \rangle$ ~ &~ $(-3.32\to 3.52)  \times 10^{-3}$~&~$(-3.32\to 3.52) \times 10^{-3}$~ \\

$\langle A_{FB}^\tau \rangle$ ~ &~ $0.255-0.272$~&~$0.058-0.26$~ \\
\hline

${\rm BR}(\bar B \to \pi^+  \mu^-\bar  \nu_\mu)$   ~ &~ $(1.39-1.49) \times 10^{-4}$ ~&~ $(1.39-1.49) \times 10^{-4}$ ~\\
 
${\rm BR}(\bar B \to \pi^+  \tau^- \bar \nu_\tau)$ ~  &~ $(0.82-1.93) \times 10^{-4}$ ~ &~ $(0.66-1.02) \times 10^{-4}$ ~\\

 $\langle A_{FB}^\mu \rangle$  ~ &~ $(-3.86\to 3.51) \times 10^{-3}$~&~$(-3.86 \to 3.51) \times 10^{-3}$~ \\

$\langle A_{FB}^\tau \rangle$ ~& $0.25-0.27$~&~ $0.0264-0.2468$~ \\
\hline

${\rm BR}(B^- \to \eta^0  \mu^-\bar  \nu_\mu)$   ~ &~ $(3.28-3.44) \times 10^{-5}$ ~&~ $(3.28-3.44) \times 10^{-5}$ ~\\
 
${\rm BR}(B^- \to \eta^0  \tau^- \bar \nu_\tau)$ ~  &~ $(1.74-3.82) \times 10^{-5}$ ~&~ $(1.32-2.12) \times 10^{-5}$ ~ \\

$\langle A_{FB}^\mu \rangle$ ~ &~ $(-3.39\to 4.0)  \times 10^{-3}$~&~$(-3.39\to 4.0) \times 10^{-3}$~ \\

$\langle A_{FB}^\tau \rangle$ ~ &~ $0.27 -0.277$~&~$0.085-0.272$~ \\
\hline

${\rm BR}(B^- \to \eta^{\prime 0}  \mu^- \bar  \nu_\mu)$   ~ &~ $(1.49-1.55) \times 10^{-5}$ ~&~ $(1.49-1.55) \times 10^{-5}$ ~\\
 
${\rm BR}(B^- \to \eta^{\prime 0}  \tau^- \bar \nu_\tau)$ ~  &~ $(0.7-1.46) \times 10^{-5}$ ~&~ $(5.0-8.33) \times 10^{-6}$ ~ \\

$\langle A_{FB}^\mu \rangle$ ~ &~ $(-2.82\to 4.68)\times 10^{-3}$~&~$(-2.92\to 4.68) \times 10^{-3}$~ \\

$\langle A_{FB}^\tau \rangle$ ~ &~ $0.287-0.31$~&~$0.153-0.298$~ \\

\hline 

\end{tabular}
\end{center}
\end{table}

\begin{table}[htb]
\begin{center}
\caption{ Same as Table III  in the presence of   $S_{L, R}$ NP couplings. }
\begin{tabular}{| c |c | c |  }
\hline
Observables~ &~ Values for $S_L$ coupling~&~ Values for $S_R$ coupling \\
 \hline
 \hline

$R_K^{\tau \mu}$ &~ $0.537-1.17$~ &~ $0.45-0.645$ ~ \\
 
 $R_\pi^{\tau \mu}$ &~ $0.55-1.38$~ &~ $0.47-0.685$ ~  ~\\

$R_\eta^{\tau \mu}$ &~ $0.5-1.16$~ &~ $0.4-0.62$ ~ ~ \\

 $R_{\eta^\prime}^{\tau \mu}$ ~ &~ $0.448-0.976$ ~ &~ $0.33-0.538$ ~ ~\\
 
 \hline
 
 $R_{\pi K}^{\mu }$ ~  &~ $1.263-1.3$ ~ &~ $1.263-1.3$ ~ ~\\
  
$R_{\pi \eta}^{\mu}$   ~&~ $4.238-4.33$ ~&~ $4.238-4.33$ ~ ~\\
 
$R_{\pi \eta^\prime}^{\mu }$ ~ &~ $9.329-9.61$ ~&~ $9.329-9.61$ ~  ~\\
 
\hline

$R_{\pi K}^{\tau}$   ~ &~ $1.32-1.5$ ~&~ $1.328-1.378$ ~ ~\\

$R_{\pi \eta}^{\tau}$   ~ &~ $4.71-5.05$ ~&~ $4.81-5.0$ ~~\\
 
$R_{\pi \eta^\prime}^{\tau }$  ~&~ $11.71-13.22$ ~ &~ $12.45-13.2$ ~~\\
 
 \hline
\end{tabular}
\end{center}
\end{table}

\subsection{Case D: Effect of $S_R$ only}

\begin{figure}[h]
\centering
\includegraphics[scale=0.4]{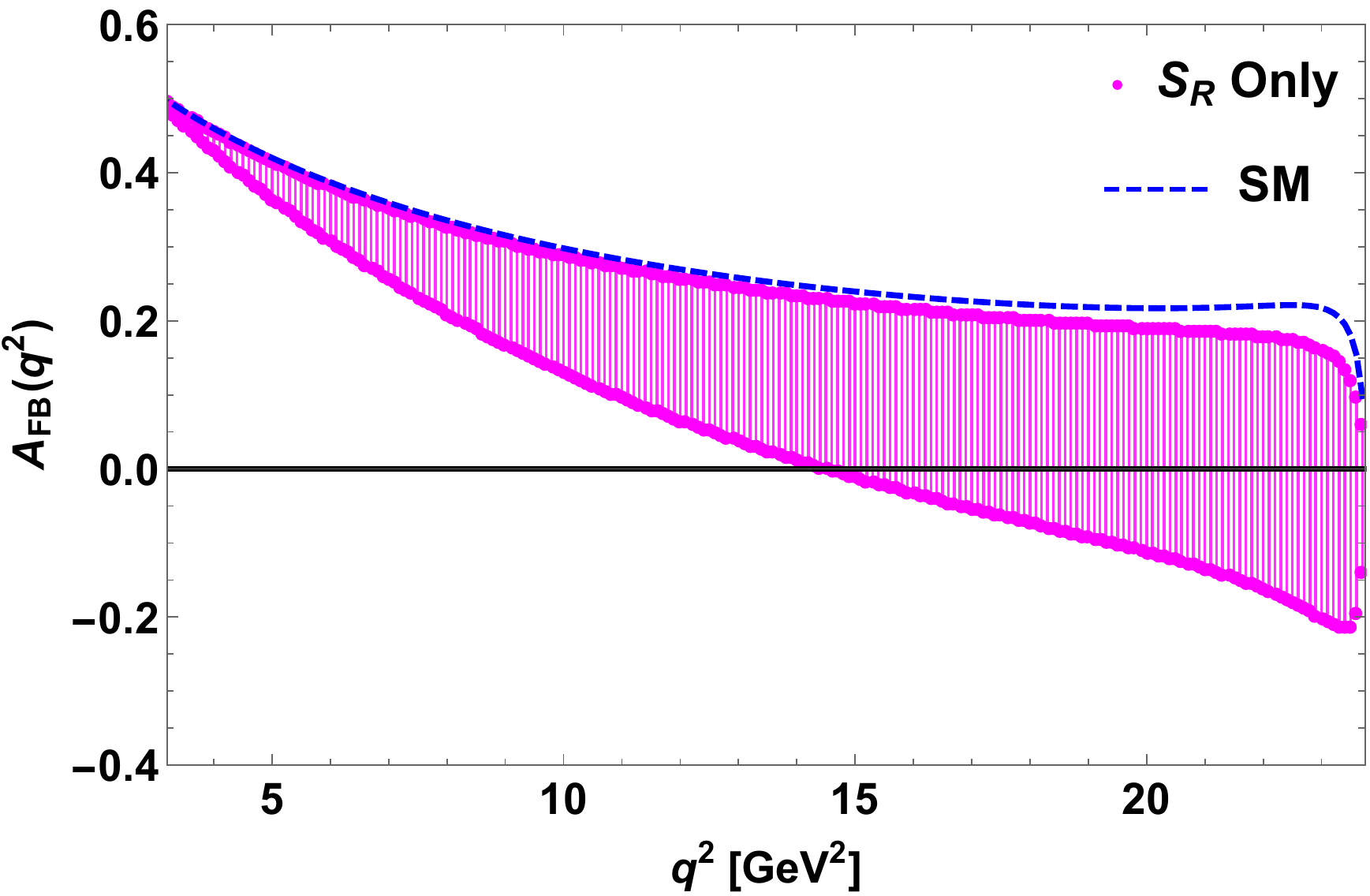}
\quad
\includegraphics[scale=0.4]{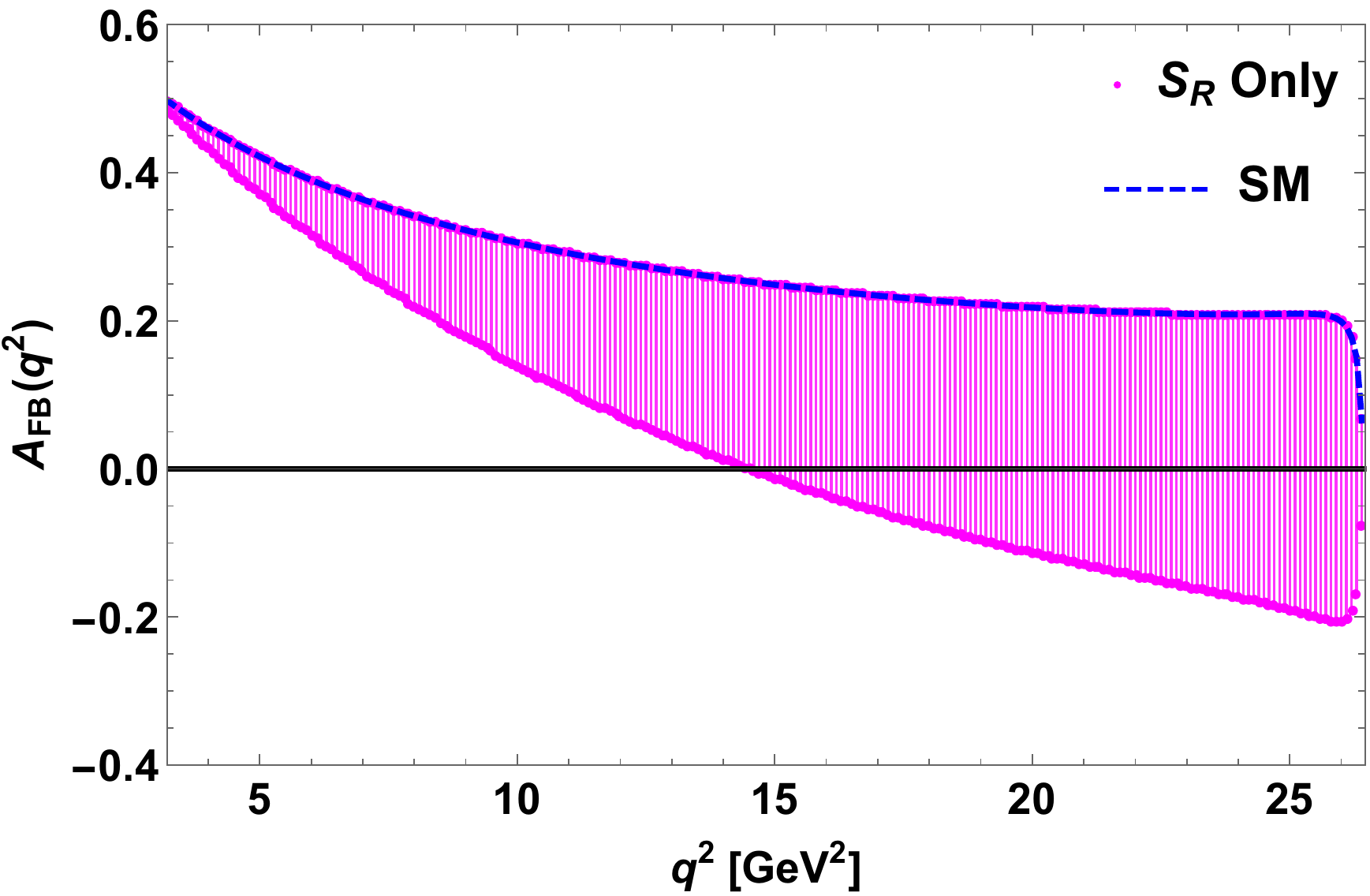}
\quad
\includegraphics[scale=0.4]{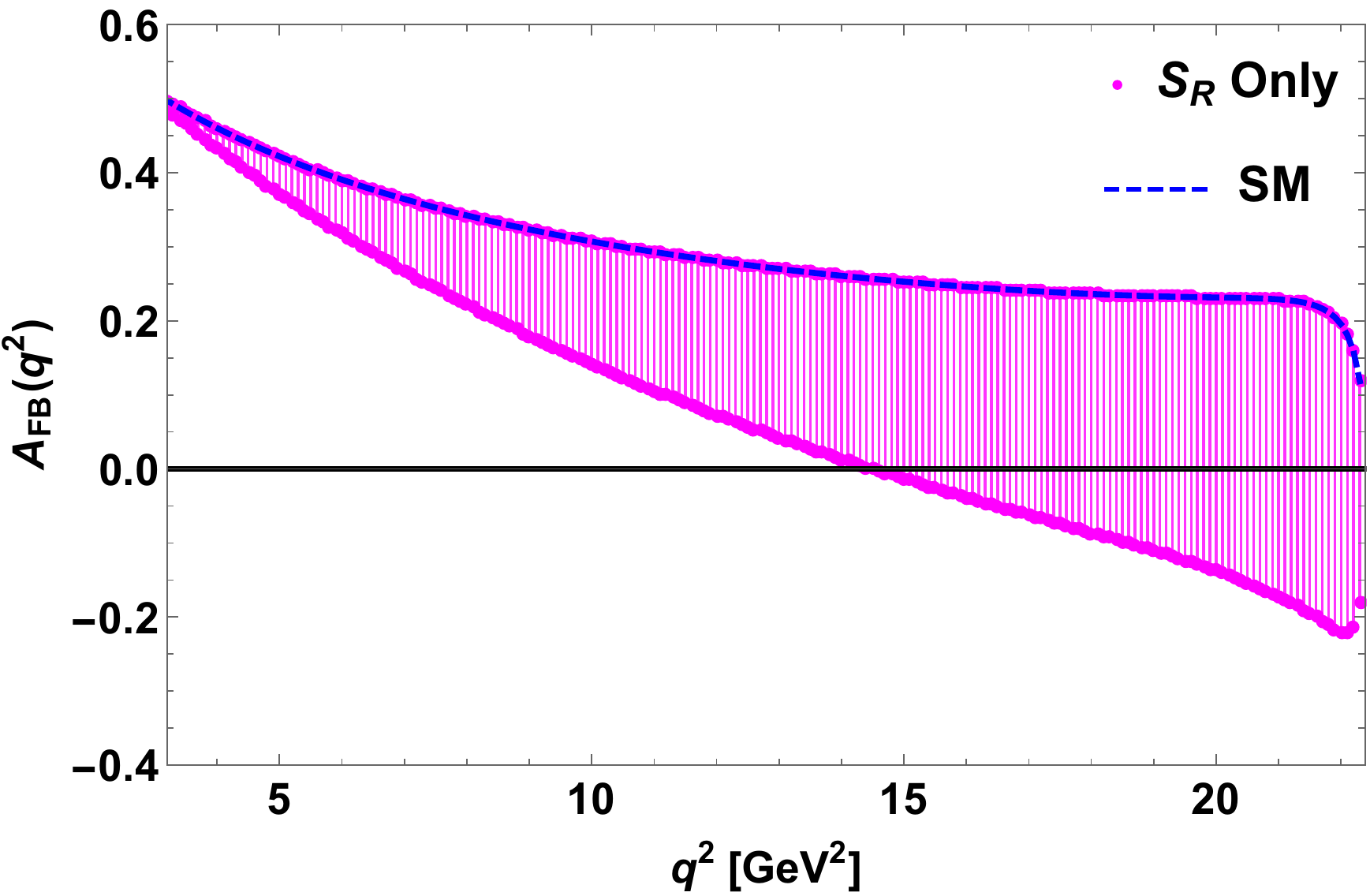}
\quad
\includegraphics[scale=0.4]{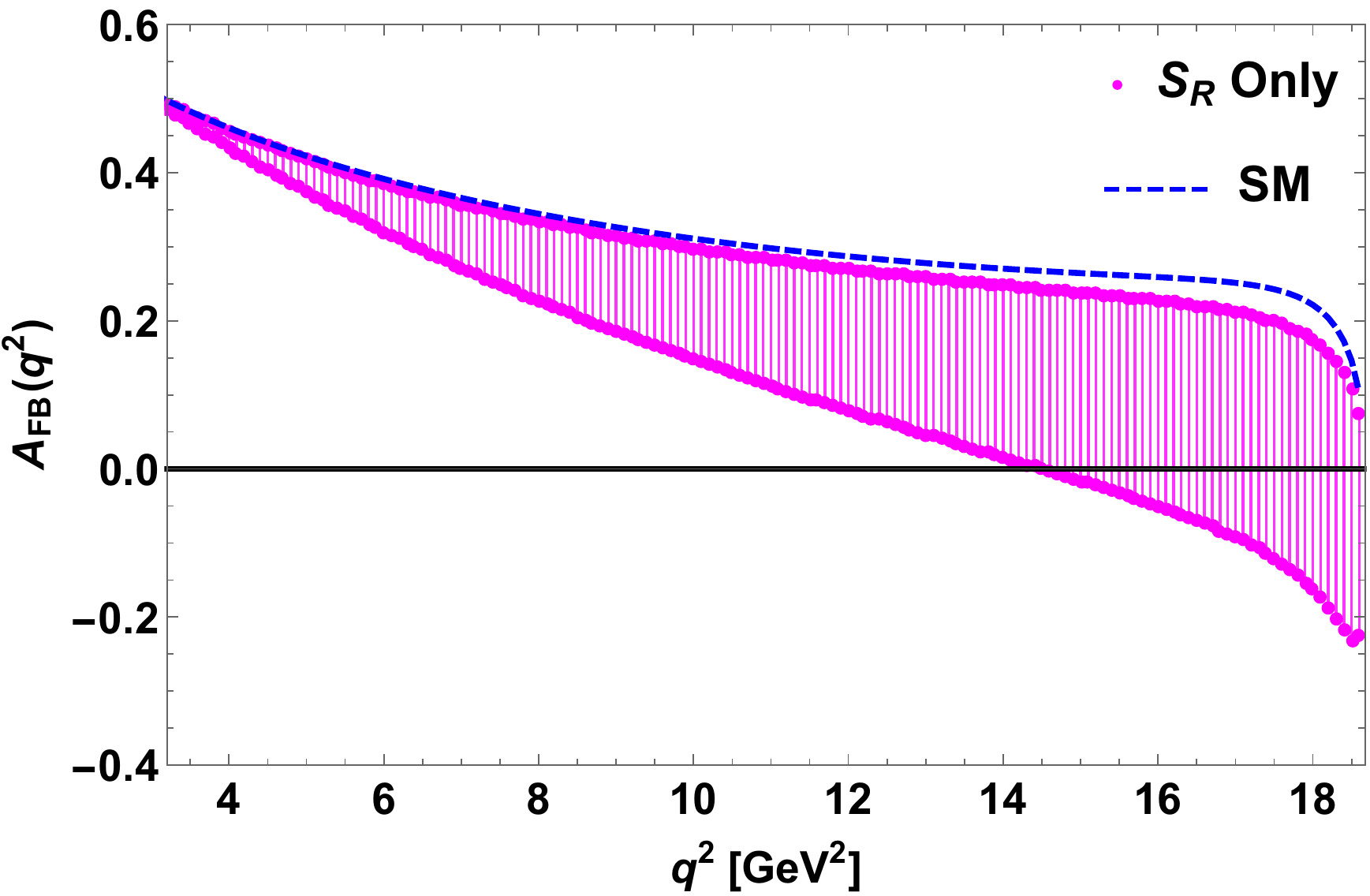}
\caption{The plots for the $q^2$ variation of  forward-backward asymmetry  of  $\bar B_s \to K^+ \tau^- \bar \nu_\tau$  (top-left panel),  $\bar B^0 \to \pi^+ \tau^- \bar \nu_\tau$  (top-right panel), $B^- \to \eta \tau^- \bar \nu_\tau$  (bottom-left panel) and $B^- \to \eta^{\prime } \tau^- \bar \nu_\tau$  (bottom-right panel) processes.} \label{fbtau-SR}
\end{figure}

\begin{figure}[h]
\centering
\includegraphics[scale=0.4]{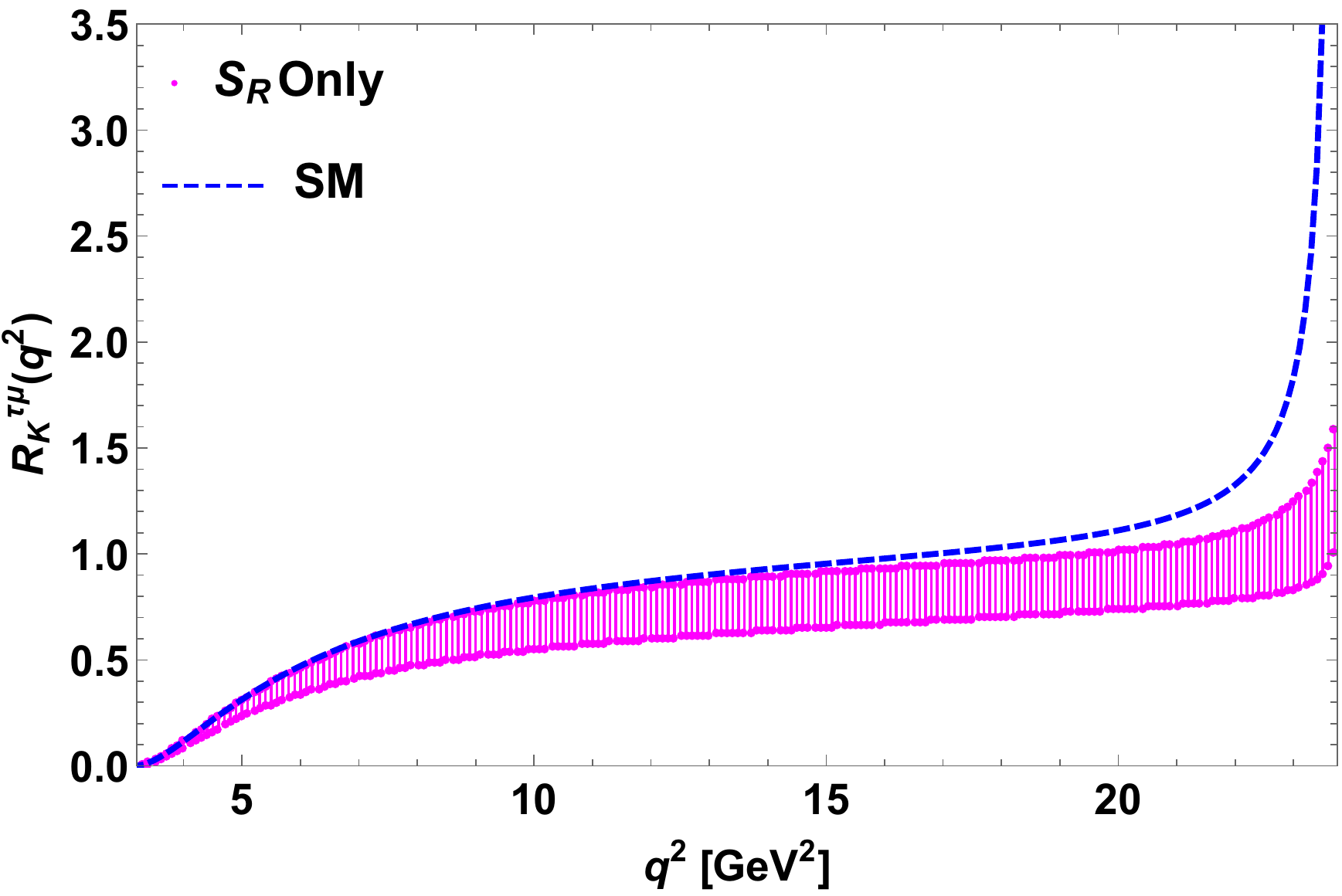}
\quad
\includegraphics[scale=0.4]{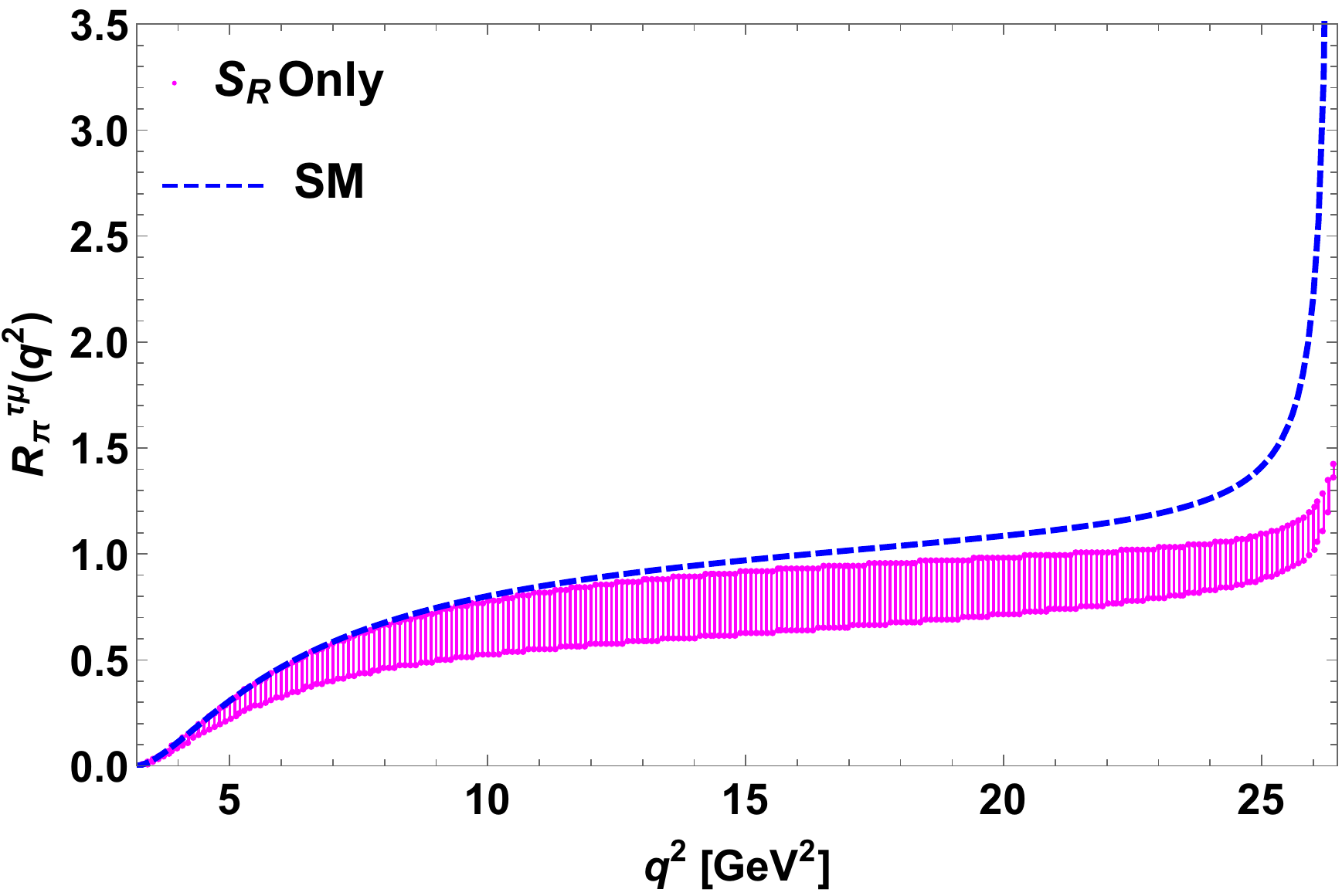}
\quad
\includegraphics[scale=0.4]{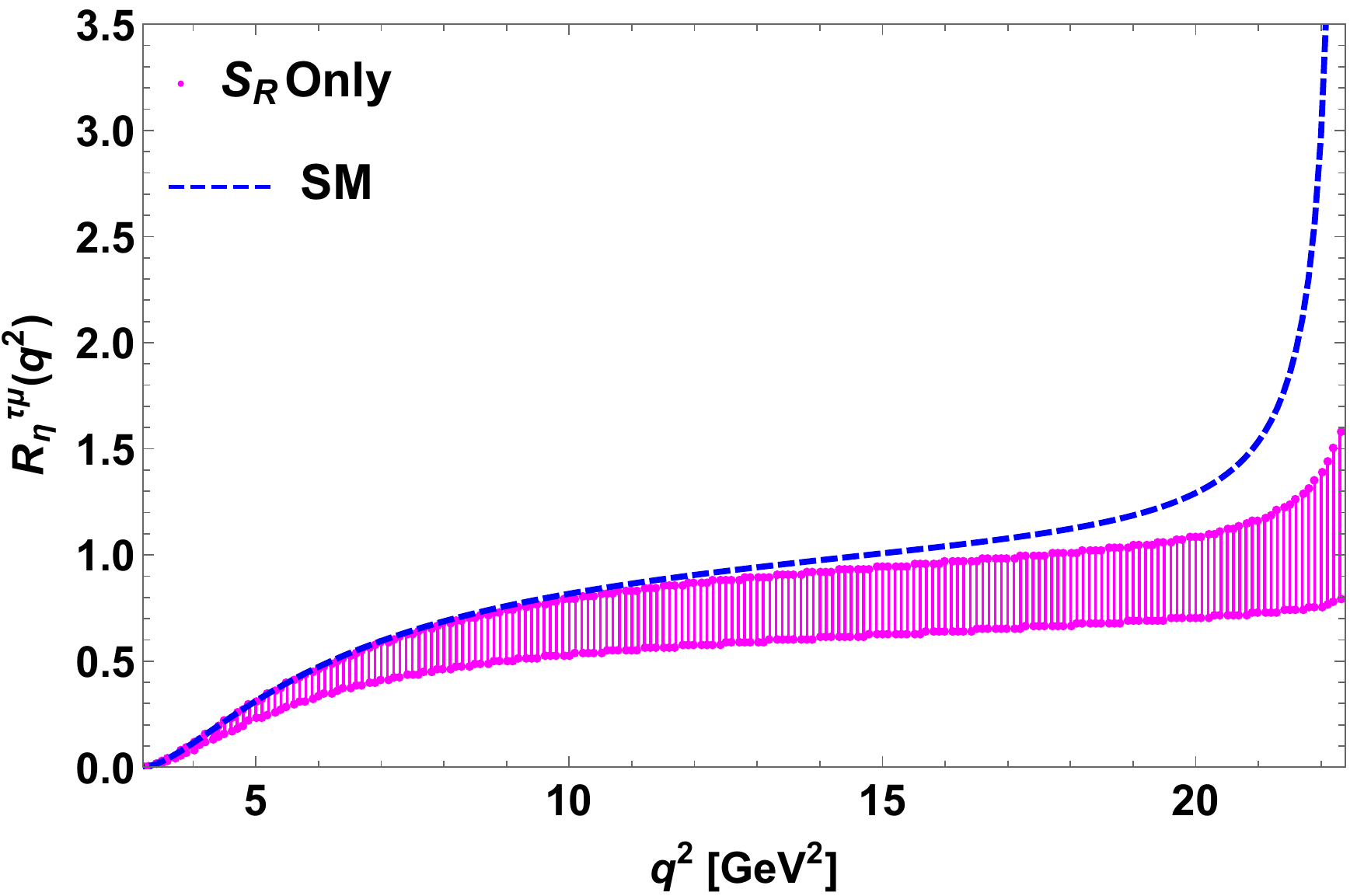}
\quad
\includegraphics[scale=0.4]{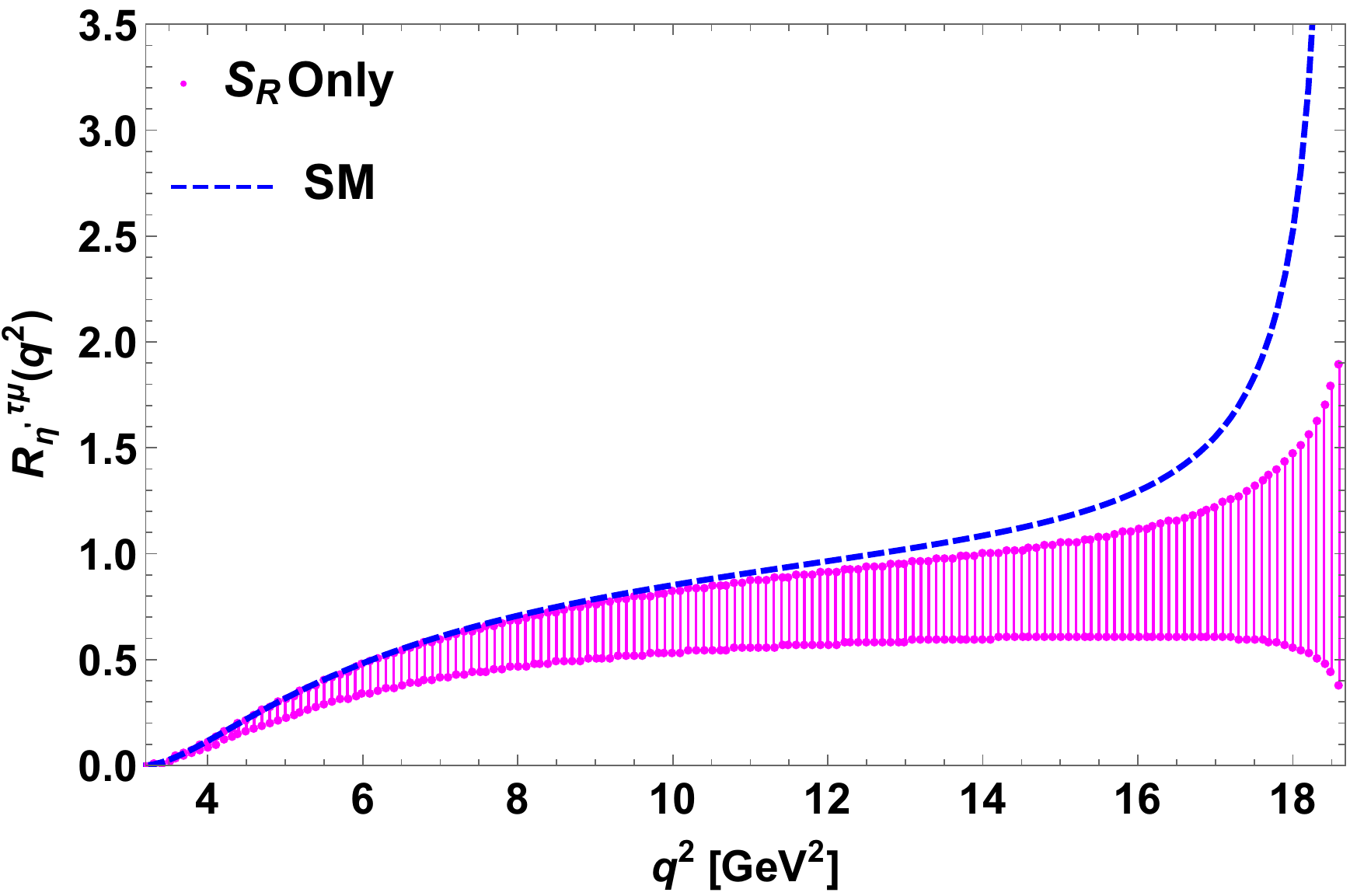}
\caption{The plots for the LNU parameters  $R_K^{\tau \mu}(q^2)$ (top-left panel), $R_\pi^{\tau \mu}(q^2)$ (top-right panel), $R_\eta^{\tau \mu}(q^2)$ (bottom-left panel) and $R_{\eta^\prime}^{\tau \mu}(q^2)$ (bottom-right panel) due to $S_R$ coupling. } \label{RK-SR}
\end{figure}

Here we perform an analysis of $ B \to P l^- \bar \nu_l$ processes with the additional $S_R$ coupling. As discussed in section II, the real part of $S_R$ coupling differs from the real part of $S_L$ by a negative sign while their imaginary parts are same.  The minimum and maximum values of $S_R$ parameter are found as
  $({\rm Re}[S_R], {\rm Im}[S_R])^{\rm max}=(0.003,  0.268)$ and 
$({\rm Re}[S_R], {\rm Im}[S_R])^{\rm min}=( -0.54, -0.03)$
for $b \to u \tau \bar  \nu_\tau$ process. For $b\to u \mu \nu$ the constraints on $S_R$ couplings are same as $S_L$.
Using these value
the $q^2$ variation of the forward-backward asymmetries  for   $B^- \to P^0 \tau^- \bar \nu_\tau$ processes are shown in Fig. \ref{fbtau-SR}. The branching ratios and forward-backward asymmetries  of these processes  are presented in Table IV.  Fig. \ref{RK-SR} represents the  variation of the LNU parameters ($R_{K, \pi, \eta, \eta^\prime}^{\tau \mu}$) due to only $S_R$ coupling.  The variation of   $R_{P P'}^\tau$  parameters are similar to those with $S_L$ coupling. 
Table V  contains the numerical values of these parameters.

The rare semileptonic  $B_s \to K l \bar \nu_l$ and $B \to \pi l  \bar \nu_l$ processes are investigated in Refs. \cite{Wang, soni}.  The analysis of  $B \to \pi l \bar \nu_l$ processes using the light cone QCD sume rule approach \cite{Ruckl} and 2HDM \cite{Bern} are also studied in the literature. In  Ref. \cite{eta, etaprime, eta-etaprime},  $B \to \eta^{(\prime)} l \bar \nu_l$ processes are studied by using various model-dependent approaches. The model independent analysis of $b \to u l \bar \nu_l$ processes can be found in \cite{Dutta1}. Our predicted  SM values of the branching ratios of $\bar  B_{(s)} \to P^+ l^- \bar \nu_l$ processes are found to be consistent with the predicted results in the literature, though due to updated input parameters, the central values of the branching ratios of these processes  have slight  deviations.
\section{$B \to V l \bar \nu_l$  processes}
\begin{figure}[h]
\centering
\includegraphics[scale=0.4]{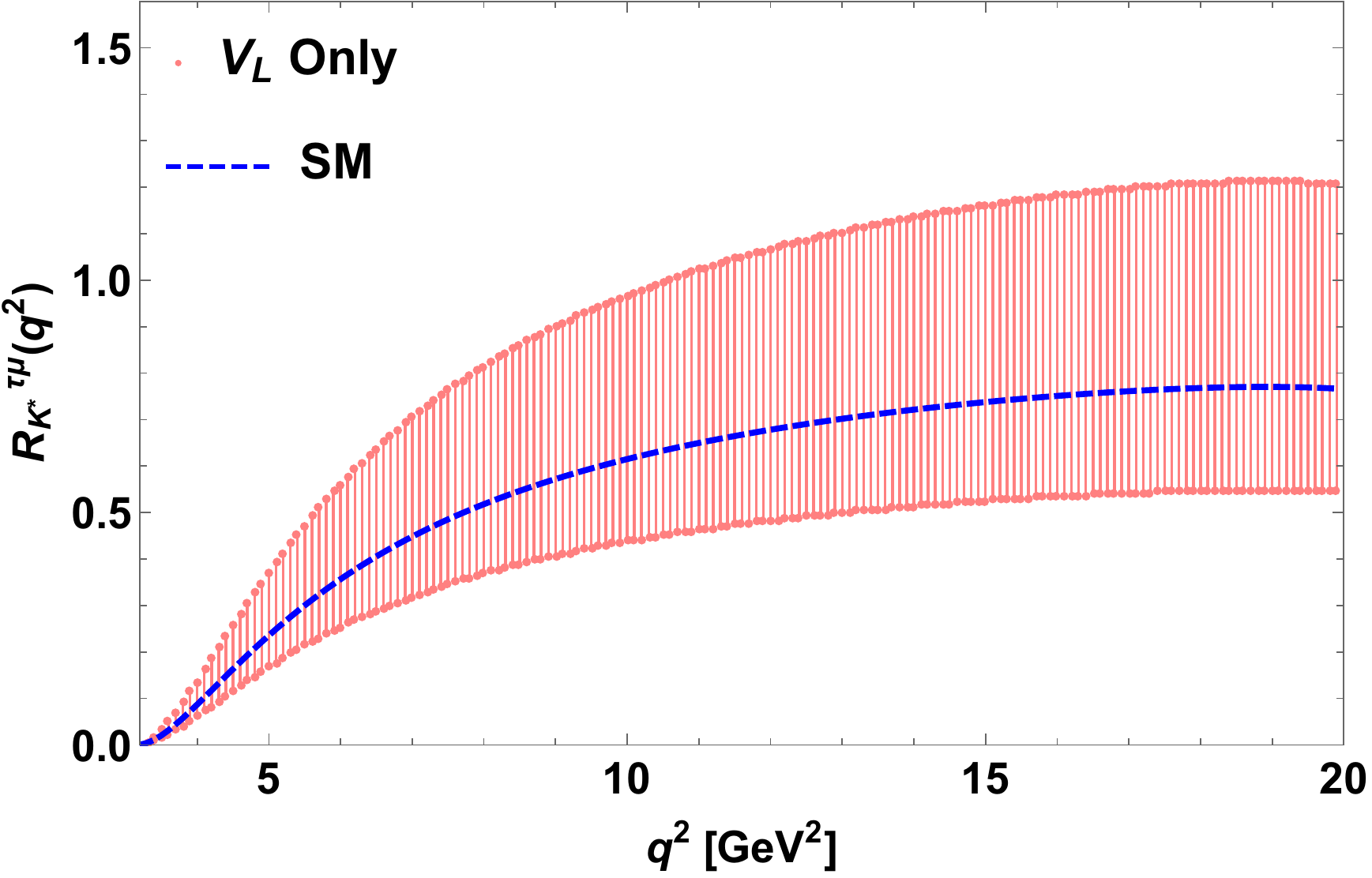}
\quad
\includegraphics[scale=0.4]{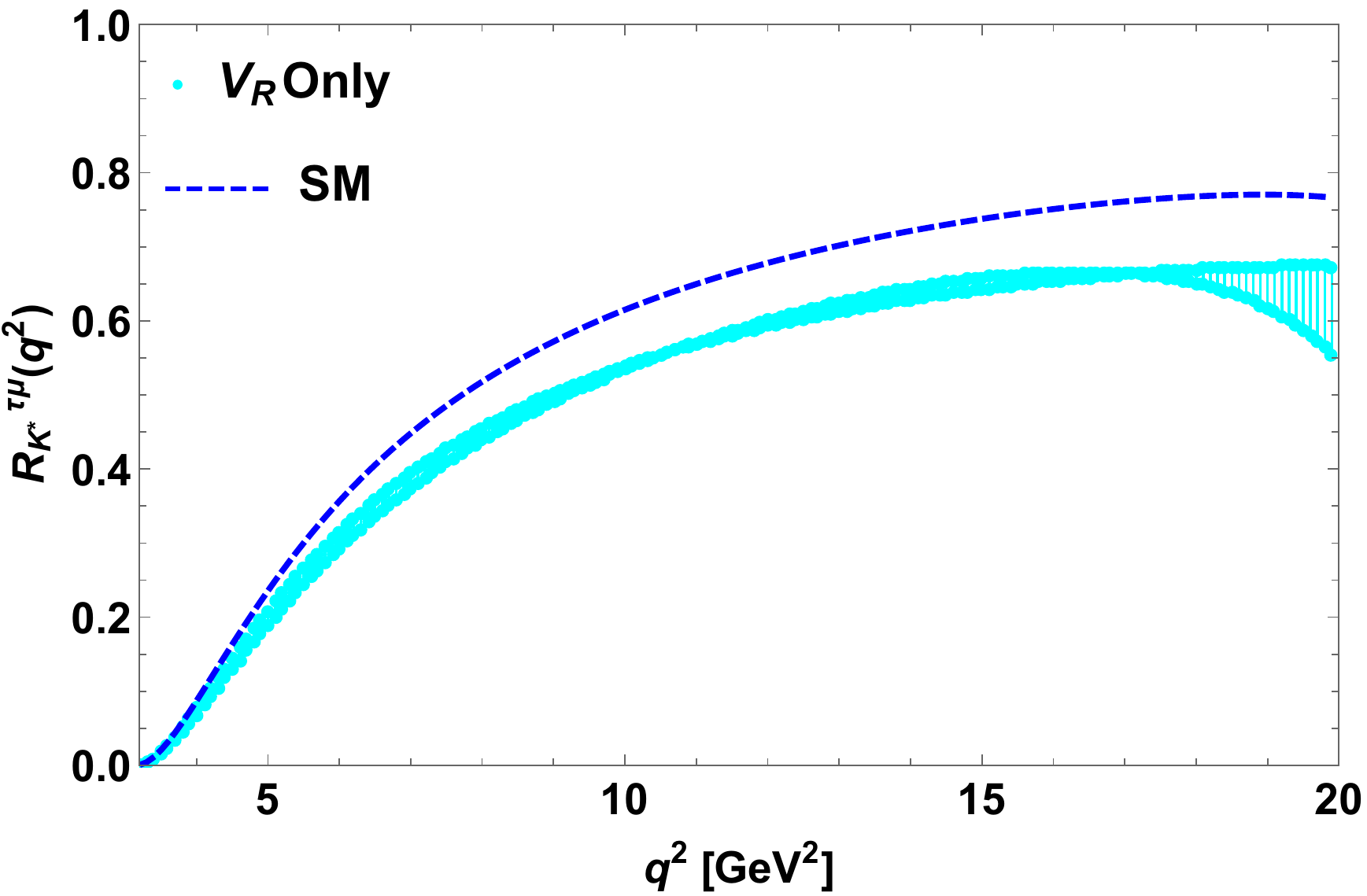}
\quad
\includegraphics[scale=0.4]{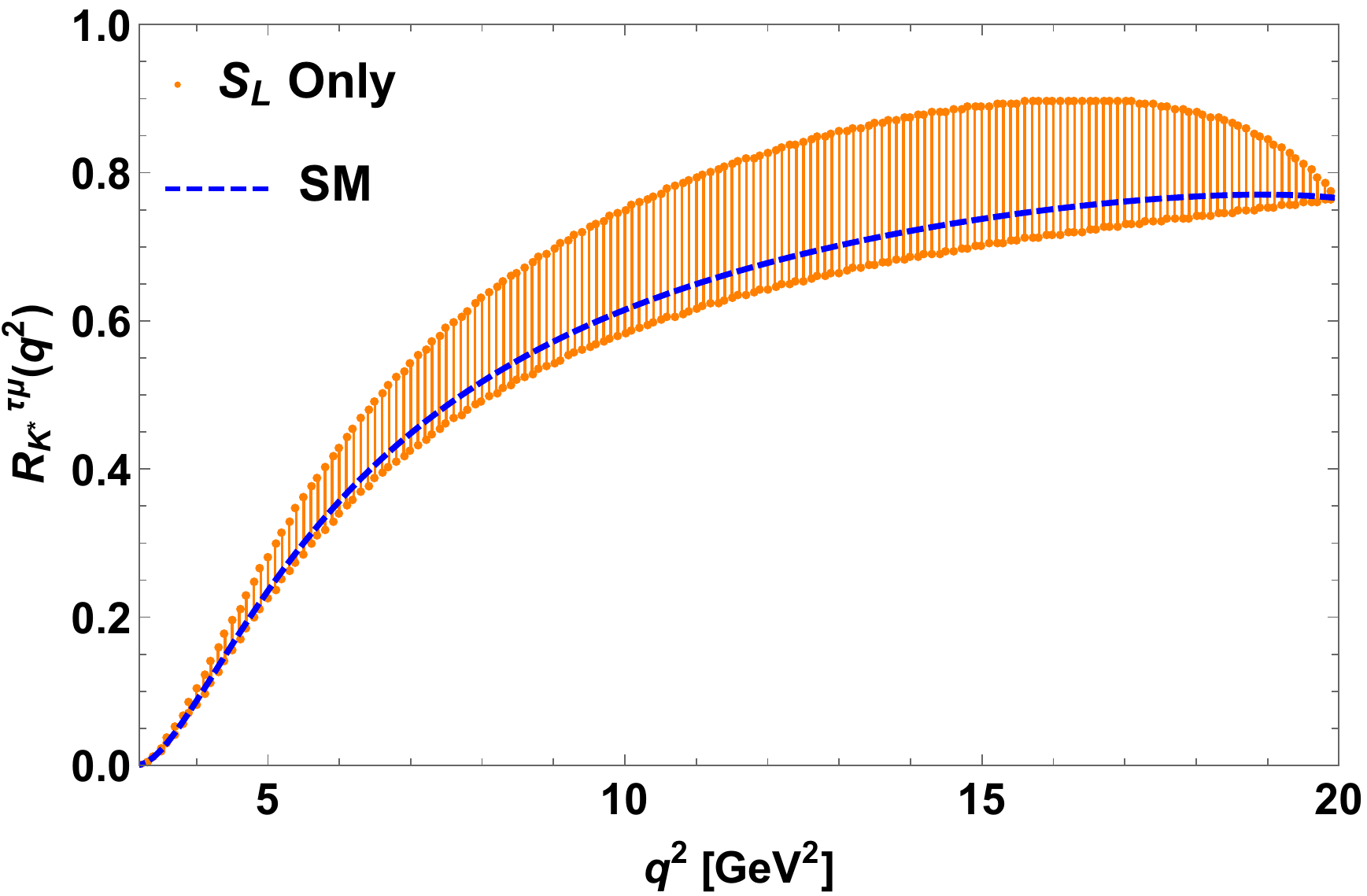}
\quad
\includegraphics[scale=0.4]{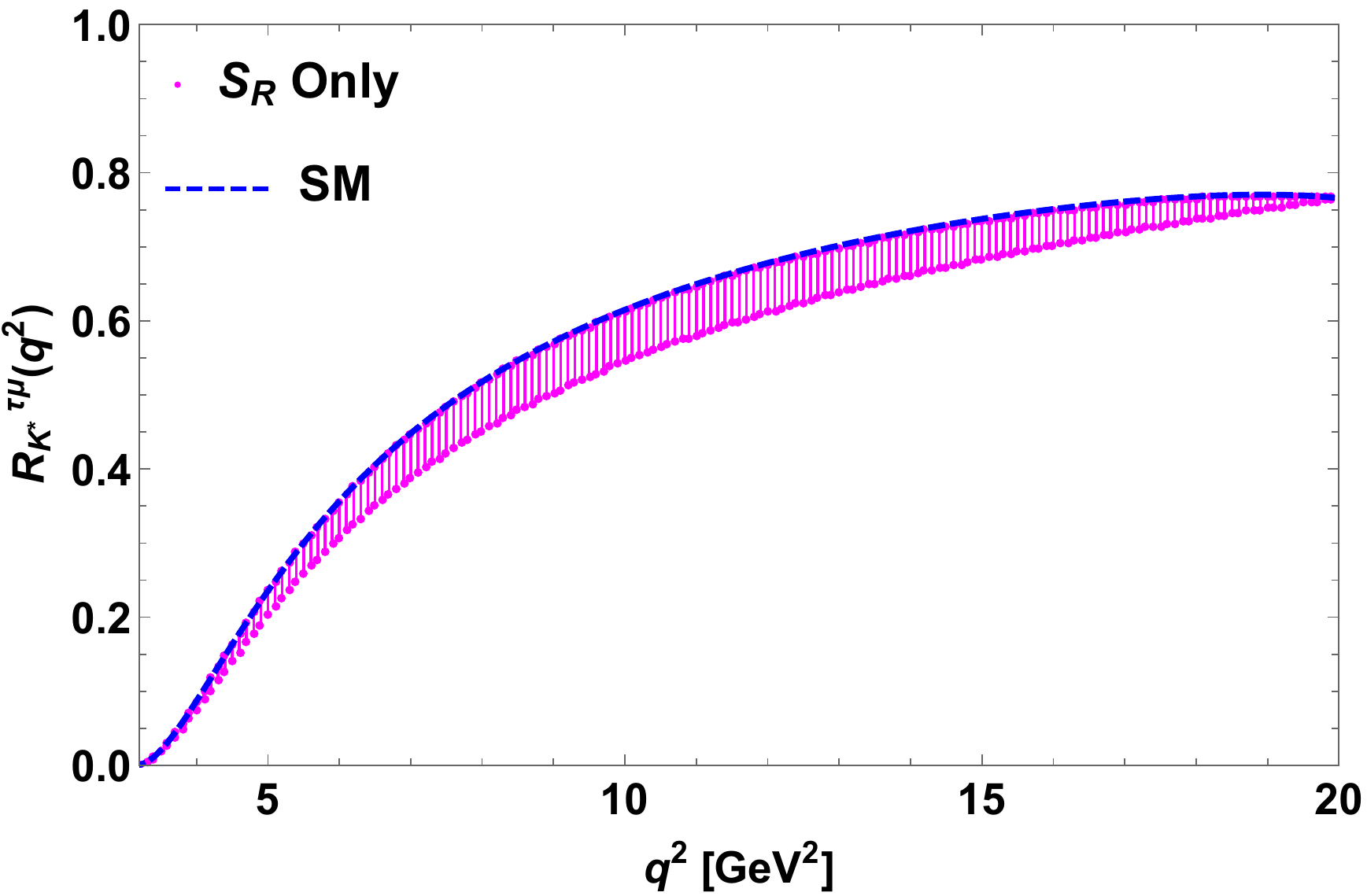}
\caption{The plots for $R_{K^*}^{\tau \mu}(q^2)$  parameters verses $q^2$ for only $V_L$ (top-left panel), $V_R$ (top-right panel), $S_L$ (bottom-left panel) and $S_R$ (bottom-right panel) couplings.}\label{Rkstar}
\end{figure}

In this section, we study the $B \to V l \bar \nu_l$ processes, where  $V=K^*, \rho$. The hadronic  matrix element of  the $B \to V l  \bar \nu_l$ processes  can be parametrized as  \cite{Sakaki}
\bea
\Big \langle V (k, \varepsilon) | \bar{u} \gamma_\mu b| \bar{B} (p_B) \Big \rangle &=& -i \epsilon_{\mu \nu \rho \sigma} {\varepsilon^\nu}^* p_B^\rho k^\sigma \frac{2V(q^2)}{M_B+M_{V}}\;, \nn \\
\Big \langle V (k, \varepsilon) | \bar{u} \gamma_\mu \gamma_5 b| \bar{B} (p_B) \Big \rangle &=& {\varepsilon^{\mu }}^* \left( M_B + M_{V}\right) A_1(q^2) -\left(p_B + k \right)_\mu (\varepsilon^* \cdot q) \frac{A_2 (q^2)}{M_B + M_{V}} \nn \\ &-& q_\mu  (\varepsilon^* \cdot q) \frac{2M_{V}}{q^2} \left[A_3(q^2) - A_0 (q^2) \right],
\eea
where 
\bea
A_3 (q^2) = \frac{M_B + M_{V}}{2 M_{V}} A_1 (q^2) - \frac{M_B - M_{V}}{2 M_{V}} A_2 (q^2)\;. 
\eea 
The differential decay rate  of $B \to V l \nu_l$ processes  with respect to $q^2$ is given by  \cite{Sakaki}
\bea
 {d\Gamma(B \to V  l \bar \nu_l) \over dq^2} &=& {G_F^2 |V_{ub}|^2 \over 192\pi^3 M_B^3} q^2 \sqrt{\lambda_V (q^2)} \left( 1 - {m_l^2 \over q^2} \right)^2 \bigg \{  ( |1 + V_L|^2 + |V_R|^2 ) \nn  \\ &\times&\left[ \left( 1 + {m_l ^2 \over 2q^2} \right) \left( H_{V, +}^2 + H_{V,-}^2 + H_{V,0}^2 \right) + {3 \over 2}{m_l^2 \over q^2} \, H_{V,t}^2 \right]  \nn \\ &-&  2{\rm Re}[(1+ V_L) V_R^{*}] \left[ \left( 1 + {m_l^2 \over 2q^2} \right) \left( H_{V,0}^2 + 2 H_{V,+} H_{V,-} \right) + {3 \over 2}{m_l^2 \over q^2} \, H_{V,t}^2 \right] \nn \\ &+&  {3 \over 2} |S_L - S_R|^2 \, H_S^2  + 3{\rm Re}[ ( 1 + V_L- V_R) (S_L^{*} - S_R^{*} ) ] {m_l \over \sqrt{q^2}} \, H_S H_{V,t}  \bigg \}, 
\eea
where  $\lambda_V = \lambda (M_B^2, M_V^2, q^2)$ and the  hadronic amplitudes in terms of  the form factors are given as
\bea
&&H_{V, \pm} (q^2)  = \left(M_B + M_{V} \right) A_1(q^2) \mp \frac{\sqrt{\lambda_{V}(q^2)}}{M_B + M_{V}} V(q^2), \nn \\
&&H_{V, 0} (q^2) = \frac{M_B + M_{V}}{2M_{V} \sqrt{q^2}} \Big [ - \left(M_B^2 - M_{V}^2-q^2 \right) A_1(q^2)  + \frac{\lambda_{V}(q^2)}{( M_B + M_{V})^2} A_2(q^2) \Bigg ], \nn \\
&&H_{V, t} (q^2)  = - \sqrt{ \frac{\lambda_{V} (q^2)}{q^2}} A_0 (q^2), \nn \\
&&H_{S} (q^2)  =- H_{S_2}^0 (q^2) \simeq -\frac{\sqrt{\lambda_{V} (q^2)}}{m_b + m_u} A_0 (q^2). 
\eea 
\begin{figure}[h]
\centering
\includegraphics[scale=0.4]{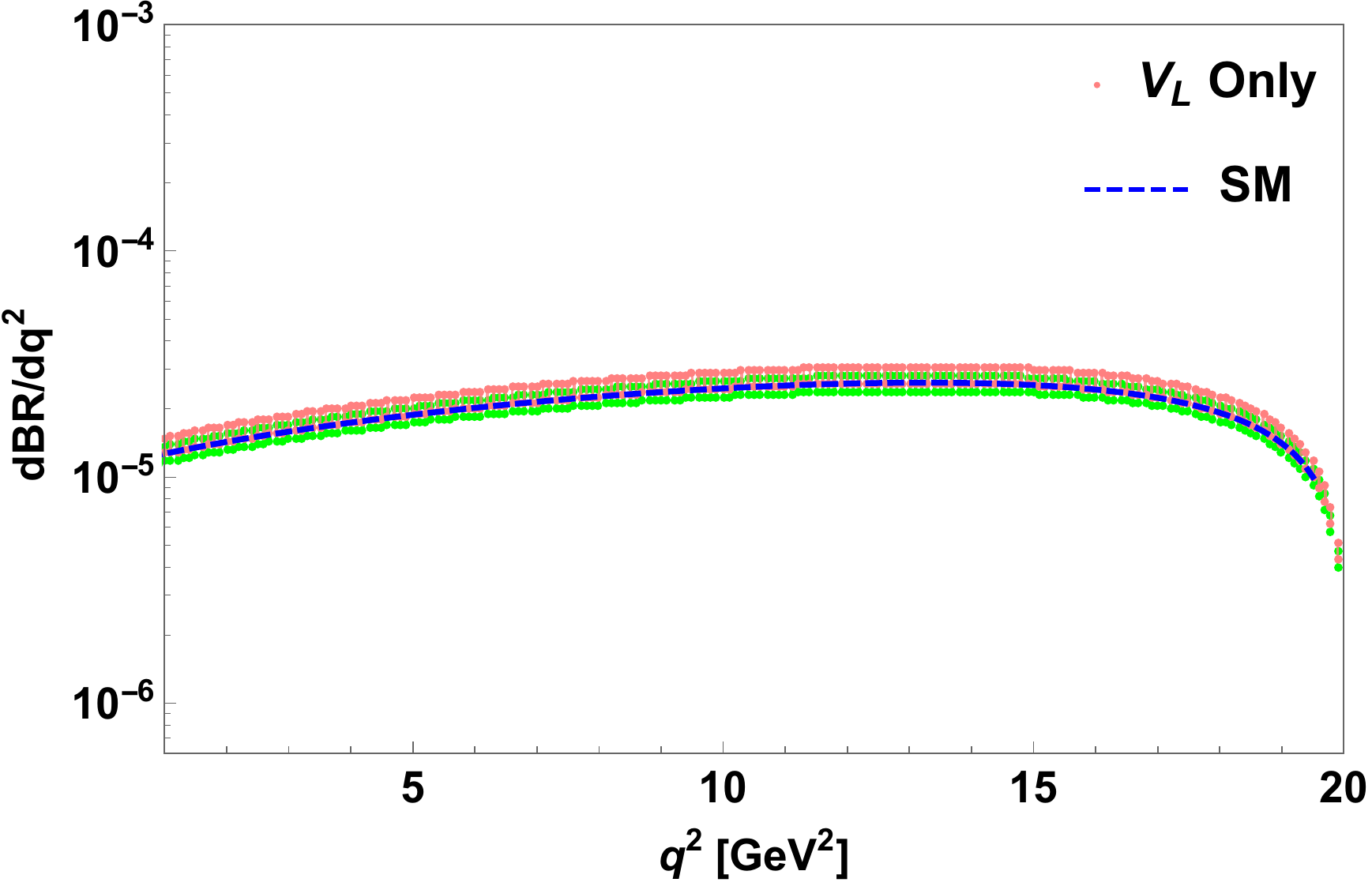}
\quad
\includegraphics[scale=0.4]{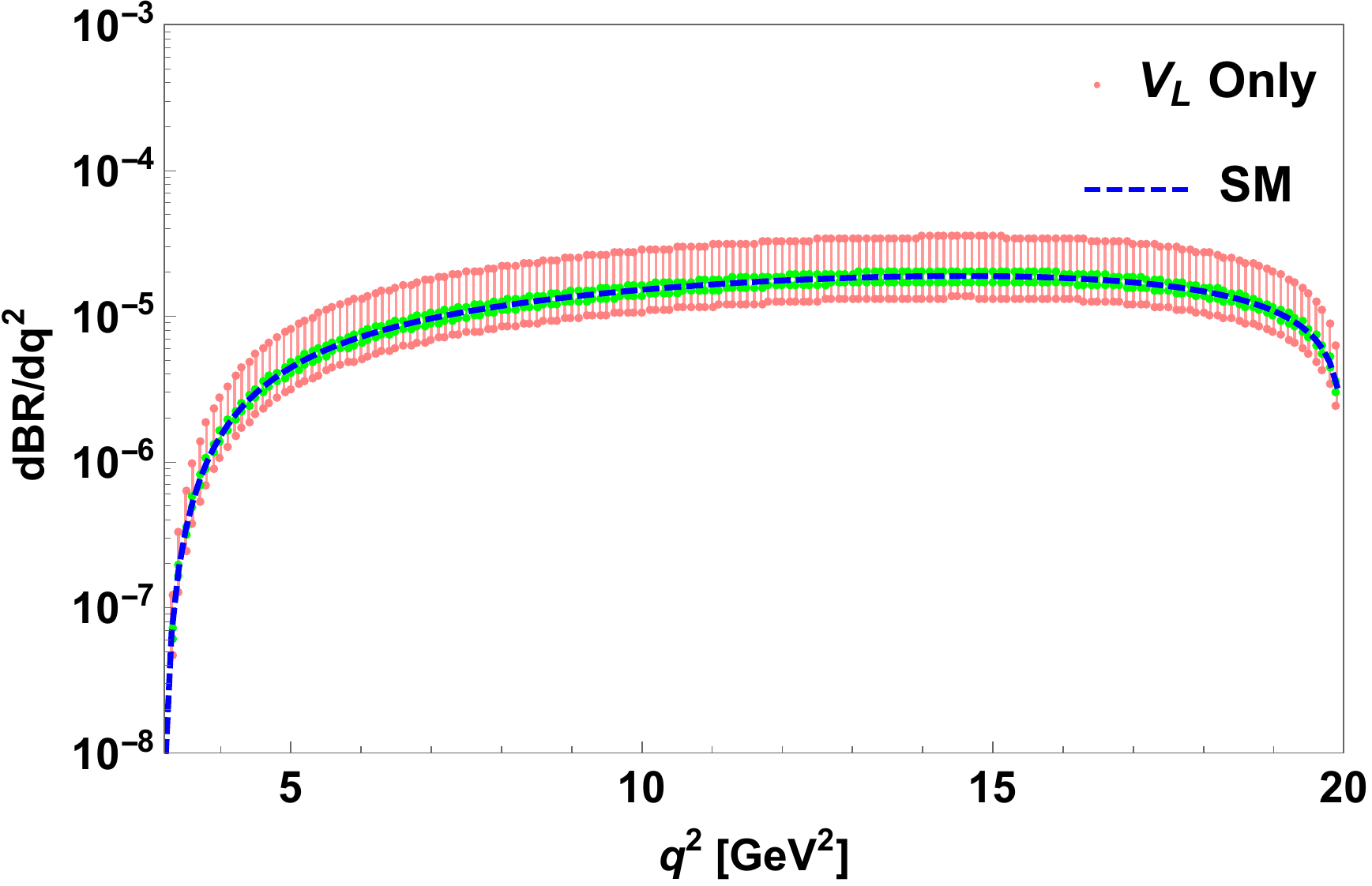}
\quad
\includegraphics[scale=0.4]{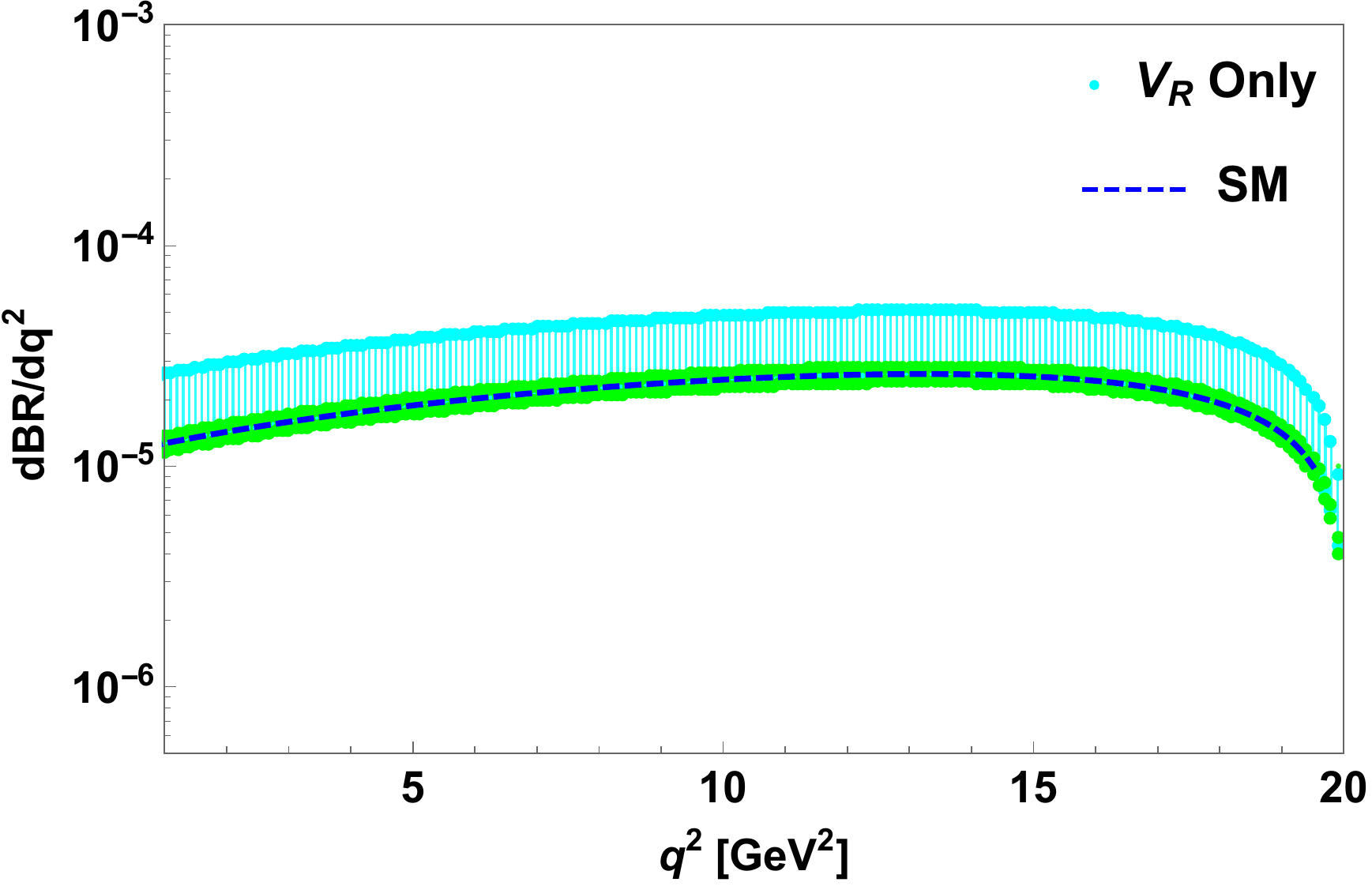}
\quad
\includegraphics[scale=0.4]{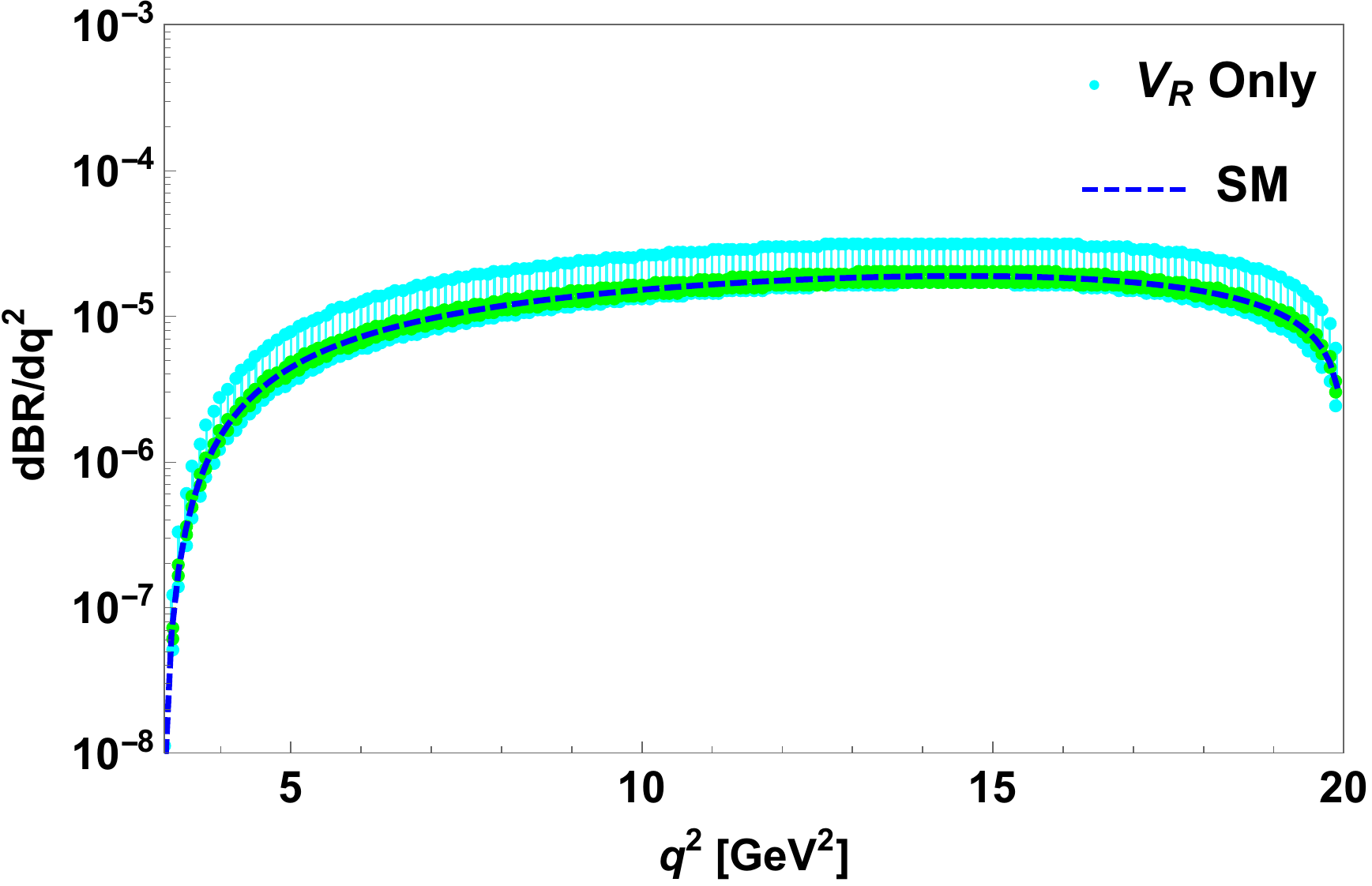}
\caption{The plots in the top panel represent  the $q^2$ variation of the branching ratios of $\bar B_s \to K^{* +} \mu^- \bar \nu_\mu$ (top-left panel) and $\bar B_s \to K^{* +} \tau^-  \bar \nu_\tau$ (top-right panel) processes for only $V_L$ coupling. The corresponding plots  for only $V_R$ coupling are shown in the bottom   panel. } \label{Bkstar-br}
\end{figure}
For the momentum transfer dependence of the form factors,  we consider the most intuitive and the simplest parametrization of the $B_{(s)} \to (K^*) \rho$ form factors, $(V(q^2), A_{0,1,2} (q^2))$ from  Ref. \cite{LCSR}. The masses of all the particles are taken from \cite{pdg}. Using these input values and the bounds on $V_{L}$ coupling  obtained from $B_u^+ \to \tau^+ \nu_\tau$ and $B^- \to \pi^0 \mu^- \bar \nu_\mu$ processes (discussed in sections II and III), we show the plots for the $q^2$ variation of branching ratios for $\bar B_s \to K^{* +} \mu^- \bar \nu_\mu $ (top-left panel) and $\bar B_s \to K^{* +} \tau^- \bar \nu_\tau$ (top-right panel) processes in the presence of $V_L$  in  Fig. \ref{Bkstar-br}. The corresponding plots in the bottom panel of this figure are for  $V_R$ coupling.  In the presence  of $V_R$ coupling, we found reasonable deviation of the branching ratios from  the SM predictions, whereas $V_L$ affects mainly  $\bar B_s \to K^{* +} \tau^- \bar \nu_\tau$ process.
\begin{figure}[h]
\centering
\includegraphics[scale=0.4]{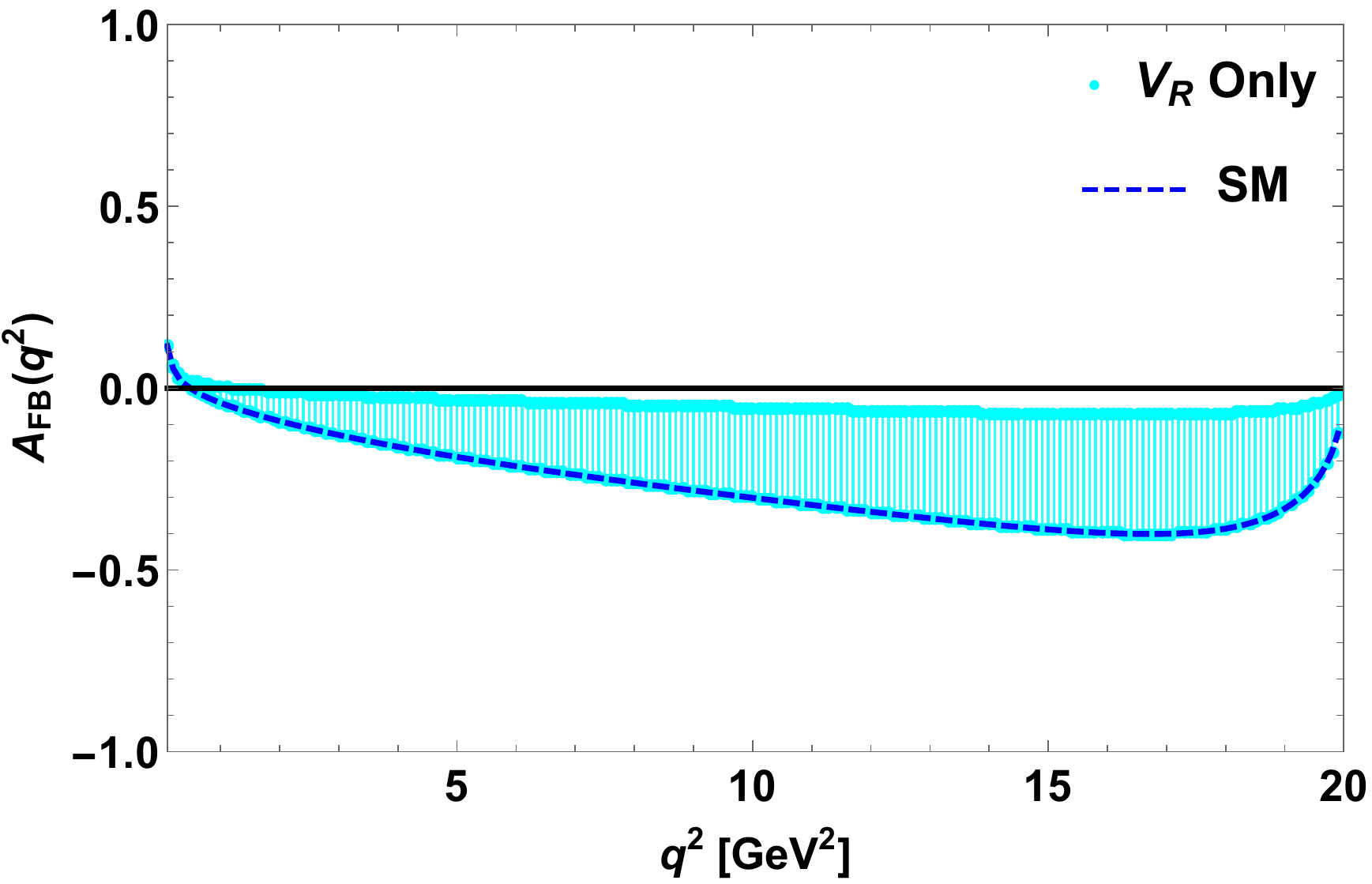}
\quad
\includegraphics[scale=0.4]{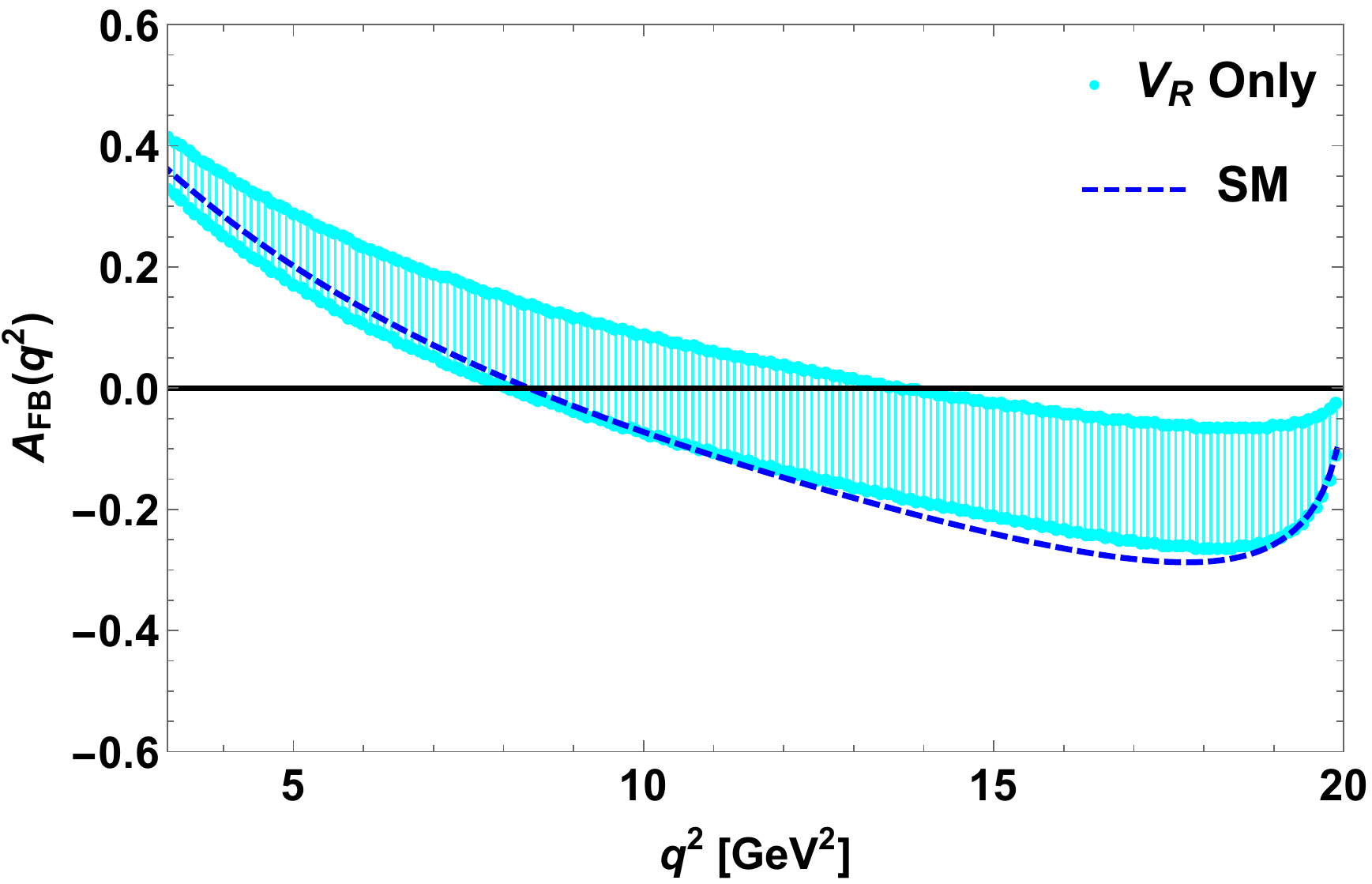}
\quad
\includegraphics[scale=0.4]{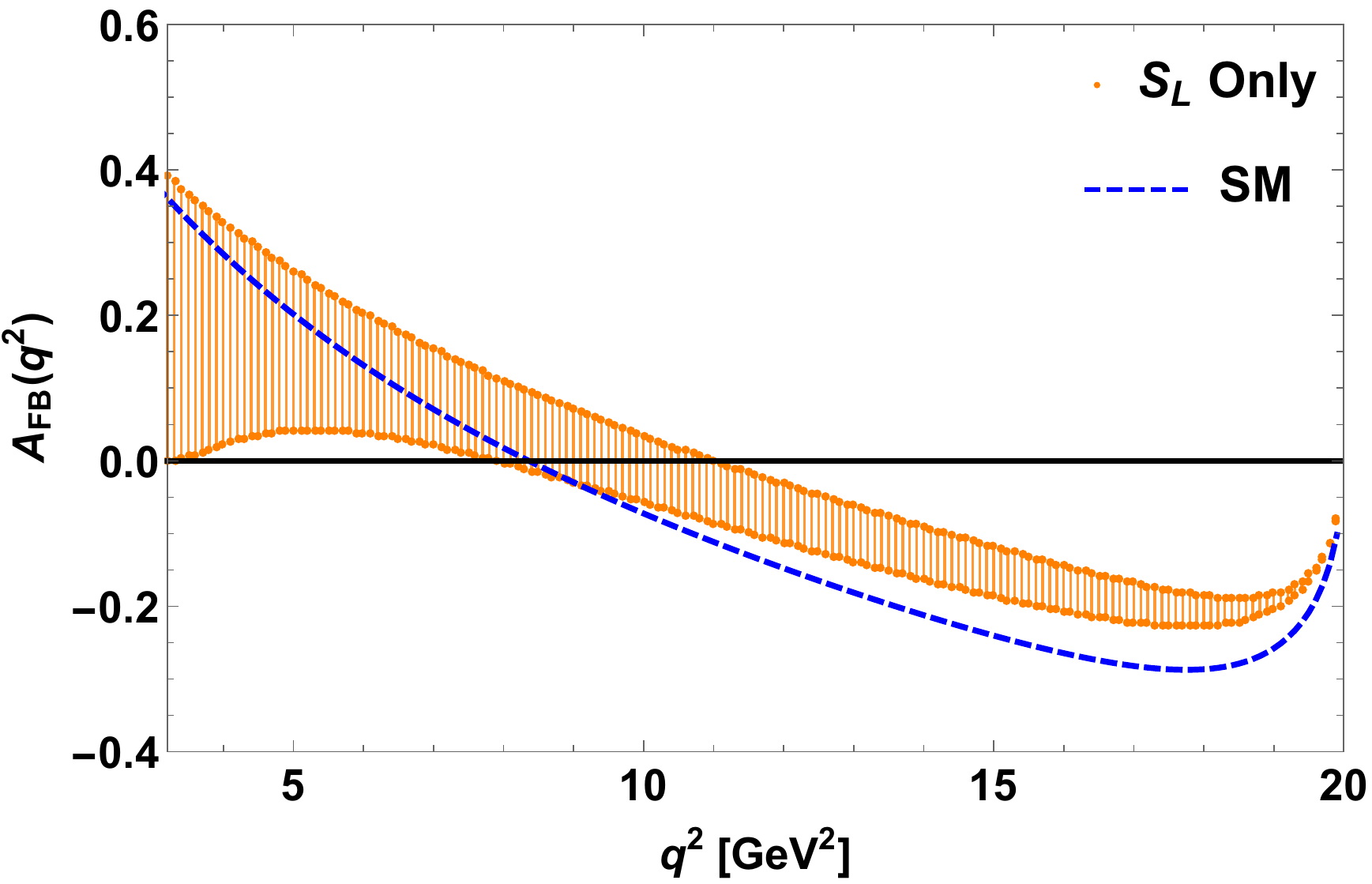}
\quad
\includegraphics[scale=0.4]{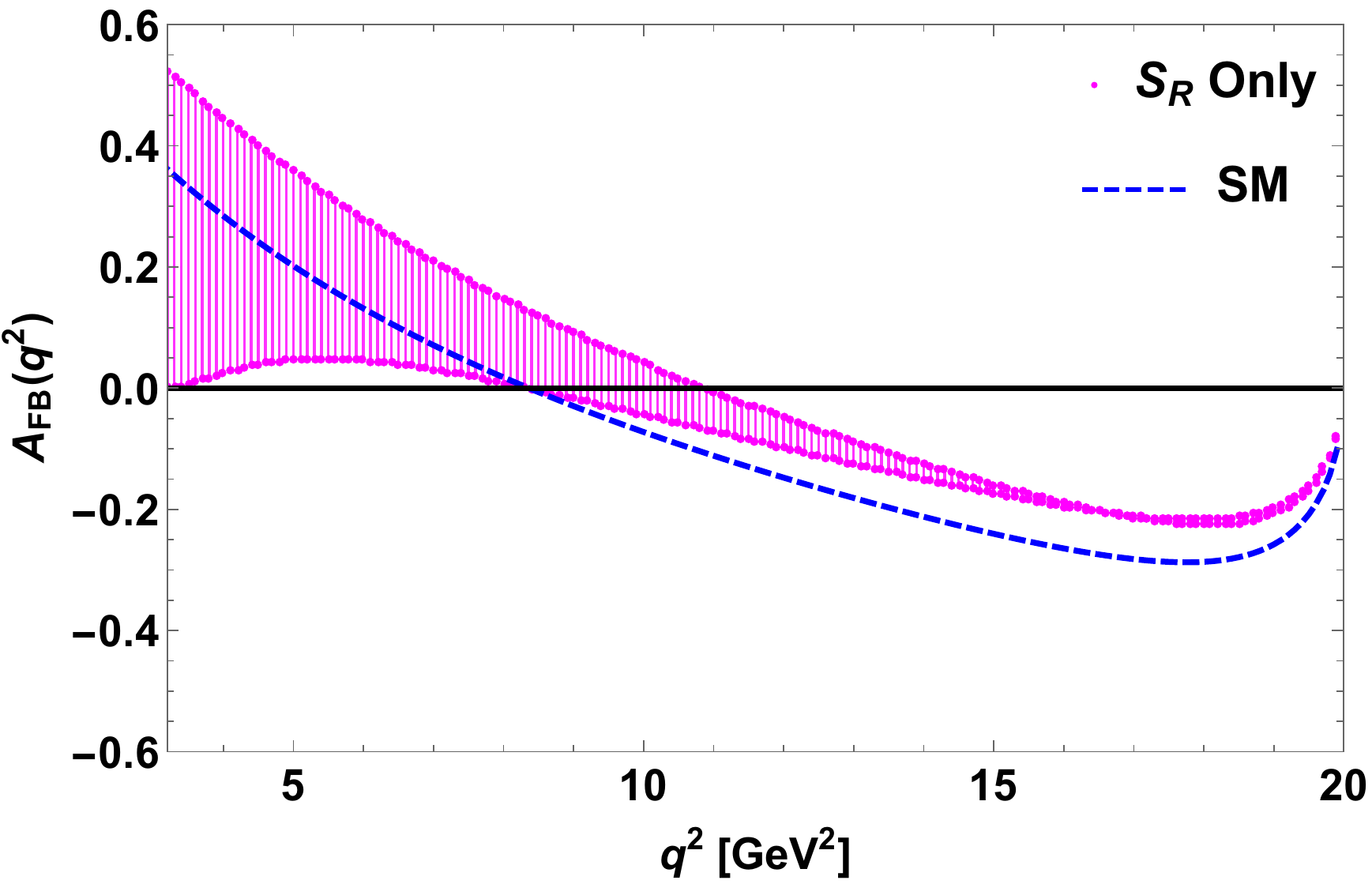}
\caption{The plots for the $q^2$ variations of the forward-backward asymmetry of $\bar B_s \to K^{* +} \tau^- \bar \nu_\tau$  processes for only $V_R$ (top-right panel), $S_L$ (bottom-left panel) and $S_R$ (bottom-right panel) couplings. The top-left panel represents the  plots for the forward-backward asymmetry of  $\bar B_s \to K^{* +} \mu^- \bar \nu_\mu$  processes for only $V_R$ coupling.}\label{Bkstar-fb}
\end{figure}
In the top-left panel of Fig. \ref{Bkstar-fb}, we show the $q^2$ variation of  forward-backward asymmetries of $B_s \to K^{* +} \mu^- \bar \nu_\mu$ processes  for $V_R$ coupling. The forward-backward asymmetry of  $B_s \to K^{* +} \tau^- \bar \nu_\tau$  processes  for $V_R$ (top-right panel), $S_L$ (bottom-left panel) and $S_R$ (bottom-right panel) couplings are presented in Fig. \ref{Bkstar-fb}. We found significant deviation in the forward-backward asymmetry parameters from  SM values due to the additional $V_R$ and $S_{L, R}$ couplings. The presence of $V_L$ coupling does not affect the forward-backward asymmetry parameters. As seen from the figure, due to $S_{L, R}$ couplings, the forward-backward asymmetry of $ \bar B_s \to K^{* +} \tau^- \bar \nu_\tau$ process receives significant deviation from its SM values, whereas  the deviation is negligible for $ \bar B_s \to K^{* +} \mu^- \bar \nu_\mu$ process. The integrated values of the branching ratios and the forward-backward asymmetries for $V_{L, R}$ and $S_{L, R}$ couplings are presented in Table VI and VII respectively.  
 In Fig. \ref{Rkstar}, we present the plots for the $R_{K^*}^{\tau \mu}(q^2)$ parameters for $V_L$ (top-left panel), $V_R$ (top-right panel), $S_L$ (bottom-left panel) and $S_R$ (bottom-right panel) couplings and the corresponding integrated values are presented in Table VIII. 
\begin{figure}[h]
\centering
\includegraphics[scale=0.4]{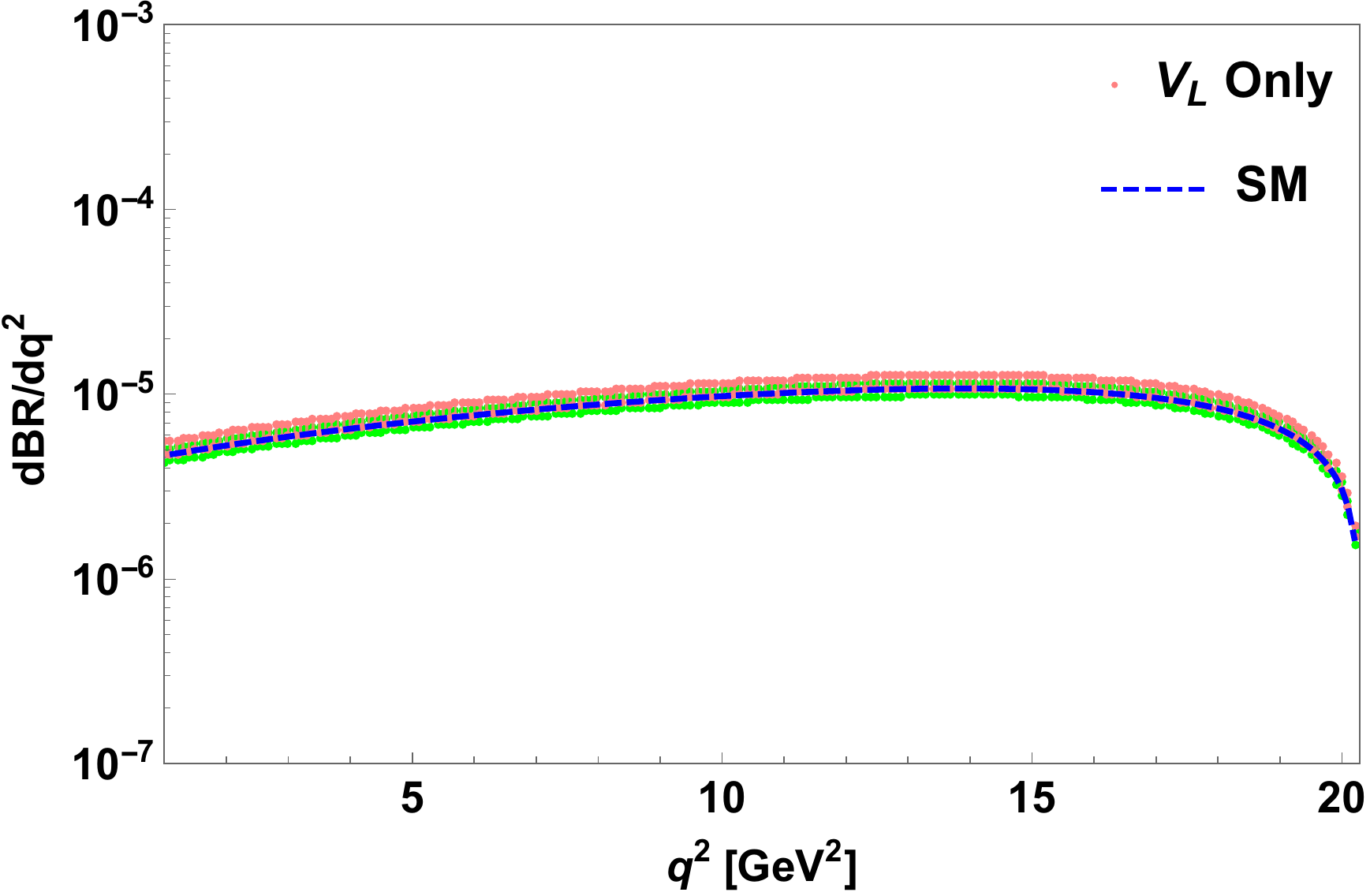}
\quad
\includegraphics[scale=0.4]{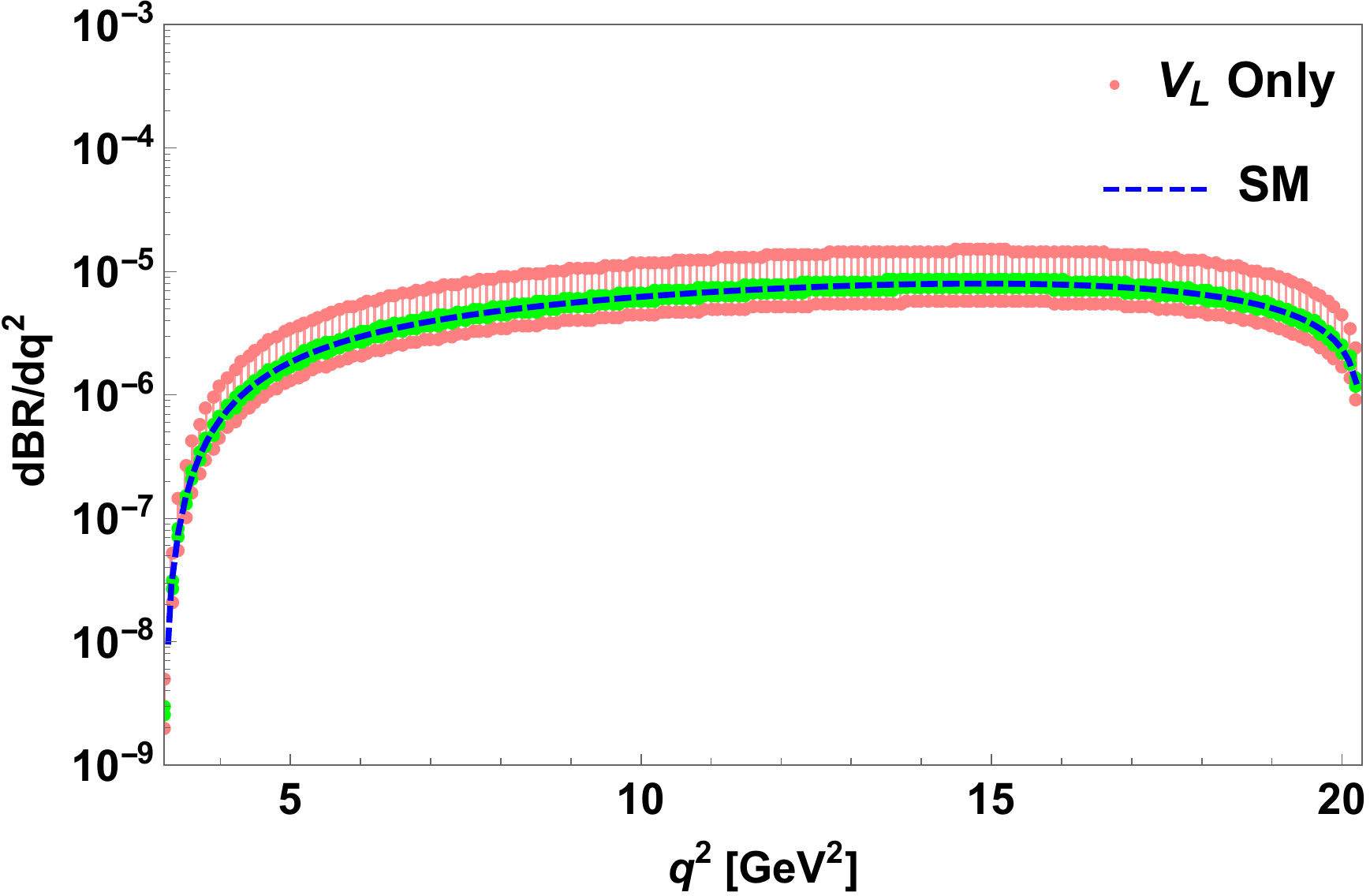}
\quad
\includegraphics[scale=0.4]{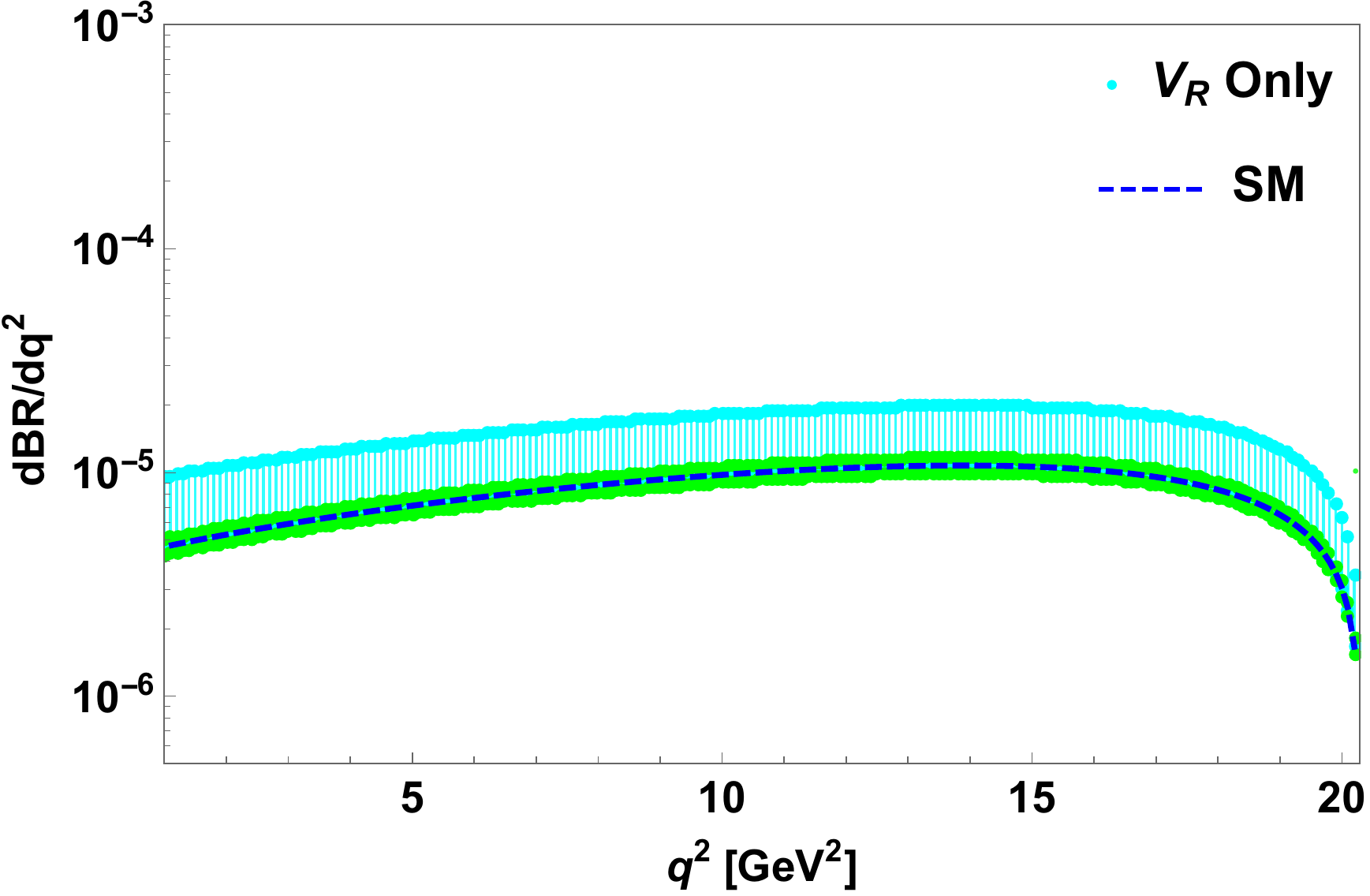}
\quad
\includegraphics[scale=0.4]{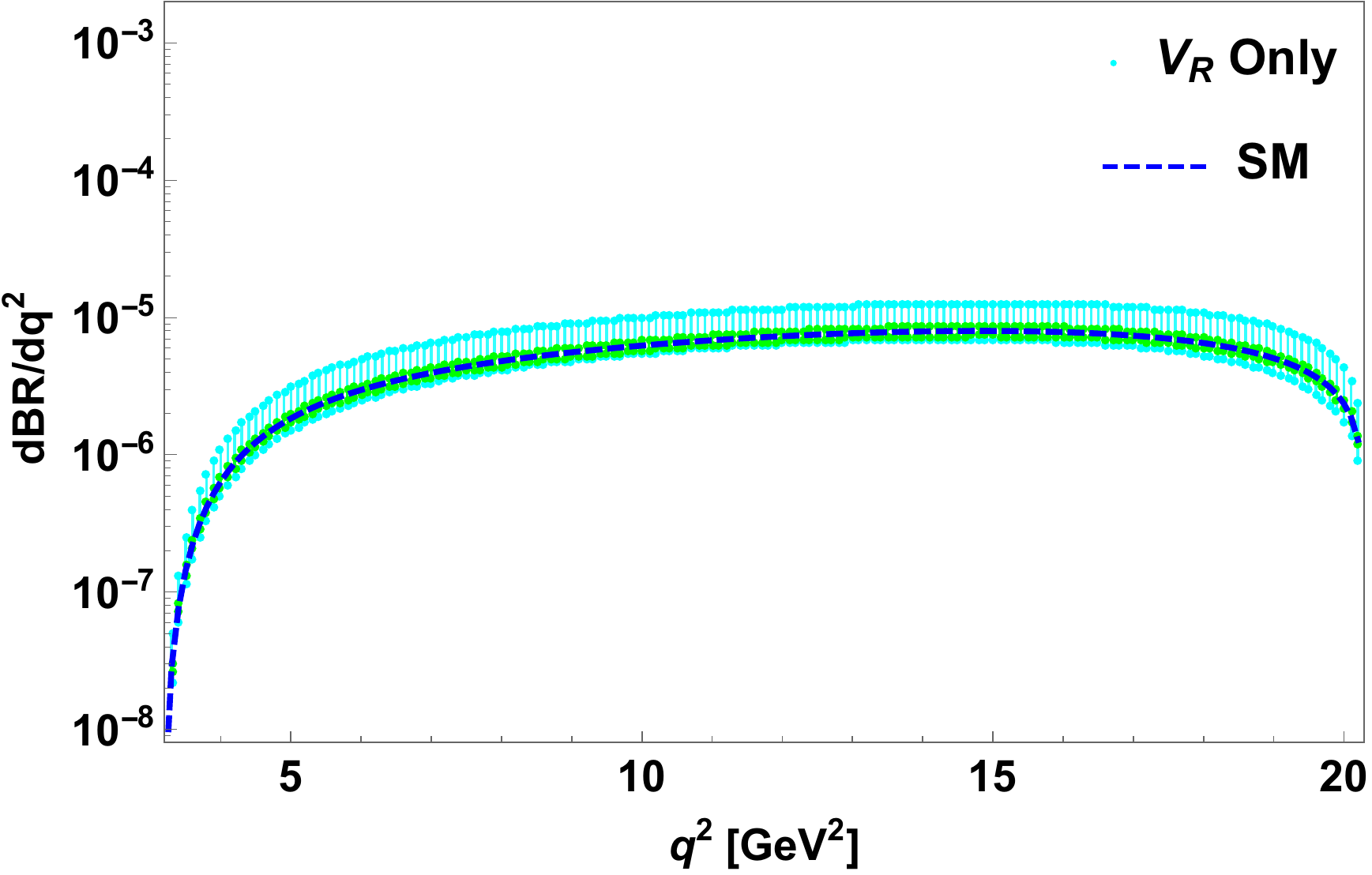}
\caption{Same as Fig. \ref{Bkstar-br} for $B^- \to \rho^0 l^- \bar \nu_l$ processes.} \label{Brho-br}
\end{figure}
\begin{figure}[h]
\centering
\includegraphics[scale=0.4]{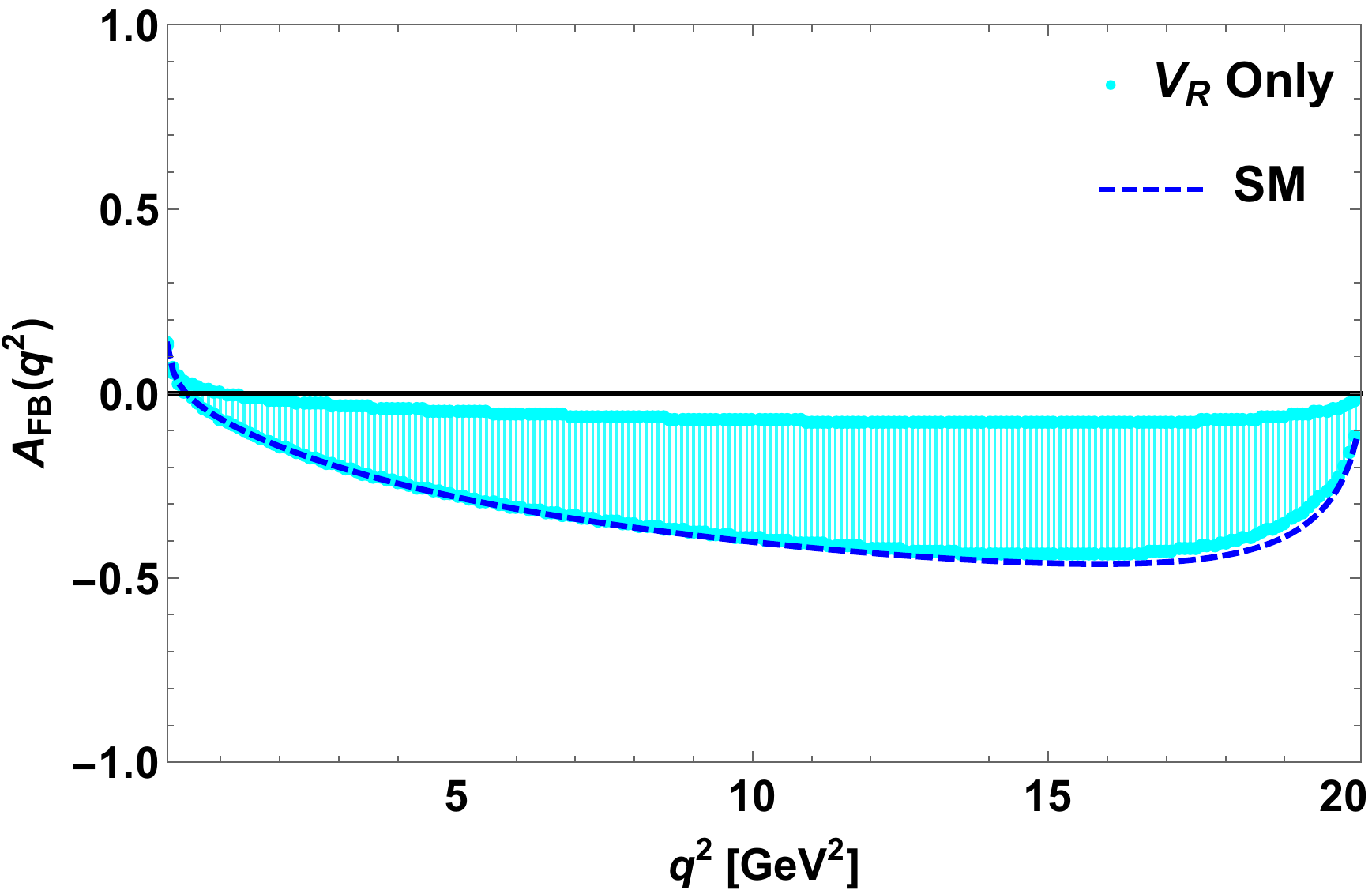}
\quad
\includegraphics[scale=0.4]{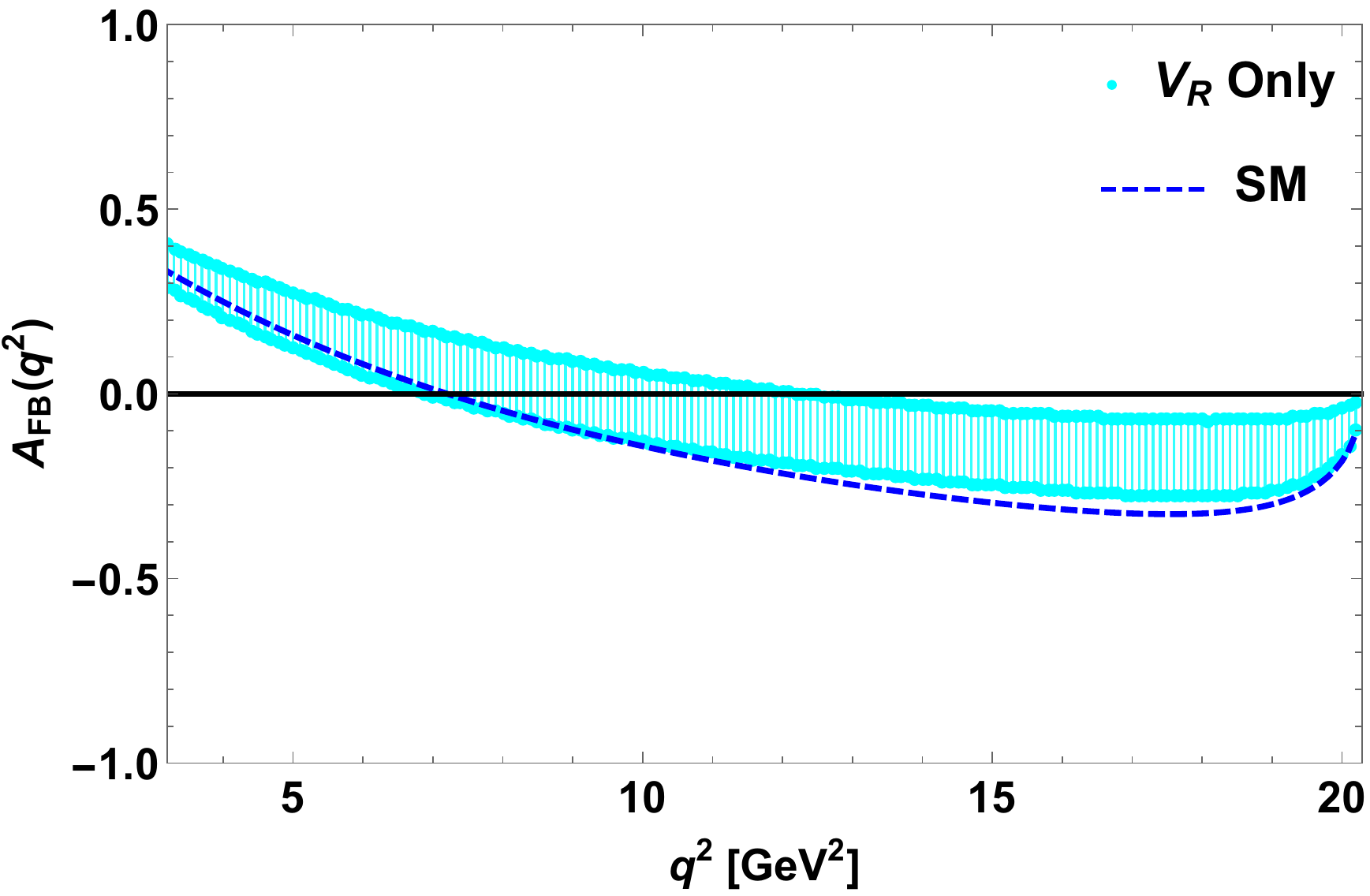}
\quad
\includegraphics[scale=0.4]{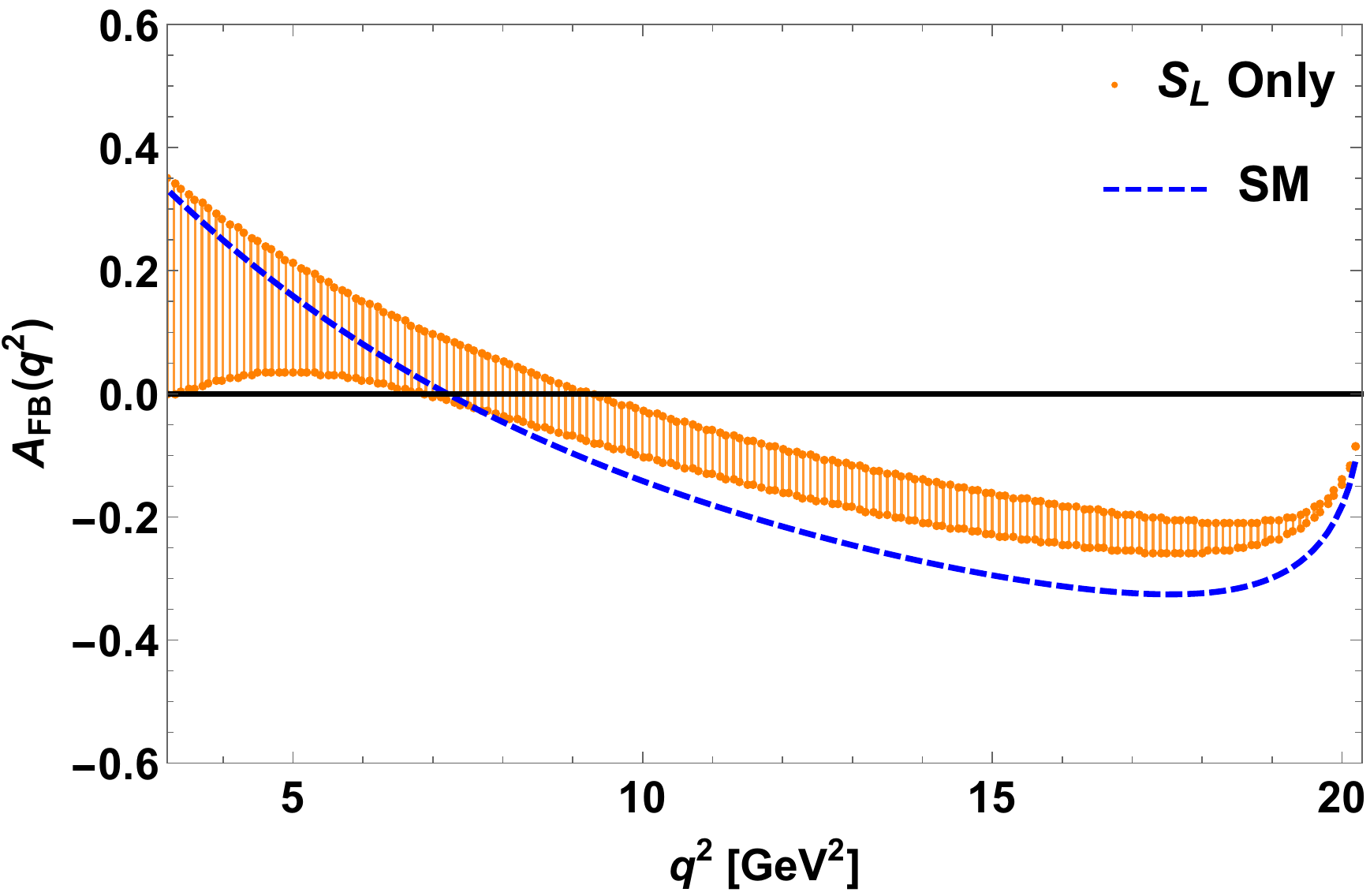}
\quad
\includegraphics[scale=0.4]{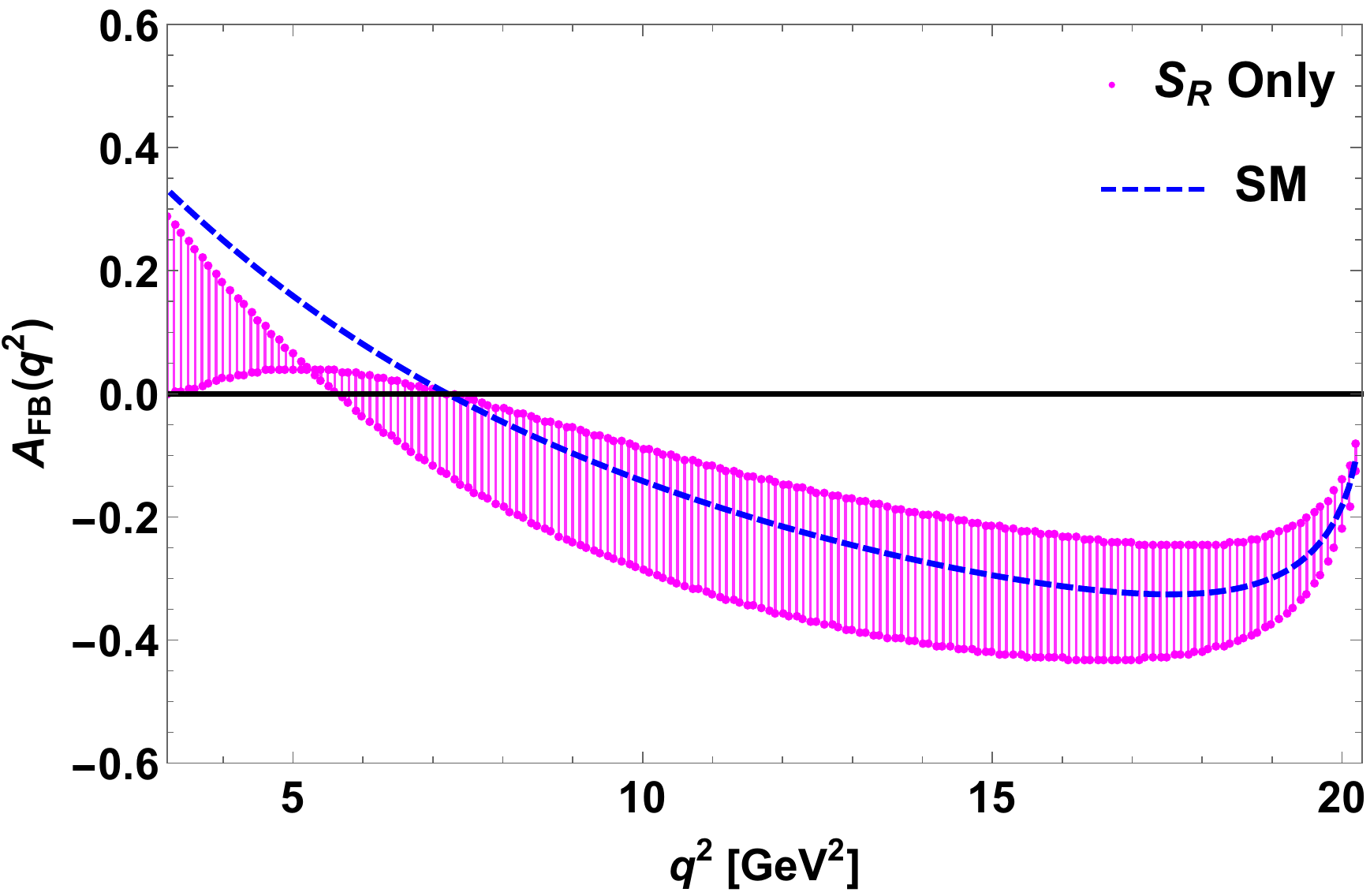}
\caption{Same as Fig. \ref{Bkstar-fb} for $B^- \to \rho^0 l^- \bar \nu_l$ processes.} \label{Brho-fb}
\end{figure}
 \begin{table}[htb]
\begin{center}
\caption{The predicted branching ratios, forward-backward asymmetries  of  $\bar B_{(s)} \to V^+ l^- \bar \nu_l$ processes, where $V=K^*, \rho$ and  $l=\mu, \tau$ in the SM and for the case of  $V_{L, R} $ NP couplings. }
\begin{tabular}{| c |c | c | c| c | }
\hline
Observables~ & Values in the SM~ &~ Values for $V_L$ coupling~&~ Values for $V_R$ coupling~ \\
 \hline
 \hline
 
${\rm BR}(B_s \to K^{*+}  \mu^- \bar \nu_\mu)$   ~&~ $(3.97 \pm 0.32) \times 10^{-4}$~ &~ $(3.97-4.68) \times 10^{-4}$ ~&~ $(3.97-8.05) \times 10^{-4}$ ~\\
 
${\rm BR}(B_s \to K^{* +}  \tau^- \bar \nu_\tau)$ ~ &~ $(2.16 \pm 0.173) \times 10^{-4}$~ &~ $(1.54-4.0) \times 10^{-4}$ ~&~ $(1.92-3.8) \times 10^{-4}$ ~ \\

$\langle A_{FB}^\mu \rangle$ ~&~$-0.293 \pm 0.023$ ~&~$-0.293$~&~$-0.293 \to -0.052$~ \\

$\langle A_{FB}^\tau \rangle$ ~&~$-0.146 \pm 0.012$~&~$-0.146$~&~$-0.138\to 0.037$~ \\

\hline

${\rm BR}( B^- \to \rho^0 \mu^- \bar \nu_\mu)$   ~&~ $(1.56 \pm 0.124) \times 10^{-4}$~ &~ $(1.56-1.85) \times 10^{-4}$ ~&~ $(1.56-3.0) \times 10^{-4}$ ~\\
 
${\rm BR}( B^- \to \rho^0  \tau^- \bar \nu_\tau)$ ~ &~ $(8.97\pm 0.71) \times 10^{-5}$~ &~ $(0.64-1.67) \times 10^{-4}$ ~ &~ $(0.8-1.52) \times 10^{-4}$ ~\\

$\langle A_{FB}^\mu \rangle$ ~&~$-0.362 \pm 0.028$ ~&~$-0.362$~&~$-0.362 \to -0.065$~ \\

$\langle A_{FB}^\tau \rangle$ ~&~$-0.184 \pm 0.015$~&~$-0.184$~&~$-0.168 \to 0.024$~ \\
\hline
\end{tabular}
\end{center}
\end{table}

 \begin{table}[htb]
\begin{center}
\caption{Same as Table VI in the presence of $S_{L, R}$ couplings. }
\begin{tabular}{| c |c |  c| c | }
\hline
Observables~  &~ Values for $S_L$ coupling~&~ Values for $S_R$ coupling~ \\
 \hline
 \hline

${\rm BR}(B_s \to K^{*+}  \mu^- \bar \nu_\mu)$   ~ &~ $(3.97-4.0) \times 10^{-4}$ ~&~ $(3.97-4.0) \times 10^{-4}$ ~\\
 
${\rm BR}(B_s \to K^{* +}  \tau^- \bar \nu_\tau)$ ~  &~ $(2.1-2.58) \times 10^{-4}$ ~&~ $(1.99-2.2) \times 10^{-4}$ ~ \\

$\langle A_{FB}^\mu \rangle$ ~&~$-0.293\to -0.291$~&~$-0.293\to -0.286$~ \\

$\langle A_{FB}^\tau \rangle$ ~&~$-0.169 \to -0.043$~&~$-0.144 \to -0.056$~ \\

\hline

${\rm BR}(B^- \to \rho^0  \mu^- \bar \nu_\mu)$   ~ &~ $(1.57-1.6) \times 10^{-4}$ ~&~ $(1.57-1.6) \times 10^{-4}$ ~\\
 
${\rm BR}( B^- \to \rho^0  \tau^- \bar \nu_\tau)$ ~  &~ $(0.87-1.12) \times 10^{-4}$ ~ &~ $(8-9.2) \times 10^{-5}$ ~\\

$\langle A_{FB}^\mu \rangle$ ~&~$-0.36\to -0.35$~&~$-0.36\to -0.35$~ \\

$\langle A_{FB}^\tau \rangle$ ~&~$-0.21 \to -0.07 $~&~$-0.32 \to -0.18 $~ \\
\hline

\end{tabular}
\end{center}
\end{table}
\begin{figure}[h]
\centering
\includegraphics[scale=0.4]{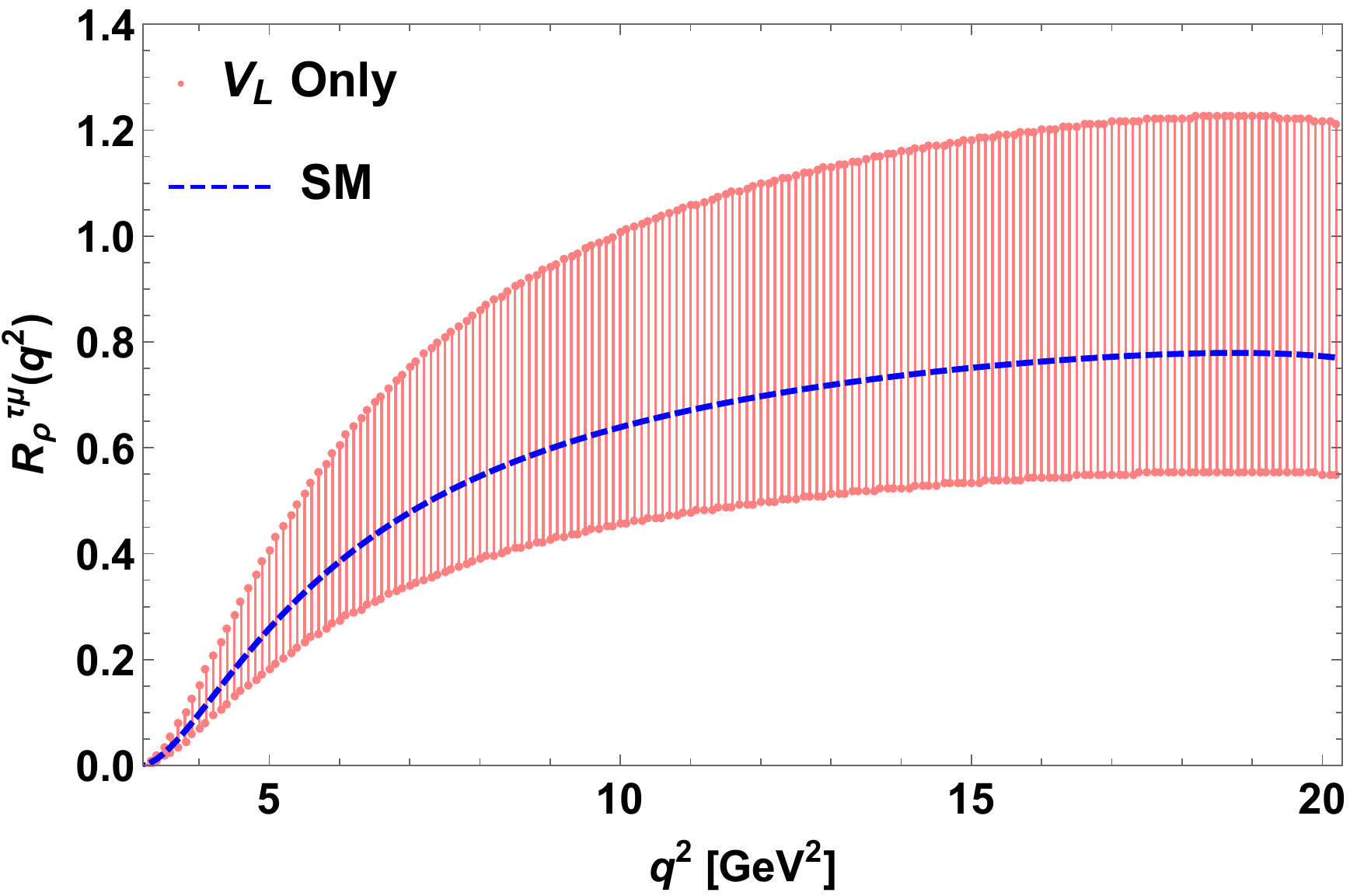}
\quad
\includegraphics[scale=0.4]{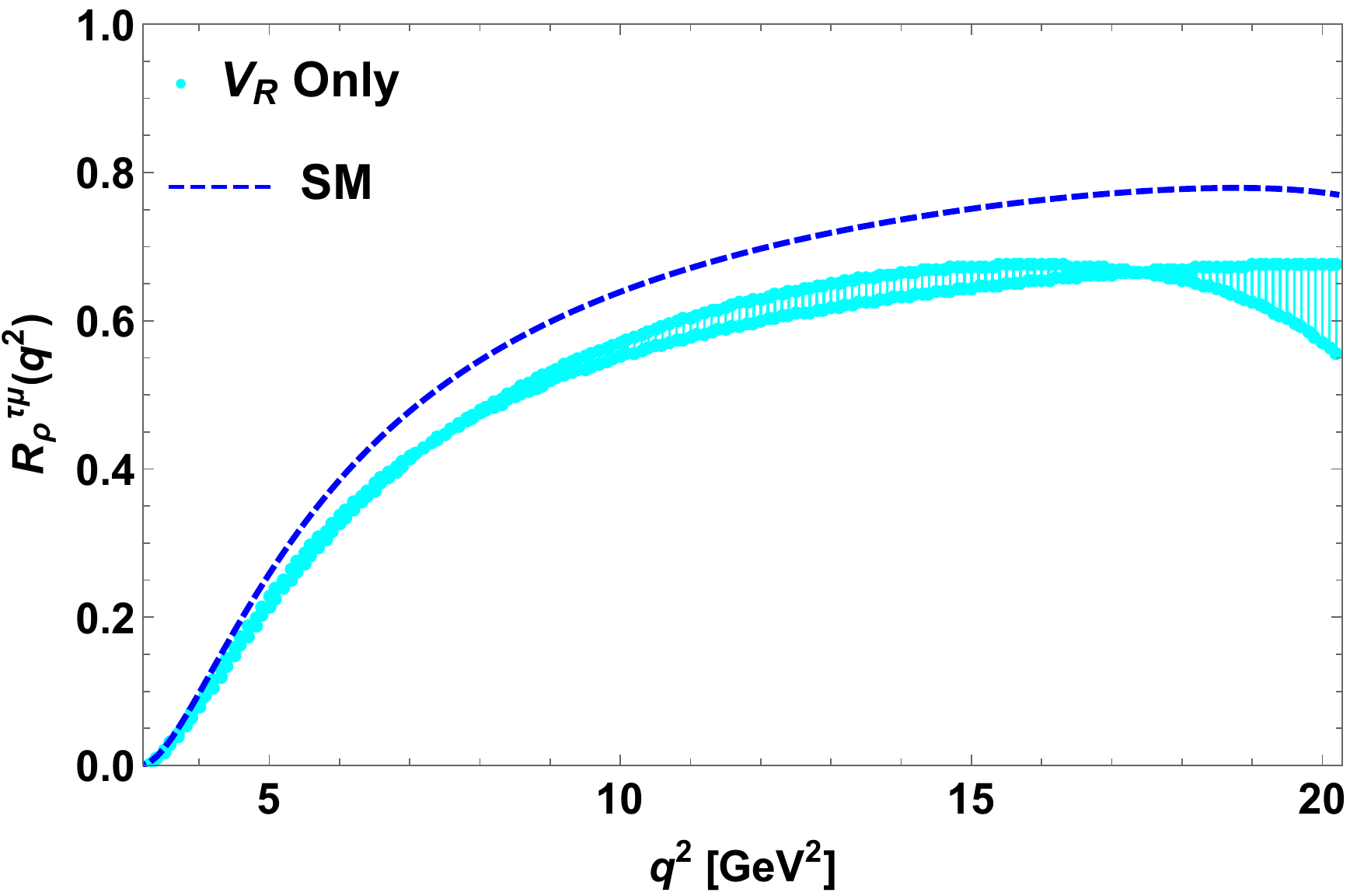}
\quad
\includegraphics[scale=0.4]{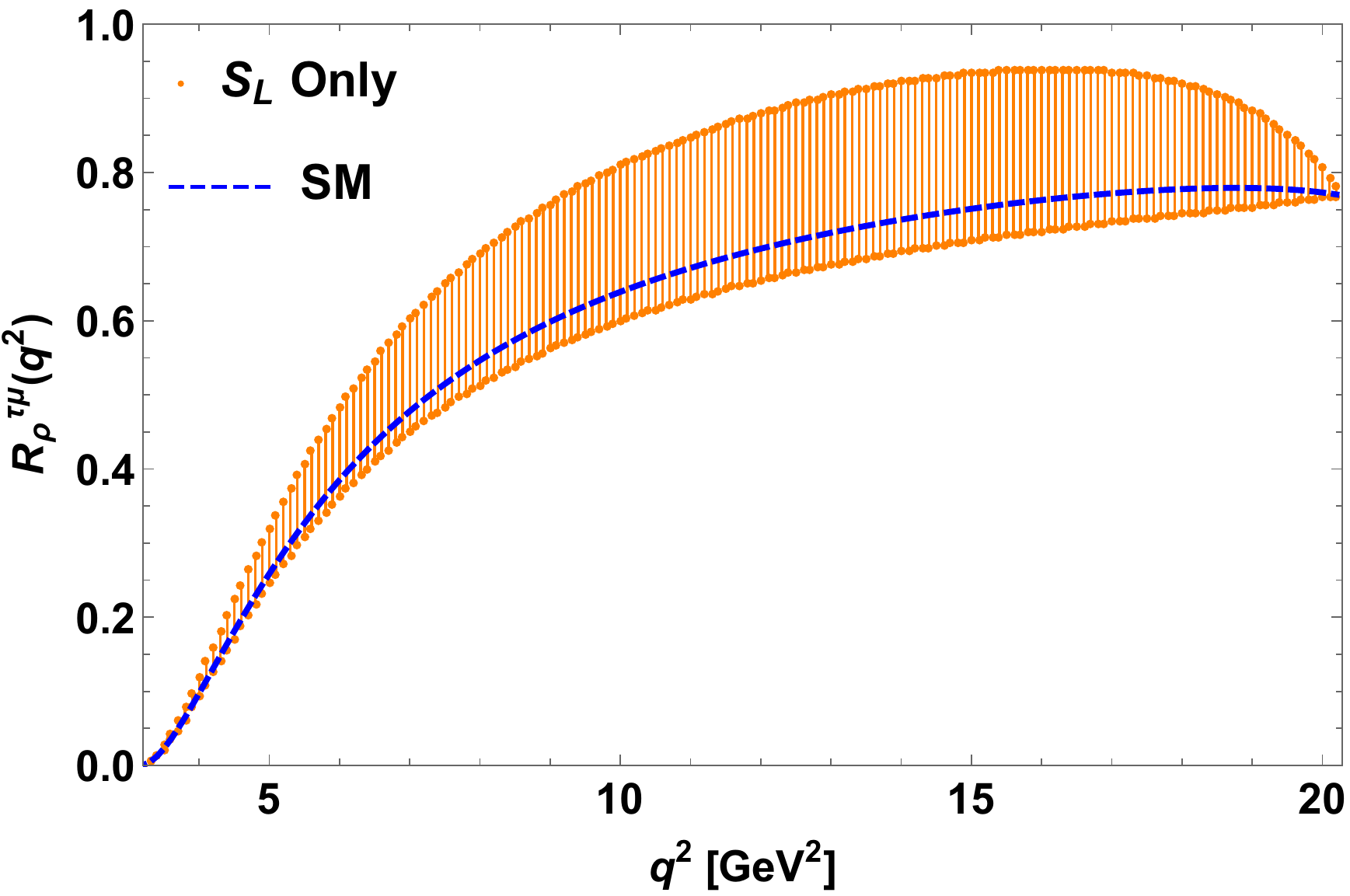}
\quad
\includegraphics[scale=0.4]{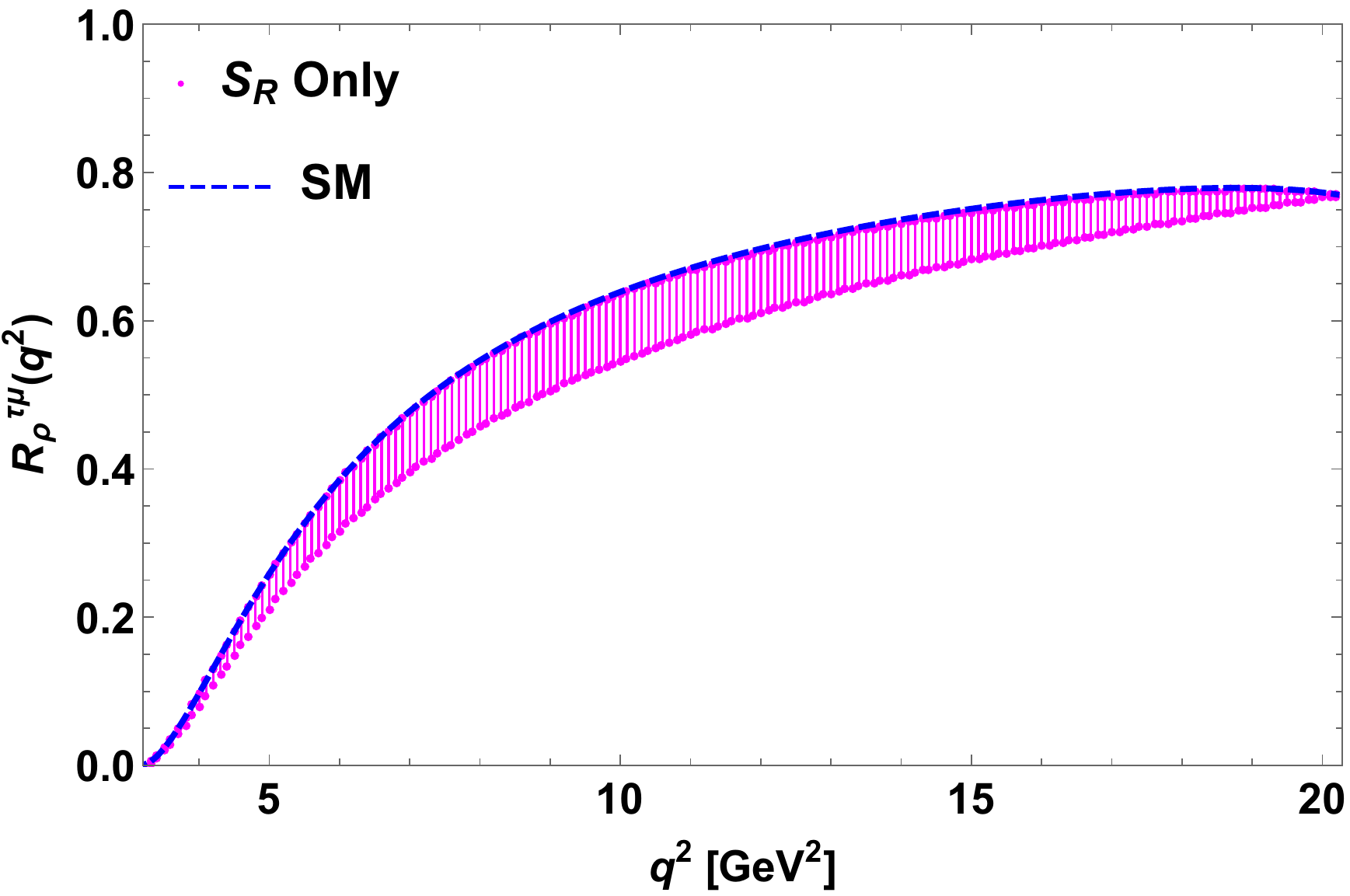}
\caption{Same as Fig. \ref{Rkstar} for $B^- \to \rho^0 l^- \bar \nu_l$ processes.}\label{Rrho}
\end{figure}
\begin{figure}[h]
\centering
\includegraphics[scale=0.4]{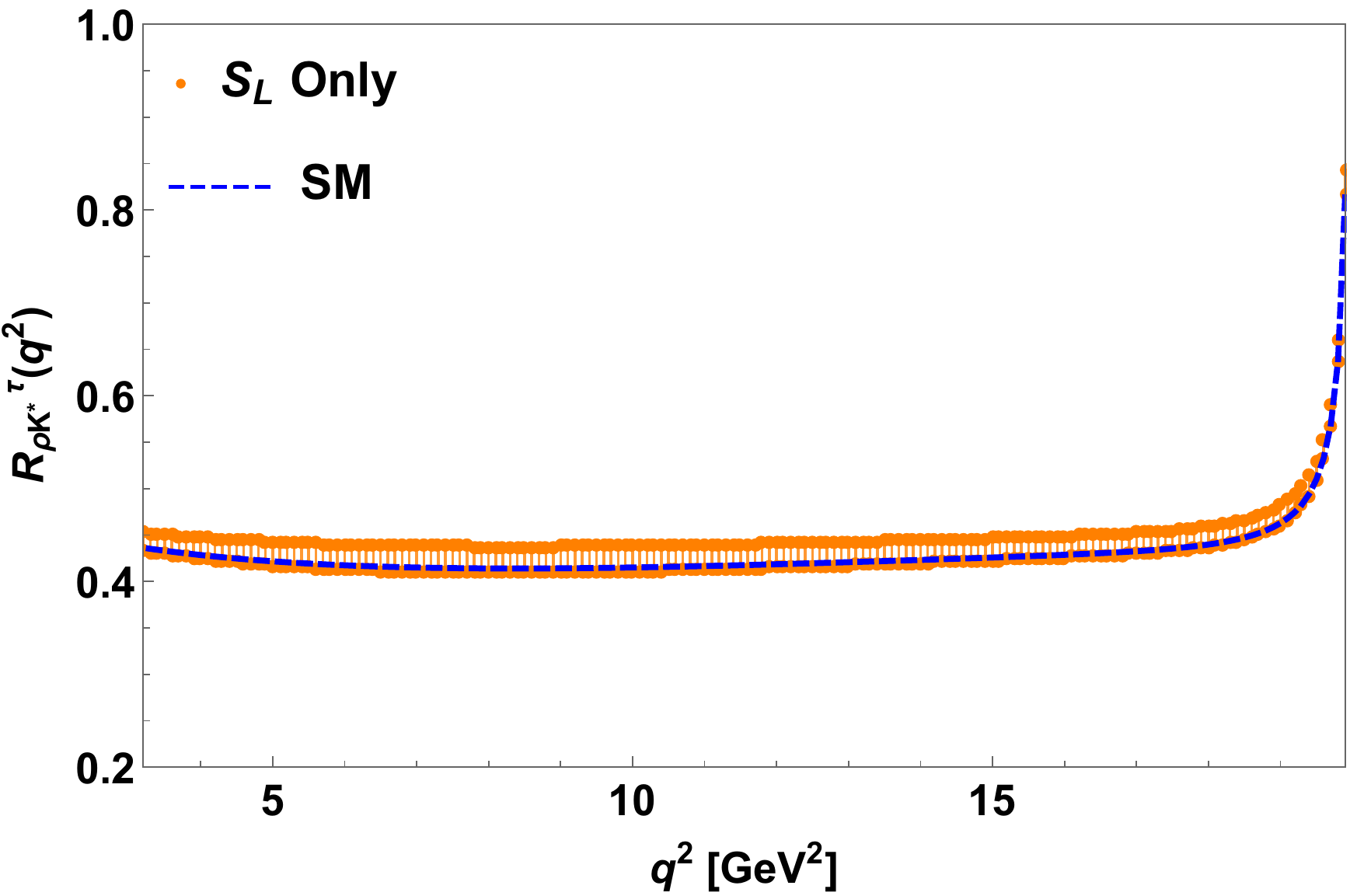}
\quad
\includegraphics[scale=0.4]{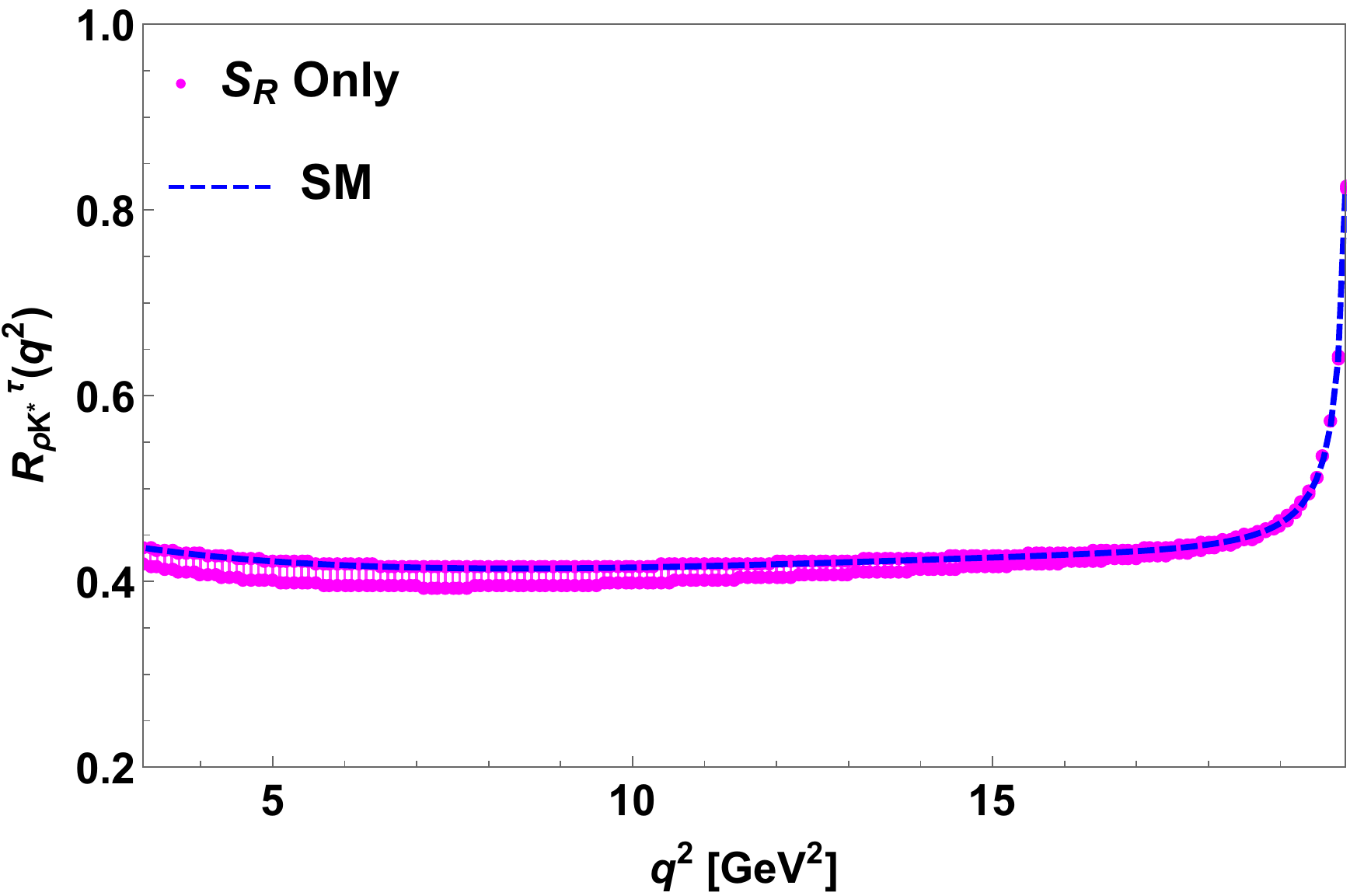}
\caption{The plots for $R_{\rho K^*}^\tau (q^2)$  parameters verses $q^2$  for only  $S_L$ (left panel) and $S_R $ (right panel)  couplings.  }\label{Rrhokstar}
\end{figure}
\begin{table}[htb]
\begin{center}
\caption{Values of  $R_{K^*}^{\tau \mu}$, $R_{\rho}^{\tau \mu}$, $R_{\rho K^*}^{\mu}$ and  $R_{\rho K^*}^{\tau}$  parameters for different cases of NP couplings.}
\begin{tabular}{|c | c| c| c| c|}
\hline
Model~&~$R_{K^*}^{\tau \mu}$~& ~$R_\rho^{\tau \mu}$~&~$R_{\rho K^*}^\mu$ ~& ~$R_{\rho K^*}^\tau$~ \\
\hline
\hline

SM~&~$0.544$~&~$0.573$~&~$0.393$~&$0.415$~ \\

$V_L$~&~$0.388-0.856$~&~$0.41-0.9$~&~$0.393-0.395$~&~$0.415-0.42$ \\

$V_R$~&~$0.47-0.474$~&~$0.5-0.51$~&~$0.373-0.393$~&~$0.4-0.42$ \\

$S_L$~&~$0.522-0.646$~&~$0.542-0.712$~&~$0.393-0.4$~&~$0.414-0.434$ \\

$S_R$~&~$0.497-0.544$~&~$0.5-0.573$~&~$0.393-0.4$~&~$0.4-0.42$ \\
\hline

\end{tabular}
\end{center}

\end{table}
The $q^2$ variation of  the branching ratios   of $\bar B \to \rho^+ l^- \bar \nu_l$ processes for $V_{L,R}$ couplings are presented in Fig. \ref{Brho-br}.  In the presence of $S_{L, R}$ couplings, the branching ratios of $\bar B \to \rho^+ l^- \bar \nu_l$ processes  have negligible deviation from the SM predictions.  The  predicted  values of the branching ratios  of these processes  are given in Table  VI and VII  respectively.  The experimental  branching ratio of $B^+ \to \rho^0 l^+ \nu_l$ process is \cite{pdg}
\bea
{\rm BR}(B^+ \to \rho^0 l^+ \nu_l)^{\rm Expt} &=& (1.58 \pm 0.11) \times 10^{-4}. 
\eea
Our predicted results for $B^- \to \rho^0 \mu^- \bar \nu_\mu$ process is consistent with the above experimental data (though a part of the allowed parameter space of $V_{L,R}$ and $S_{L, R}$ give  values on the higher side of the observed central value).
The forward-backward asymmetry plots for $\bar B \to \rho^+ l^- \bar  \nu_l$ are presented in Fig. \ref{Brho-fb} and the corresponding numerical values are given in Table VI and VII.  Fig. \ref{Rrho} represents the  plots of LNU parameter $R_{\rho}^{\tau \mu}(q^2)$ for $V_L$ (top-left panel), $V_R$ (top-right panel), $S_L$ (bottom-left panel) and $S_R$ (bottom-right panel) couplings.
In Fig. \ref{Rrhokstar}, we show the variation of the parameter $R_{\rho K^*}^\tau (q^2)$ with respect to $q^2$ for only $S_L$ (left panel) and $S_R$ (right panel) couplings. The integrated values of these  parameters  are given in Table VIII. The additional $V_{L, R}$ couplings don't affect the $R_{\rho K^*}^l$  parameters. 

In the literature, the $B \to V l \nu_l$ processes are investigated in both model-dependent and independent ways \cite{Dutta1, Feldmann}.  Our findings on these processes are consistent with these predictions. 

\section{Conclusion}

Inspired by the recent measurement of $R_{K^{(*)}}$ parameter at LHCb and the observed $R_{D^{(*)}}$ anomalies in $b \to s l^+ l^-$ and $b \to c l \bar \nu_l$ processes,    we performed a model independent analysis of  the rare semileptonic $b \to u l \bar{\nu}_l$ processes in this paper. We considered the generalized effective Lagrangian in the presence of  new physics, which contributes additional   coefficients to the SM. In our work the new coefficients are considered to be complex and we have taken into account the effect of one Wilson coefficient at a time to compute the allowed parameter space  of these new  coefficients.  Using the experimental branching ratios of $B_u^+ \to \tau^+ \nu_\tau$ and $B^- \to \pi^0 \mu^- \bar \nu_\mu$ processes, we have  constrained the new couplings.  We  then calculated the branching ratios, forward-backward asymmetries of $B \to P l \bar \nu_l$ processes, where  $P=K, \pi, \eta^{(\prime)}$ for all possible cases of new couplings.  In the presence of $V_{L, R}$ couplings, we found reasonable deviation in the branching ratios  of these processes from the corresponding SM predictions, but the corresponding forward-backward asymmetry parameters don't show any deviation. In the case of $S_{L, R}$ couplings, the branching ratios have slight  deviation from the SM predictions. However, the forward-backward asymmetry parameters have comparatively large deviations from the SM values.    We then computed the lepton non-universality parameters, in order to test the presence of the violation of lepton universality in $b \to u l \bar \nu_l$ processes.

 Besides the semileptonic decays of $B$ meson to a pseudoscalar meson, we also studied the $B \to V l \bar \nu_l$ processes, where $V$ is a vector meson and $V=K^*, \rho$. We calculated the branching ratios, 
 forward-backward asymmetries and the lepton non-universality parameters for these processes.    The presence of additional $V_{L, R}$ Wilson coefficients result larger deviation in the branching ratios and other observables in the  $B \to V l \bar \nu_l$ processes. The effect of $S_{L, R}$ couplings on branching ratios of these processes is almost negligible. However, the forward-backward asymmetry of $B \to V \tau \bar \nu_\tau$ process deviates significantly  from  SM.   We also observe that, the rare semileptonic $b \to u l \bar \nu_l$ processes also violate the lepton flavour universality. Thus, the study of $b \to u l \bar \nu_l$ processes are  necessary in both  theoretical and experimental point of view in order to search new physics.

{\bf Acknowledgments}

SS and RM  would like to thank Science and Engineering Research Board (SERB), Government of India for financial support through grant No. SB/S2/HEP-017/2013. AR   acknowledges  University Grants Commission for financial support.

\end{document}